%% file: counts_planck.tex
\documentclass[traditabstract,longauth]{aa} 

\input PlanckCommands.tex

\usepackage{graphicx,natbib,pjournal,epsfig} 
\usepackage[breaklinks, colorlinks, citecolor=blue]{hyperref} 
\usepackage{color} 
\usepackage{float} 
\usepackage[english]{babel} 
\usepackage{txfonts} 
\usepackage{url} 
 
\begin{document} 
 
\input{PIP_38_authors_and_institutes.tex}

\title{\textit{Planck\/} intermediate results}

\subtitle{VII. Statistical properties of infrared and radio
  extragalactic sources \\ from the \textit{Planck\/} Early Release
  Compact Source Catalogue at frequencies between 100 and
  857\,GHz\thanks{online material at \url{http://www.aanda.org} and
    counts at
    \url{http://www.ias.u-psud.fr/irgalaxies/planck_hfi_counts_2013/}}\thanks{corresponding
    author \email{herve.dole@ias.u-psud.fr}} }

\date{Submitted 19-Jul-2012 / Accepted 27-Nov-2012} 

\abstract{ We make use of the \Planck\ all-sky survey to derive number
  counts and spectral indices of extragalactic sources -- infrared and
  radio sources -- from the \Planck\ Early Release Compact Source
  Catalogue (ERCSC) at 100 to 857\,GHz (3\,mm to 350\micron). Three
  zones (deep, medium and shallow) of approximately homogeneous
  coverage are used to permit a clean and controlled correction for
  incompleteness, which was explicitly not done for the ERCSC, as it
  was aimed at providing lists of sources to be followed up.  Our
  sample, prior to the 80\,\% completeness cut, contains between 217
  sources at 100\,GHz and 1058 sources at 857\,GHz over about 12,800
  to 16,550\,deg$^2$ (31 to 40\,\% of the sky). After the 80\,\%
  completeness cut, between 122 and 452 and sources remain, with flux
  densities above 0.3 and 1.9\,Jy at 100 and 857\,GHz. The sample so
  defined can be used for statistical analysis.  Using the
  multi-frequency coverage of the \Planck\ High Frequency Instrument,
  all the sources have been classified as either dust-dominated
  (infrared galaxies) or synchrotron-dominated (radio galaxies) on the
  basis of their spectral energy distributions (SED). Our sample is
  thus complete, flux-limited and color-selected to differentiate
  between the two populations. We find an approximately equal number
  of synchrotron and dusty sources between 217 and 353\,GHz; at
  353\,GHz or higher (or 217\,GHz and lower) frequencies, the number
  is dominated by dusty (synchrotron) sources, as expected.  For most
  of the sources, the spectral indices are also derived.  We provide
  for the first time counts of bright sources from 353 to 857\,GHz and
  the contributions from dusty and synchrotron sources at all HFI
  frequencies in the key spectral range where these spectra are
  crossing.  The observed counts are in the Euclidean regime. The
  number counts are compared to previously published data (from
  earlier \Planck\ results, {\it Herschel}, BLAST, SCUBA, LABOCA, SPT,
  and ACT) and models taking into account both radio or infrared
  galaxies, and covering a large range of flux densities. We derive
  the multi-frequency Euclidean level -- the plateau in the normalised
  differential counts at high flux-density -- and compare it to {\it
    WMAP}, {\it Spitzer\/} and {\it IRAS\/} results. The submillimetre
  number counts are not well reproduced by current evolution models of
  dusty galaxies, whereas the millimetre part appears reasonably well
  fitted by the most recent model for synchrotron-dominated sources.
  Finally we provide estimates of the local luminosity density of
  dusty galaxies, providing the first such measurements at 545 and
  857\,GHz.}
\keywords{Cosmology: observations -- Surveys -- Galaxies: statistics - 
  Galaxies: evolution -- Galaxies: star formation -- Galaxies: active} 
 
\titlerunning{\Planck\ statistical properties of extragalactic IR and radio 
ERCSC sources 100--857\,GHz}\authorrunning{Planck Collaboration}

\maketitle 
 

\section{Introduction}

\begin{figure}[!ht] 
   \centering 
  \includegraphics[width=0.30\textwidth,angle=90]{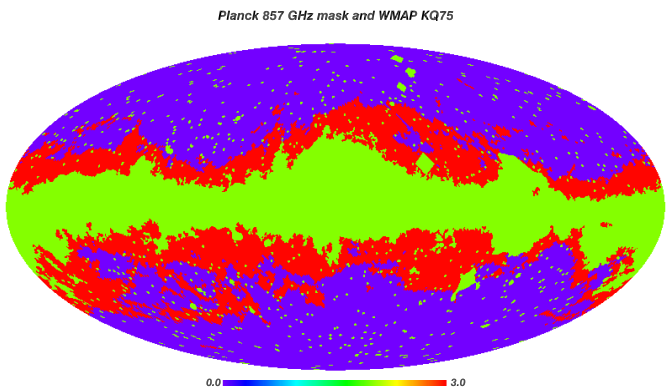} 
  \caption{Comparison between our \Planck\ 857\,GHz mask (red indicates 
    regions removed from the analysis) and the {\it WMAP} 7-year KQ75 mask (green 
    means removed); unlike the case for the mask employed in this 
    paper, the {\it WMAP} mask excludes some point sources. The 
    background (blue) is the sky area used for our analysis. This map 
    is a Mollweide projection of the sky in Galactic coordinates.} 
   \label{fig:mask857} 
\end{figure}

Among other advantages, all-sky multifrequency surveys have the
benefit of probing rare and/or bright objects in the sky. One reason
to probe bright objects is to study the number counts of extragalactic
sources and their spectral shapes.  In the far-infrared (FIR) the
sources detected by these surveys are usually dominated by
low-redshift galaxies with $z<0.1$, as found by {\it IRAS} at 60\micron\
\citep{ashby96} but a few extreme objects like the lensed F10214 source
\citep{rowan-robinson91} also appear.  However, the
population in the radio band is dominated by synchrotron sources (in
particular, blazars) at higher redshift \citep[see][for a recent
review]{de_zotti2010}.  Previous multifrequency all-sky surveys were
carried out in the infrared (IR) range by the {\it IRAS} satellite
(between 12 and 100\micron; \citealt{neugebauer84}), and more recently
by {\it Akari} (between 2 and 180\micron; \citealt{murakami2007}) and
{\it WISE} (between 3.4 and 22\micron; \citealt{wright2010}). Early, limited sensitivity
surveys were carried out in the IR and microwave range by {\it COBE} (between
1.2\micron\ and 1\,cm; \citealt{boggess92}), and more recently in the
microwave range by {\it WMAP} (between 23 and 94\,GHz;
\citealt{bennett2003,wright2009,massardi2009,de_zotti2010}).  

\Planck's\ all-sky, multifrequency surveys offer several advantages to
all of the above.  The frequency range covered is wide, extending from
30 to 857\,GHz; observations at all frequencies were made
simultaneously, reducing the influence of source variability; the
calibration is uniform; and the delivered catalogue of sources used in
this paper was carefully constructed and validated.

\begin{figure}[!ht] 
   \centering 
  \includegraphics[width=0.30\textwidth,angle=90]{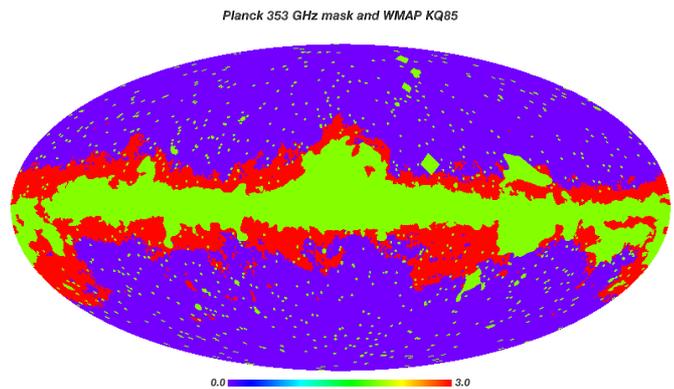} 
  \caption{Comparison between our \Planck\ 353\,GHz mask (red means 
    removed) and {\it WMAP} 7-year KQ85 mask (green). The 
    background (blue) is the sky area used for the analysis.} 
   \label{fig:mask353} 
\end{figure}

\begin{figure}[!hb] 
   \centering 
  \includegraphics[width=0.50\textwidth]{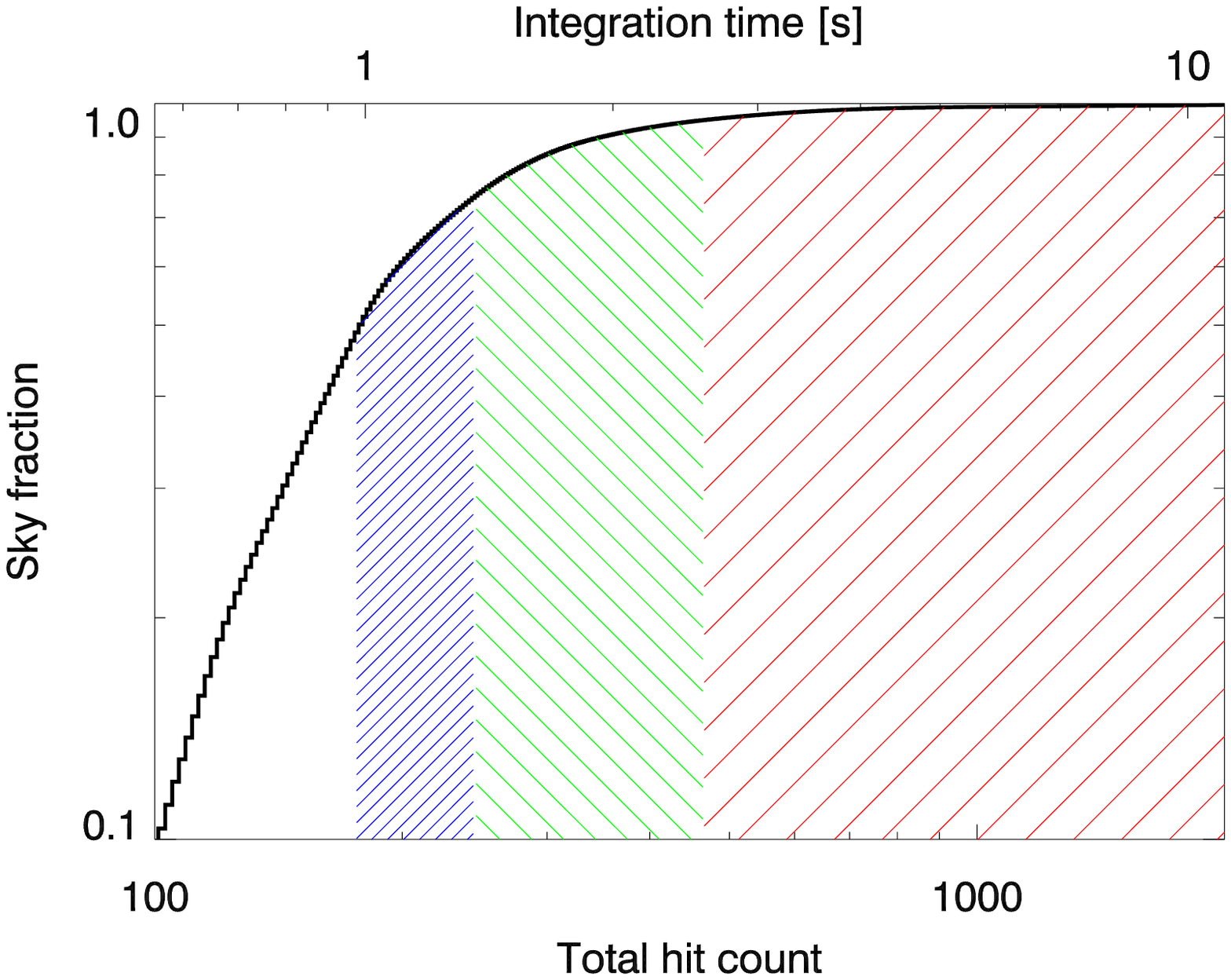} 
  \caption{Cumulative distribution of \Planck\ hit counts on the sky
    (here at 857\,GHz with Nside=2048), with the corresponding
    integration time per sky pixel (given on the top axis). The three
    sky zones used in the analysis are defined at 857\,GHz: shallow
    (50 to 75\,\% of the hit count distribution, short-spaced lines,
    blue); medium (75 to 95\,\% of the hit counts, medium-spaced
    lines, green); and deep (95\,\% and above hits, widely-spaced
    lines, red).}
   \label{fig:hitcounts} 
\end{figure}

The \Planck\ frequency range fully covers the transition between the
dust emission dominated regime (tracing star formation), and the
synchrotron regime (tracing active galactic nuclei). The statistical
analysis of the populations in this spectral range has never been done
before. At large flux densities (typically 1\,Jy and above), number
counts from all-sky surveys of extragalactic FIR sources show a
Euclidean component, i.e. a distribution of the number of source per
flux density bin $S_{\nu}$ at observed frequency $\nu$ ($dN/dS_{\nu}$
in Jy$^{-1}$\,sr$^{-1}$) proportional to $S_{\nu}^{-2.5}$ (see
Eq. \,\ref{eq:p}).  This result is in line with expectations from a
local Universe uniformly filled with non-evolving galaxies
\citep{lonsdale89,hacking91,bertin97,massardi2009,wright2009}. In the
radio range, the Euclidean part is modified by the presence of
higher-redshift sources. At flux densities smaller than typically 0.1
to 1\,Jy, an excess in the number counts compared to the Euclidean
level is an indication of evolution in luminosity and/or density of
the galaxy populations. This effect is clearly seen in deeper surveys
in the FIR
(e.g.~\citealt{genzel2000,dole2001,dole2004a,frayer2006,soifer2008,bethermin2010});
in the submillimetre (submm) range
(e.g.~\citealt{barger99,blain99,ivison2000,greve2004,coppin2006,weiss2009,patanchon2009,clements2010,lapi2011});
-- and in the millimetre and radio ranges
(e.g.~\citealt{de_zotti2010,vieira2010,vernstrom2011}).

\begin{figure}[!ht] 
   \centering 
  \includegraphics[width=0.30\textwidth,angle=90]{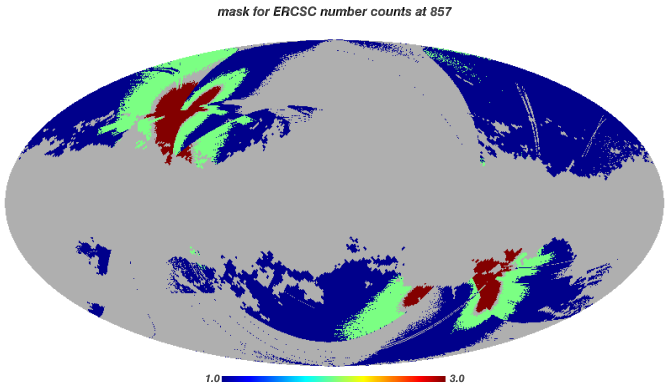} 
  \caption{The three sky zones used in the analysis at 857\,GHz: deep 
    (red); medium (green); and shallow (blue); These are based on the 
    857\,GHz hit counts.} 
   \label{fig:surveys857} 
\end{figure}

The \Planck\footnote{\Planck\ (\url{http://www.esa.int/Planck}) is a 
  project of the European Space Agency (ESA) with instruments provided 
  by two scientific consortia funded by ESA member states (in 
  particular the lead countries France and Italy), with contributions 
  from NASA (USA) and telescope reflectors provided by a collaboration 
  between ESA and a scientific consortium led and funded by Denmark.} 
all-sky survey covers nine bands between 30 and 857\,GHz. It gives us 
for the first time robust extragalactic counts over a wide area of sky 
at these wavelengths, and the first all-sky coverage between 3\,mm 
({\it WMAP}) and 160\micron\ ({\it Akari}) -- e.g. see Table\,1 of 
\citet{planck2011-1.10}. The counts in turn give us powerful constraints on the 
long-wavelength spectral energy distribution (SED) of the dusty 
galaxies investigated e.g. by {\it IRAS}, and on the short-wavelength 
SED of the active galaxies studied at radio wavelengths, e.g. by {\it 
  WMAP} or ground-based facilities.

For \Planck's\ six highest frequency bands (100 to 857\,GHz, we
present here the extragalactic number counts and spectral indices of
galaxies selected at high Galactic latitude and using identifications.
\Planck\ number counts and spectral indices of extragalactic
radio-selected sources were already published for the frequency range
30 to 217\,GHz using results from the LFI and HFI instruments
\citep{planck2011-6.1}.  The transition between synchrotron-dominated
sources and thermal dust-dominated occurs in the crucial spectral
range 200-800\,GHz . Thus our broader frequency data allow a better
spectral characterisation of sources.

We use the {\it WMAP} 7 year best-fit $\Lambda$CDM cosmology 
\citep{larson2011}, with $H_0$ = 71\kmsMpc, $\Omega_\Lambda$ = 0.734 
and $\Omega_{\rm M}$ = 0.266.

\begin{figure}[!ht] 
   \centering 
  \includegraphics[width=0.30\textwidth,angle=90]{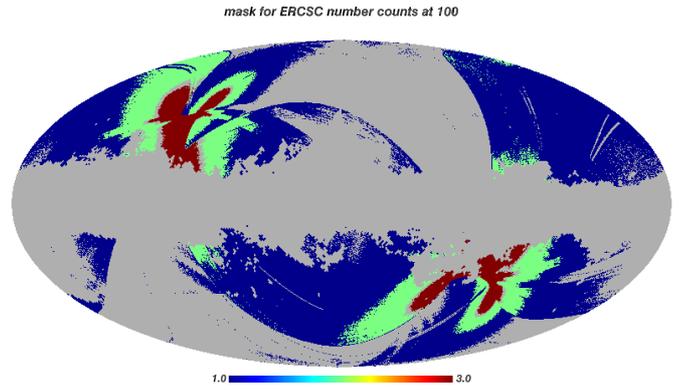} 
  \caption{The three sky zones used in the analysis at 100\,GHz: deep 
    (red); medium (green); and shallow (blue). These are based on the 
    100\,GHz hit counts.} 
   \label{fig:surveys100} 
\end{figure}

\begin{figure}[!b] 
   \centering 
  \includegraphics[width=0.50\textwidth,angle=0]{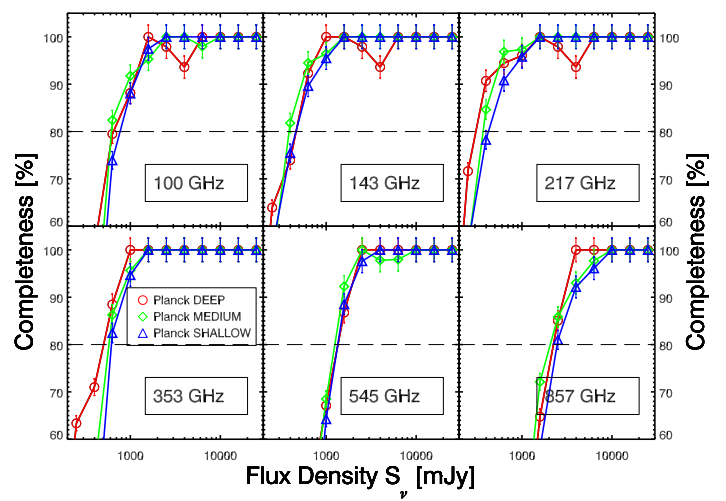} 
  \caption{Completeness (vs.~flux density of sources) coming from the 
    Monte-Carlo runs for the ERCSC and derived for each zone. The 
    horizontal dashed line represents our threshold for number count 
    analysis.} 
  \label{fig:completeness} 
\end{figure}

\section{\Planck\ data, masks and sources}

 \Planck\ \citep{tauber2010a, Planck2011-1.1} is the third generation
space mission to measure the anisotropy of the cosmic microwave
background (CMB).  It observes the sky in nine frequency bands
covering 30--857\,GHz with high sensitivity and angular resolution
from 31\arcm\ to 5\arcm.  The Low Frequency Instrument LFI;
\citep{Mandolesi2010, Bersanelli2010, Planck2011-1.4} covers the 30,
44, and 70\,GHz bands with amplifiers cooled to 20\,\hbox{K}.  The
High Frequency Instrument (HFI; \citealt{Lamarre2010,
Planck2011-1.5}) covers the 100, 143, 217, 353, 545, and 857\,GHz
bands with bolometers cooled to 0.1\,\hbox{K}.  Polarization is
measured in all but the highest two bands \citep{Leahy2010,
Rosset2010}.  A combination of radiative cooling and three mechanical
coolers produces the temperatures needed for the detectors and optics
\citep{Planck2011-1.3}.  Two Data Processing Centers (DPCs) check and
calibrate the data and make maps of the sky \citep{Planck2011-1.7,
Planck2011-1.6}.  \Planck's sensitivity, angular resolution, and
frequency coverage make it a powerful instrument for galactic and
extragalactic astrophysics as well as cosmology.  Early astrophysics
results are given in Planck Collaboration VIII--XXVI 2011, based on data
taken between 13~August 2009 and 7~June 2010.  Intermediate
astrophysics results are now being presented in a series of papers
based on data taken between 13~August 2009 and 27~November 2010.

The \Planck\ data used in this paper (unlike other intermediate
\Planck\ papers) come entirely from the Early Release Compact Source
Catalogue, or ERCSC \citep{planck2011-1.10, planck2011-1.10sup}.  This
in turn is based on \Planck's\ first 1.6 sky surveys, data taken
between 13 August 2009 and 7 June 2010.  First results from the ERCSC
are published as \Planck\ early papers
\citep{planck2011-6.1,planck2011-6.2,planck2011-6.3a,planck2011-6.4a}.
In this paper, we use only HFI data, covering the 100--857\,GHz range
in six bands.

\subsection{Galactic masks} 
 
To obtain reliable extragalactic number counts, uncontaminated by
Galactic sources, we mask out areas of the sky strongly affected by
Galactic sources, defining a set of ``Galactic masks''.  These are
based on removing a fraction of the sky above a specified level in sky
surface brightness. We use two masks, one at high frequencies (857 and
545\,GHz), and one at lower frequencies (353\,GHz and below). The use
of two different masks is motivated by the different astrophysical
components dominating the higher HFI frequencies and the lower
frequencies, which are not necessarily spatially correlated.  While
emission from Galactic dust dominates at 857\, GHz, its spectrum decreases with
decreasing frequency. On the contrary, the synchrotron and free-free
components dominate at 100 GHz.
 
Before applying a brightness cut to the maps, we degrade the angular
resolution of the maps from 1\parcm5 ($N_{\rm side}$=2048 in {\sc Healpix};
\citealt{gorski2005}) down to 55\arcm ($N_{\rm side}$=64).
The maps at low resolution are then interpolated at the 
original high angular resolution. Creating a Galactic mask using this 
procedure has the double benefit of: 1) not masking the bright sources 
(because they are smoothed away); and 2) smoothing the Galactic structure. 
We checked to make sure that even the brightest sources remained unmasked
after applying this smoothing.
 
The 857\,GHz Galactic mask keeps 48\,\% of the sky for analysis (thus 
removing 52\,\% of the sky), and this corresponds to a cut of
2.2\,MJy\,sr$^{-1}$ at 857\,GHz. This mask is applied at 857 and 545\,GHz. 
The 353\,GHz Galactic mask keeps 64\,\% of the sky for analysis (thus 
removing 36\,\% of the sky), and corresponds to a cut of 0.28 
MJy\,sr$^{-1}$ at 353\,GHz.This mask is applied at 353, 217, 143 and
100\,GHz. 
 
Figs.~\ref{fig:mask857} and \ref{fig:mask353} show the \Planck\ 
masks, comparing them with the {\it WMAP} 7 year KQ75 and KQ85 masks 
\citep{gold2011,jarosik2011}. Note that we do not use the {\it WMAP} 
masks in this work.

\subsection{Three zones in the sky: deep, medium and shallow} 
 
Three main zones are identified to ensure reasonably homogeneous
coverage of the sky by the \Planck\ detectors at each frequency,
thereby allowing a clean and simple estimate of the completeness. As
noted above, the \Planck\ data used here correspond to approximately
1.6 complete surveys of the sky; in addition each survey has
non-uniform
coverage of the sky 
\citep{Planck2011-1.1, Planck2011-1.5, Planck2011-1.7}. While performing 
statistics on sources drawn from a 
non-uniformly covered survey is feasible, both the nature of the
\Planck\ data (including scanning strategy, and masking of planets
(see the ERCSC article \citealt{planck2011-1.10}) and its
heterogeneous coverage (see Fig.~\ref{fig:hitcounts}) make it
difficult to implement. We therefore select three zones in the sky, in
each of which the observations are approximatively homogeneous in
integration time.
 
The hit count can be defined by counting the number of times a single 
\Planck\ detector observes one sky position in the sky. The hit count 
can also be defined for a particular frequency band: it is the number of times 
each sky pixel has been hit by any \Planck\ detector at a given 
frequency. We will be using this latter definition. This quantity is 
similar to $N_{\rm obs} $ in {\it WMAP} data files. 
 
The three zones have hit counts varying by not more than a factor of 
two, except in the smaller deep zone (at the ecliptic poles) where 
there is high redundancy. They are defined as (and illustrated in 
Fig.~\ref{fig:hitcounts}):

\begin{itemize} 
\item deep: $<$5\% of the best covered sky fraction (or 95\,\%
  or more of the cumulative hit count distribution at a given
  frequency);
 
\item medium: 5 to 25\% of the best covered sky fraction (or 75
  to 95\,\% of the cumulative hit count distribution at a given
  frequency);
 
\item shallow: 25 to 50\% of the best covered sky fraction (or
  50 to 75\,\% of the cumulative hit count distribution at a given
  frequency).
 
\end{itemize} 
 
Thus, pixels in the deep zone (at a given frequency) all have a hit
count value greater than or equal to the hit count value corresponding
to 95\,\% of the total distribution at this frequency.  Note that each
frequency map has different hit counts, due to the focal plane
geometry; each zone will thus have slight differences in geometry from
one frequency to another, leading to slightly different surface areas.
Table~\ref{tab:sourcenumber} summarizes the surface area of each zone;
typically, the deep zone covers 1000\,deg${}^2$, the medium zone about
3000\,deg${}^2$, and the shallow about
12000\,deg${}^2$. Fig.~\ref{fig:surveys857} (or \ref{fig:surveys100})
shows the three different zones used in this analysis: deep, medium
and shallow at 857\,GHz (100\,GHz), respectively.

\input{table_source_identifications_new.tex}


\input{table_source_zone_stat_planck_numbercounts_new.tex}


\input{table_numbers_planck_numbercounts_new.tex}

 
\begin{figure*}[!ht] 
\centering 
\epsfig{file=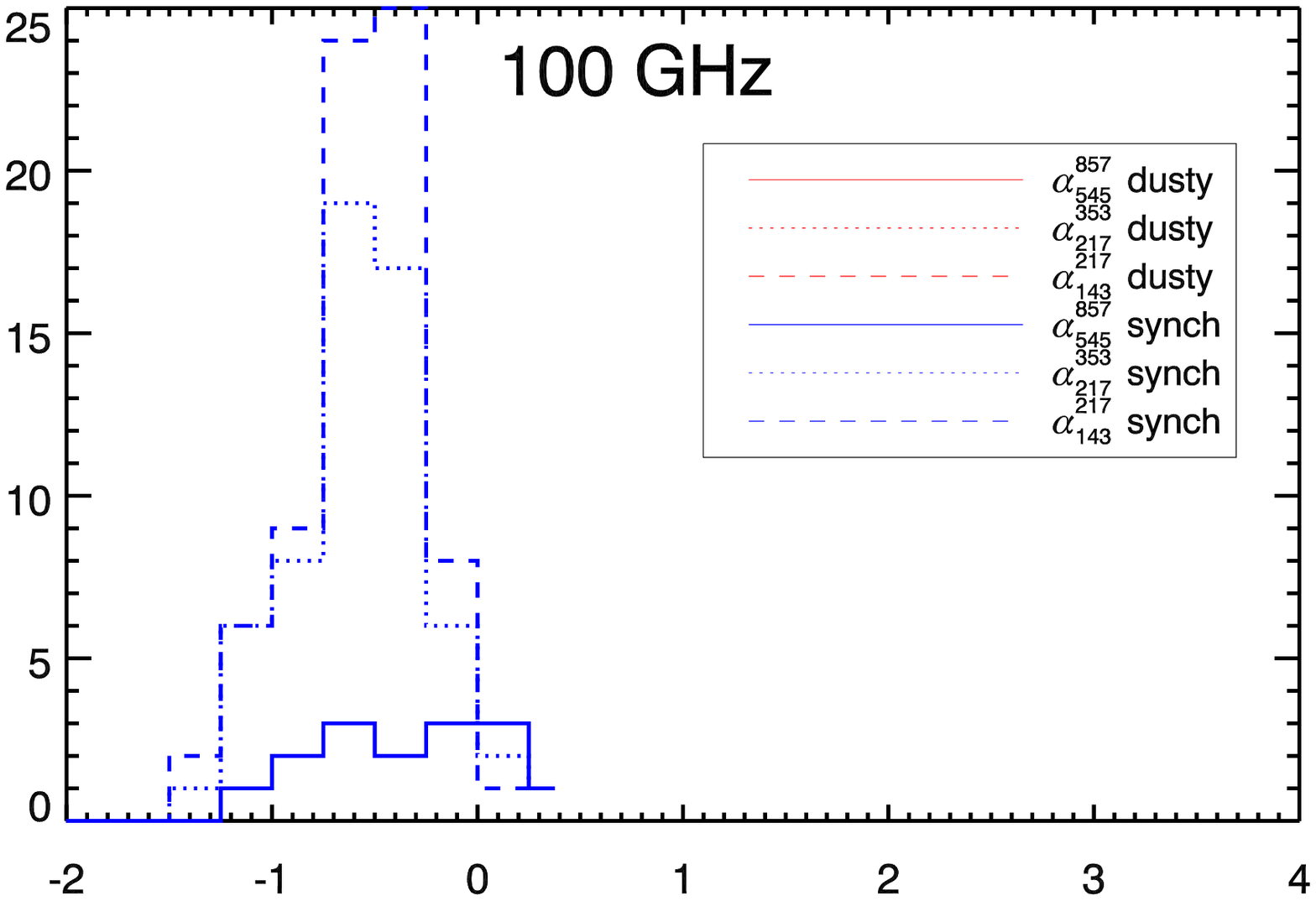, clip=, width=0.3\linewidth}  
\epsfig{file=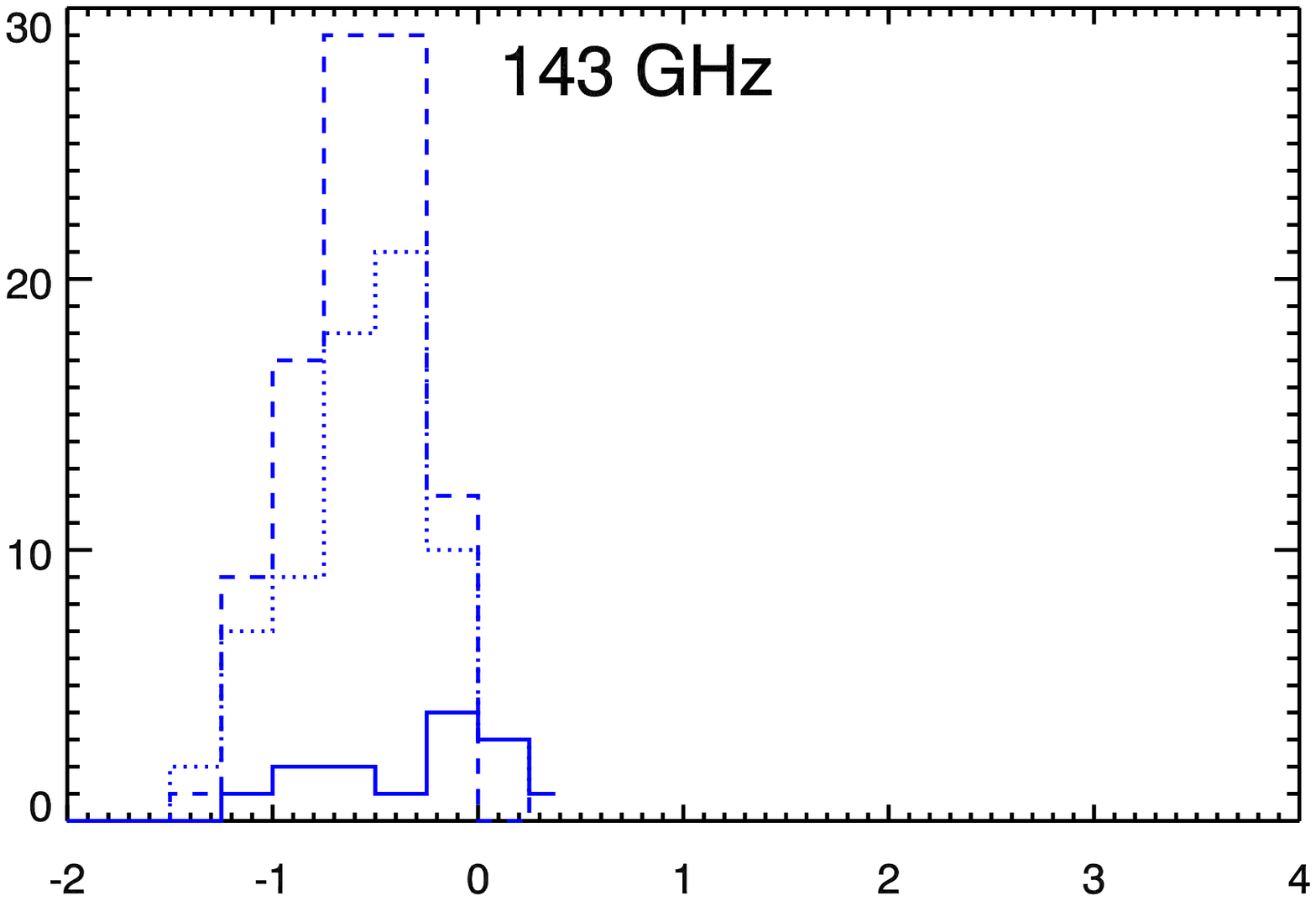, clip=, width=0.3\linewidth}  
\epsfig{file=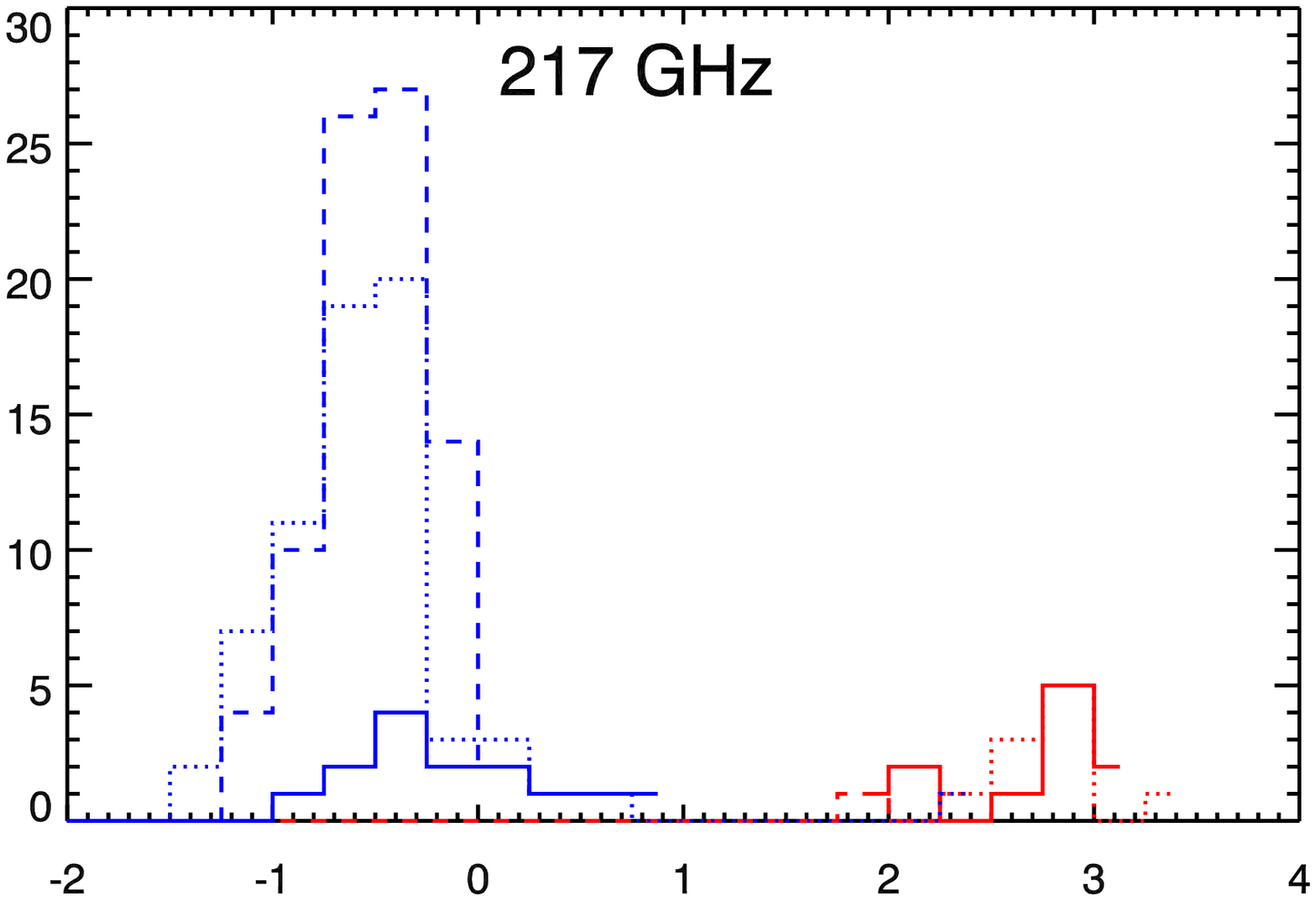, clip=, width=0.3\linewidth}\\  
\epsfig{file=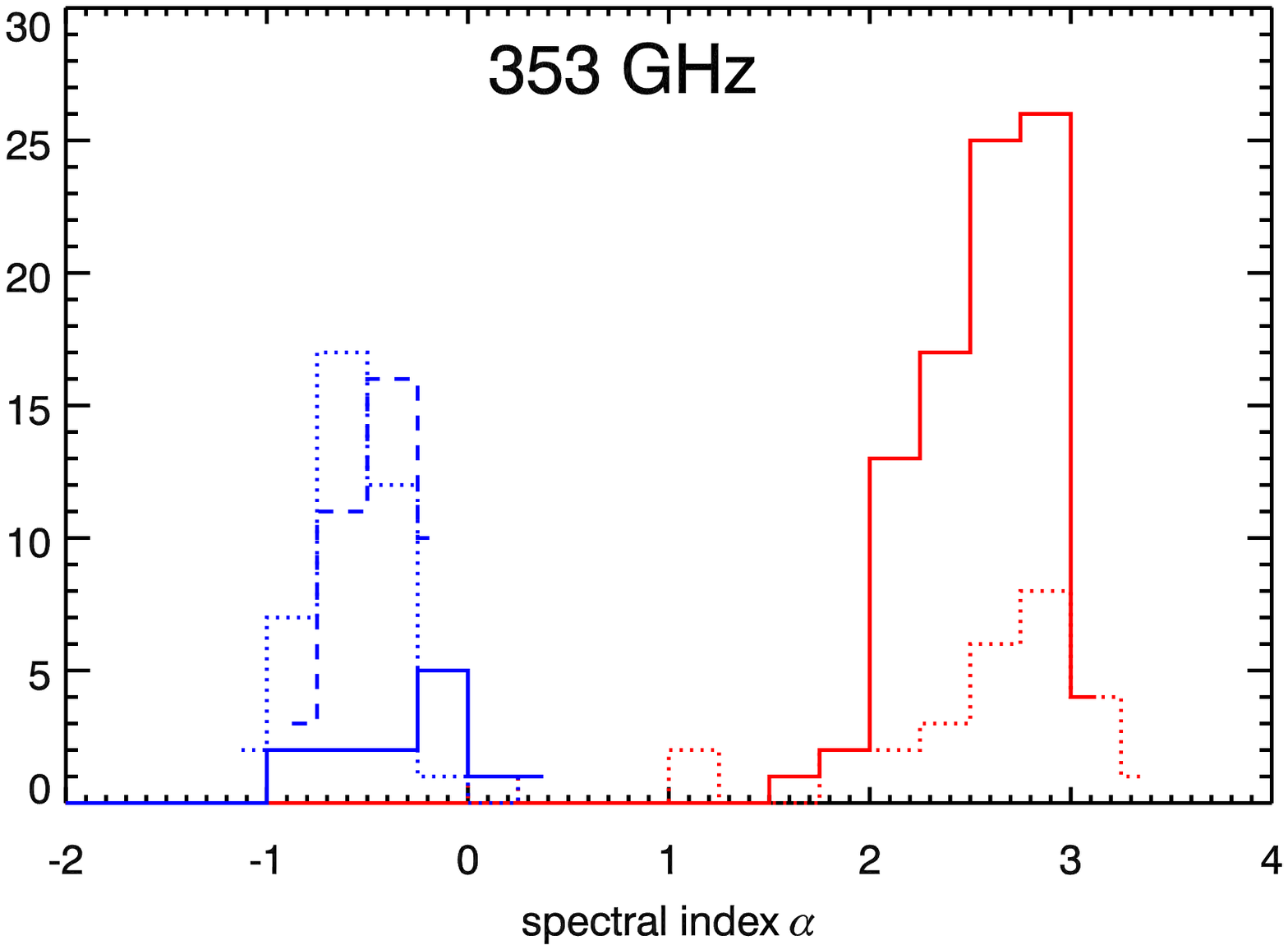, clip=, width=0.3\linewidth} 
 \epsfig{file=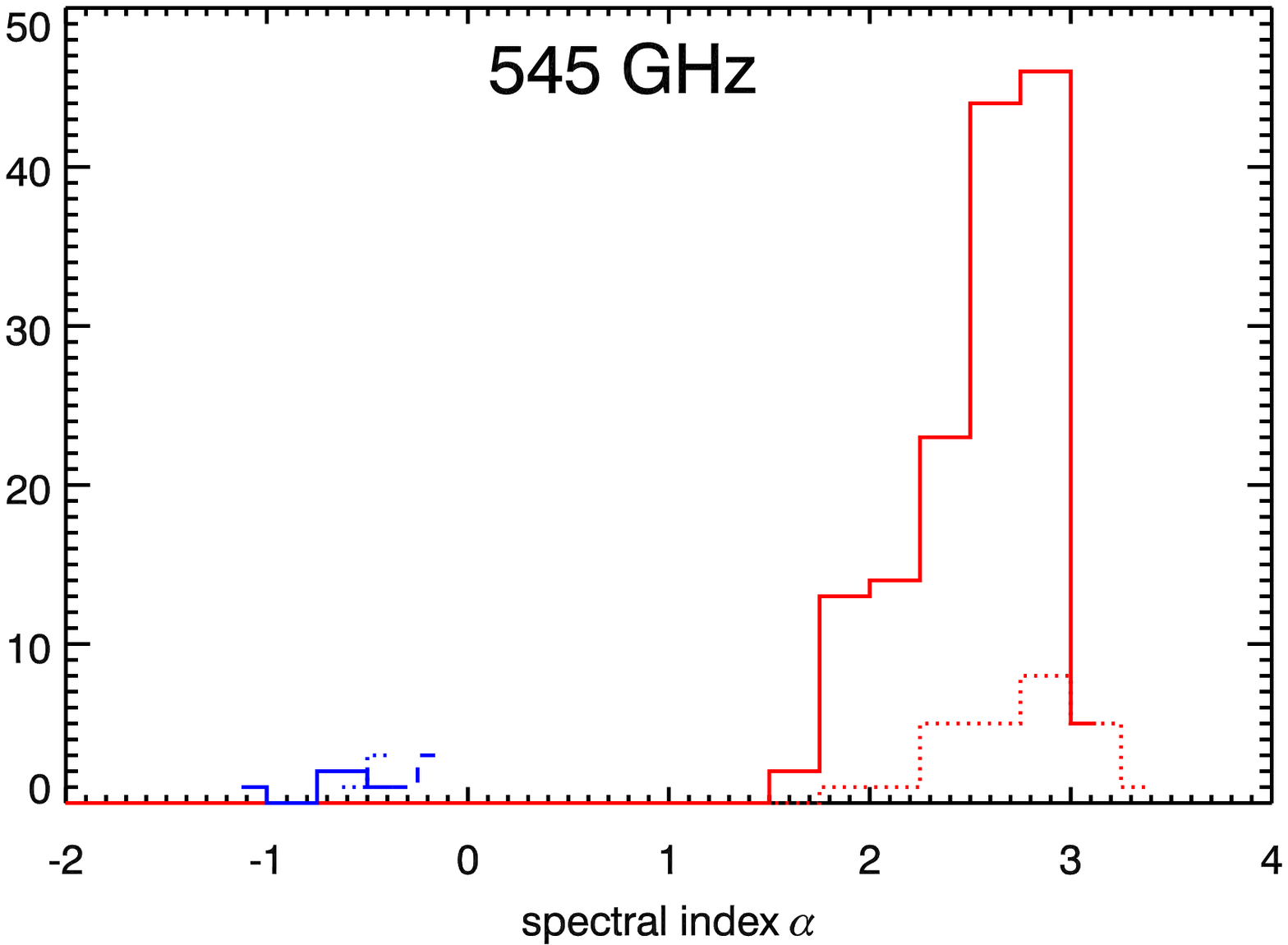, clip=, width=0.3\linewidth} 
\epsfig{file=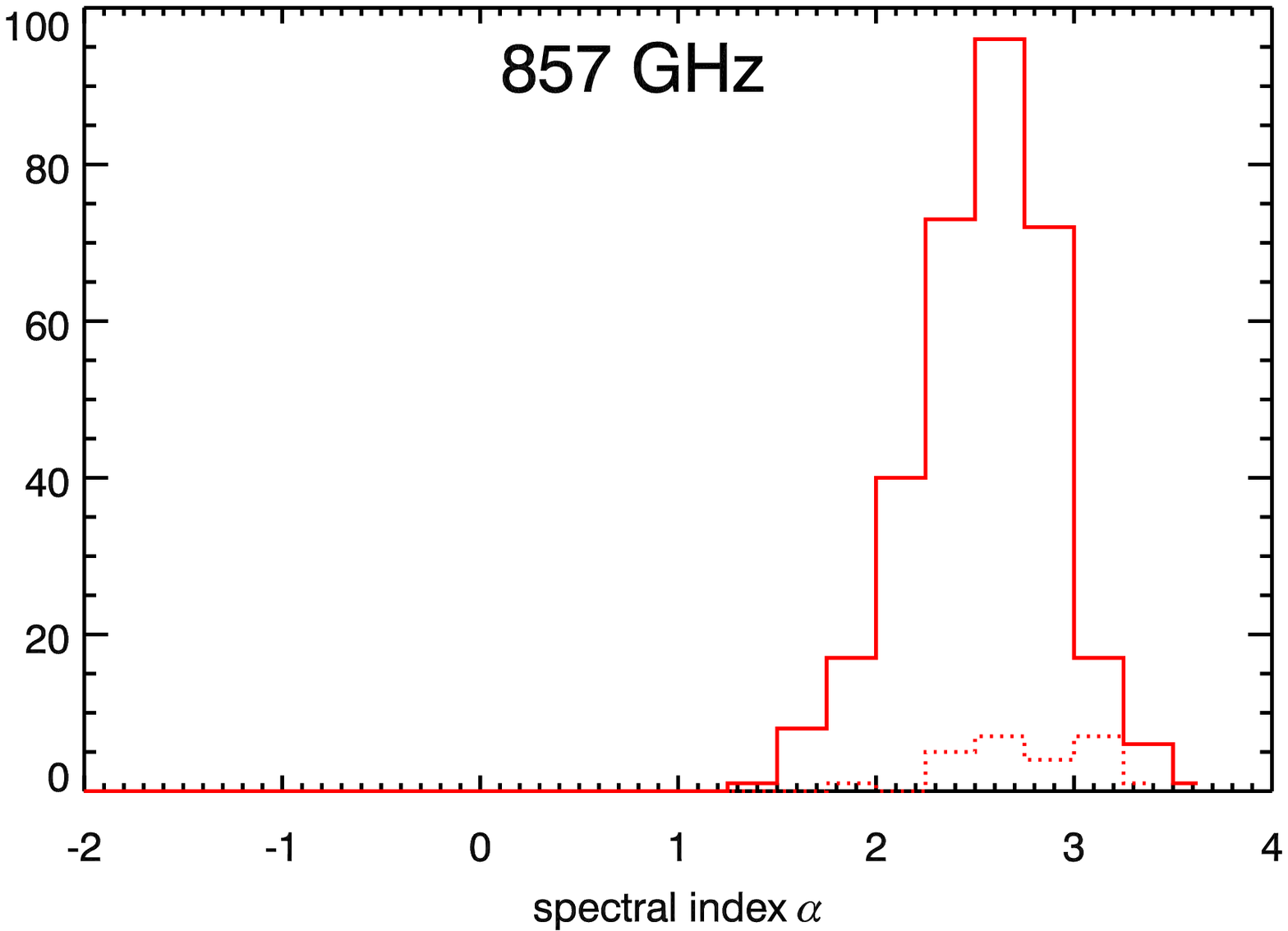, clip=, width=0.3\linewidth}  
 
\caption{ Distribution of spectral indices (for the sources present in 
  the ERCSC at a given frequency in our sample with completeness of 
  80\,\% or above). Here $\alpha^{217}_{143}$ is shown as a dotted line, 
  $\alpha^{353}_{217}$ as a dashed line, and $\alpha^{857}_{545}$ as a 
  solid line. The region $2 \le \alpha \le 4$ is typical of thermal 
  dust emission. In red we show the dusty sources, whereas in blue we 
  show the synchrotron sources. The sources in all three samples 
  (deep, medium, and shallow) are combined here. Note that, as 
  expected, the 100\,GHz, and 143\,GHz samples are dominated by radio 
  galaxies, whereas the 545\,GHz and 857\,GHz samples are dominated by 
  dusty galaxies. At 217\,GHz and 353\,GHz we observe the transition 
  between the two populations, with significant numbers of both types 
  being present in the samples. \label{fig_alphas}} 
\end{figure*}

\subsection{Sample selection and validation} 
\label{sect:sample} 
 
The sample is drawn from the ERCSC \citep{planck2011-1.10}, which was
constructed to contain high SNR sources. Notice that at high
frequency, the noise is dominated by the confusion, mainly due to
faint extragalactic sources and Galactic cirrus
\citep{condon74,hacking87,franceschini89,helou90,franceschini91,franceschini94,toffolatti98,dole2003,negrello2004,dole2006}. The
selection is performed with the following steps at each HFI frequency
independently:
 
\begin{itemize} 
 
\item select sources within each zone: deep, medium and shallow; 
 
\item select point sources, using the keyword ``EXTENDED'' set to 
  zero; 
  
\end{itemize} 
 
These criteria should favour the presence of galaxies rather than 
Galactic sources. To validate this, we make three checks in addition 
to using conservative masks. 
 
1. We measure the mid-IR to ~submm flux density ratios of known 
Galactic cold cores (from the \Planck\ Early Cold Core catalogue, ECC, 
\citealt{planck2011-7.7b}) and conversely of known galaxies in the 
ERCSC. Using {\it WISE} \citep{wright2010} W3 and W4 bands (when 
available with the first public release), and the 857\,GHz HFI band, we 
measure a factor of 100 to 200 between the submm-to-mid-infared ratios 
of galaxies and ECC sources. When measuring this ratio in our sample, 
we see that the submm-to-mid-infared colours of all sources in our sample 
are compatible with galaxy colours, and not with ECC colours. 
 
2. The CIRRUS flag in the ERCSC gives an estimate of the normalised 
neighbour surface density of sources at 857\,GHz, as a proxy for 
cirrus contamination. The median value of the CIRRUS flag in our sample is 
0.093 at 857\,GHz, a low value compatible with no cirrus contamination 
when used in conjunction with the EXTENDED=0 flag 
(e.g. \citealt{herranz2012}). 
 
3. We query the NED and SIMBAD databases at the positions of all our 
ERCSC sources using a 2\parcm5 search radius.  Each \Planck\ source 
has many matches (many of them completely unrelated, e.g. foreground 
stars), and the identification is more complex at higher 
frequencies. However, as we show later, our 
$N(>S)$ cumulative distribution of sources is always less than 200 
sources per steradian, i.e. less than $3.3 \times 10^{-4}$ ERCSC 
sources per 2\parcm5 search radius on average.  We thus search for 
the most probable match by identifying the source 
type in this order: Galactic, then extragalactic. The Galactic types 
include supernova remnants, planetary nebulae, nebulae, {\sc Hii} 
regions, stars, molecular clouds, globular/star clusters. We call a 
source ``Galactic unsecure'' when one of the two databases returns no 
identification and the other a Galactic identification.  We do not use 
``Galactic secure'' or ``Galactic un-secure'' sources in the analysis in this 
paper. The statistics of identifications is given in 
Table~\ref{tab:ids}. 
 
Our final sample is composed of confirmed galaxies, the vast majority
being NGC, {\it IRAS}, radio galasy and blazar objects, as well as
some unidentified sources (a small fraction of the total number). The
few completely unidentified sources, where no SIMBAD or NED ID was
found, are interpreted as potential galaxies, and hence are included
in our counts, because they have a small cirrus flag value (see point
2 above). Their relatively small number don't change the results
  presented in this article, wether or not we include these sources.

Table~\ref{tab:sourcenumber} summarises the source number and surface 
area of each zone (deep, medium and shallow). We find a total number 
of sources ranging from 217 at 100\,GHz to 1058 at 857\,GHz.

\subsection{Completeness} 
\label{sect:completeness} 
The ERCSC Pipeline \citep{planck2011-1.10} used extensive Monte-Carlo
simulations ( to account for systematic and sky noise) to assess
various parameters such as positional or flux density
accuracies. Here, we use the results of those runs to estimate the
completeness in each of the three zones, as presented in
Fig.~\ref{fig:completeness}. The uncertainties in completeness are at
the 5\,\% level, as discussed in \citet{planck2011-1.10} and
\citet{planck2011-1.10sup}. The correction for incompleteness is then
applied to the number counts of each zone separately.
 
We use a completeness level threshold of 80\,\% for all 
frequencies.  This ensures: 1) minimal source contamination; 2) no 
photometric biases \citep{planck2011-1.10sup}; and 3) good 
photometric accuracy \citep{planck2011-1.10sup} -- see 
Sect.~\ref{sect:photometry}.  The number of sources actually used to 
estimate the number counts is given in Tab.~\ref{tab:numbers}, which 
also includes the number of unidentified sources. In the end, we use a 
number of sources ranging from 122 at 100\,GHz to 452 at 857\,GHz 
(Tab.~\ref{tab:sourcenumber}).

\begin{figure}[!hb] 
   \centering 
  \includegraphics[width=0.5\textwidth,angle=0]{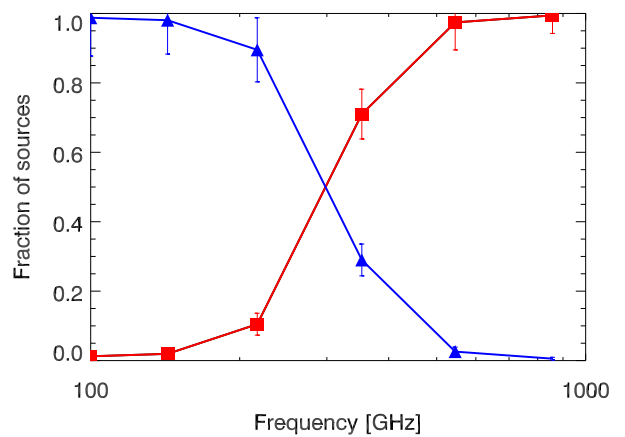} 
  \caption{Fraction of galaxy types as a function of frequency, on 
    the sample having completeness of 80\,\% and above: dusty (red 
    squares) and synchrotron (blue triangles). Error bars are 
    Poissonian.} 
  \label{fig:statsed} 
\end{figure}


\input{table_number_counts_highfreq_new.tex}

\input{table_number_counts_lowfreq_new.tex}

\subsection{Photometry} 
\label{sect:photometry} 
 
The photometry of the ERCSC is extensively detailed in 
\citet{planck2011-1.10} as well as in the explanatory supplement 
\citet{planck2011-1.10sup}. Here we use the ``FLUX'' field for flux 
densities. Notice that 100 and 217\,GHz flux densities can be affected 
by Galactic CO lines \citep{Planck2011-1.7}. 
 
We would like to emphasise that the extensive simulations performed in 
the process of generating/validating the ERCSC allow us to derive 
reliable completeness estimates for each zone (see 
Sect.~\ref{sect:completeness}) and also to estimate the quality of the 
photometry. In the faintest flux density bins that we are using 
(corresponding to 80\,\% completeness), there is no photometric offset, 
and the photometric accuracy from the Monte-Carlo simulations 
\citep[][; Fig.~7 for reference]{planck2011-1.10sup} is about: 35\,\% at 480\,mJy 
for 100\,GHz; 30\,\% at 300\,mJy for 143\,GHz; 20\,\% at 300\,mJy for 217\,GHz; 
20\,\% at 480\,mJy for 353\,GHz; 20\,\% at 1207\,mJy for 545\,GHz; and 20\,\%
at 1913\,mJy for 857\,GHz. This scatter in the faintest flux density bins 
strongly decreases at larger flux densities. Note that photometric 
uncertainties can bias the determination of the counts slope 
\citep[e.g.][]{Murdoch1973}; at our completeness level, the effect is negligible.
 
From our sample, we also create ``Band-filled catalogues''. For each 
frequency/zone sample, we take each source position from the ERCSC and 
perform aperture photometry from the corresponding images in the other 
frequencies. We adopt 4\,$\sigma$ as the detection threshold. These 
aperture photometry measurements (and upper limits) are used for the 
spectral classification of sources and in the spectral index 
determinations, but {\it not} in the number counts measurements (which rely 
only on ERCSC flux densities). We define the spectral index $\alpha$ 
by $S_{\nu}\propto \nu^{\alpha}$. 
 
The derived spectral indices are used to determine the colour 
correction of the ERCSC flux densities \citep{Planck2011-1.7}. This 
correction changes the flux densities by at most 5\,\% at 857\,GHz, 
15\,\% at 545\,GHz, 14\,\% at 353\,GHz, 12\,\% at 217\,GHz, and 1\,\% at 
143\,GHz and 100\,GHz.


\input{table_number_counts_highfreq_dusty_new.tex}

\input{table_number_counts_lowfreq_synch_new.tex}

\section{Classification of galaxies into dusty or synchrotron 
  categories} 
\label{sect:classification} 
 
For the purposes of this paper, we aim for a basic classification 
based on SEDs that separates sources into those dominated by thermal 
dust emission and those dominated by synchrotron emission. (Free-free 
emission does exist, but is not dominant, e.g., \citealt{peel2011}). 
In order to classify our sources by type, we start with the 
band-filled catalogues discussed in Section~\ref{sect:photometry}. 
Thermal dust emission is expected to show spectral indices in the 
range $\alpha$\,$\sim$\,2\,--\,4. On the other hand, colder 
temperature sources can show lower $\alpha^{857}_{545}$, if the 
857\,GHz measurement falls near the spectral peak. Also, the presence of a strong 
synchrotron component, or perhaps a free-free emission component, would 
start to flatten the SED below $\sim$\,353\,GHz.  With such issues in 
mind, we have set up the following classification algorithm: 
\begin{itemize} 
\item[$\bullet$] all sources with 
$\alpha^{857}_{545}$\,$\geq$\,2, or $\alpha^{545}_{353}$\,$\geq$\,2 
are assigned a ``dusty'' classification; 
\item[$\bullet$] all sources where both of these spectral indices are 
  lower than 2, including non-detections, are assigned ``synchrotron''
  classification. 
\end{itemize} 
The resulting classification is summarised in Fig.~\ref{fig_alphas}, 
showing the spectral index distributions for each type as a function 
of observed frequency. 
 
However, some sources are difficult to classify, and could be part of 
an ``intermediate dusty'' or ``intermediate synchrotron'' type. These 
intermediate sources can be defined as follows: 
\begin{itemize} 
\item[$\bullet$] being dusty (according to our criterion above) but 
  also having $\alpha^{857}_{100} < 1$ 
\item[$\bullet$] being synchrotron (according to our criterion above)
  but also being detected either at 857 or 545 GHz, and
    undetected at 353 and 217 and 143 GHz, i.e. sources that show
  both a significant dust component and a strong synchrotron
  component.
\end{itemize}

\begin{figure*}[!ht] 
   \centering 
  \includegraphics[width=0.98\textwidth,angle=0]{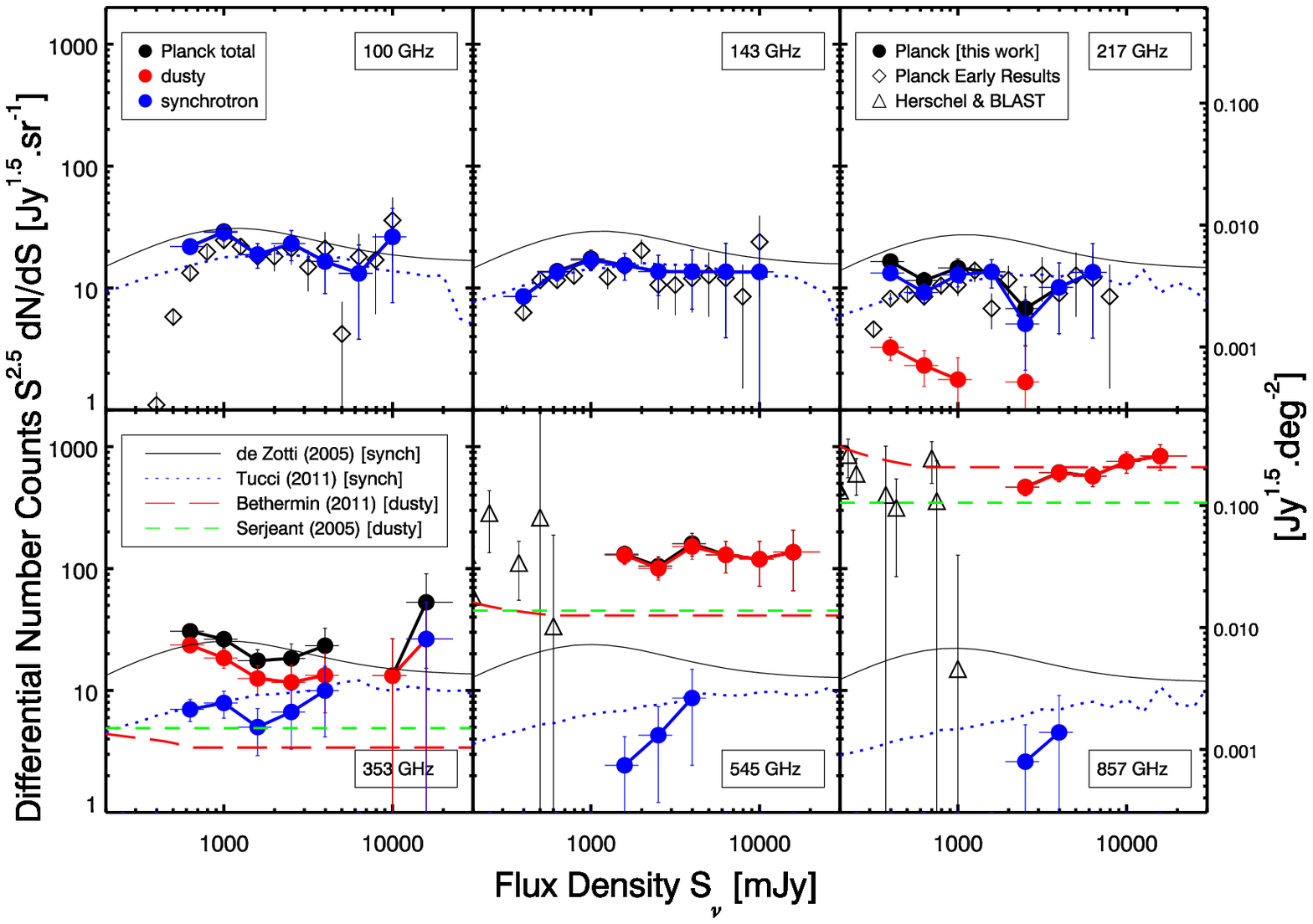} 
  \caption{\Planck\ differential number counts, normalised to the 
    Euclidean value (i.e.\ $S^{2.5} dN/dS$), compared with models and 
    other data sets. \Planck\ counts: total (black filled circles); 
    dusty (red circles); synchrotron (blue circles). Four models are 
    also plotted:
    \citet[][dealing only with synchrotron sources -- solid line]{de_zotti2005};
    \citet[][dealing only with synchrotron sources -- dots]{tucci2011};
    \citet[][dealing only with dusty sources -- long dashes]{bethermin2011a};
    \citet[][dealing only with local dusty sources -- short dashes]{serjeant2005}.
    Other data sets: 
    \Planck\ early counts for 30\,GHz-selected radio galaxies 
    \citep{planck2011-6.1} at 100, 143 and 217\,GHz (open diamonds); 
    {\it Herschel} ATLAS and HerMES counts at 350 and 500\micron\ from 
    \citet{oliver2010} and \citet{clements2010}; BLAST at the same two 
    wavelengths, from \citet{bethermin2010a}; all shown as 
    triangles. Left vertical axes are in units of 
    Jy$^{1.5}$\,sr$^{-1}$, and the right vertical axis in 
    Jy$^{1.5}$.deg$^{-2}$. } 
  \label{fig:countsdiffall}  
\end{figure*}

Among the sources included in the number counts analysis, fewer than 
10\,\% are classified as ``intermediate''. This fraction rises 
significantly if we remove the completeness cut due to the increasing 
photometric uncertainties at lower flux densities (see 
Appendix~\ref{sect:interm} for details). Examples of both ``dusty'' or 
``synchrotron'' sources with somewhat unusual SEDs are discussed in 
Appendix~\ref{sect:individualsources} .

\section{{\it Planck} extragalactic number counts between 100 and 857\,GHz} 
\label{sect:counts} 
 
The \Planck\ extragalactic number counts (differential, normalised to 
the Euclidean slope, and completeness-corrected) are presented in 
Fig.~\ref{fig:countsdiffall} and 
Tables~\ref{tab:number_counts_highfreq} and 
\ref{tab:number_counts_lowfreq}. They are obtained using a mean of the 
3 zones (weighted by the surface area of each zone). 
 
The error budget in the number counts is made up of: (i) Poisson
statistics (ii); the 5\,\% uncertainty in the completeness correction
(iii); the absolute photometric calibration uncertainty of 2\,\% at
and below 353\,GHz, and 7\,\% above 545\,GHz \citep{Planck2011-1.5,
  Planck2011-1.7}. According to e.g. Eq. 1 of \citet{bethermin2011a},
calibration uncertainties produce errors scaling as the 1.5 power in
the Euclidean, normalized, differential number counts.
 
Notice that for our bright counts, the error budget is dominated by
sample variance of nearby sources and consequently by small-number
statistics.  For instance, the small wiggle seen in the counts at the
three highest frequencies (seen at 600\,mJy at 353\,GHz, 4\,Jy at
545\,GHz and 10\,Jy at 857\,GHz) is due to a few tens of local NGC
sources in the medium zone (see
Appendices~\ref{sect:individualsources} and \ref{sect:countzone}).
 
Integral (i.e. cumulative) combined number counts are shown in
Fig.~\ref{fig:countsintegral}. Although error bars are highly
correlated, these counts provide a useful estimate of the source
surface density. The completeness correction is also applied here, and
we use the same cuts in flux density as for the differential
counts. Tab.~\ref{tab:number_counts_highfreq} and
\ref{tab:number_counts_lowfreq} also give the $N>S$ values.

\section{Further Results \& Discussion}

\subsection{Nature of the Galaxies at submillimetre and millimetre wavelengths} 
 
The change in the nature of sources (synchrotron dominated vs. dusty) with frequency
was first observed in the \Planck\ data in \cite{planck2011-1.10}. Our
new sample allows a more precise quantification because of its
completeness.  The statistics of synchrotron vs.~dusty galaxies are
summarised in Fig.~\ref{fig:statsed}, showing the fraction of galaxy
type as a function of frequency. We estimate the uncertainty in the
classification to be of the order of 10\,\% (see
Appendix~\ref{sect:interm}). The striking result is the almost equal
contribution of both source types near 300\,GHz. The high frequency
channels (545 and 857\,GHz) are, unsurprisingly, dominated ($>90\,\%$)
by dusty galaxies. The low frequency channels are, unsurprisingly,
dominated ($>95\,\%$) by synchrotron sources at 100 and 143\,GHz.  At
217\,GHz, fewer than 10\,\% of the sources show a dust-dominated SED.
 
All the sources from our complete sample have an identified spectral
type (by construction), and we can compute the number counts separately for
synchrotron and dusty galaxies.  Fig.~\ref{fig:countsdiffall},
\ref{fig:countsdiffallspt} and \ref{fig:countsintegral} show the
differential and integral number counts by type, also given in
Tables~\ref{tab:number_counts_dusty} and
\ref{tab:number_counts_synch}. We note that at 353\,GHz, about two
thirds of the number counts are made-up by dusty sources.  At 217\,GHz
(545\,GHz) there is a minor contribution (10\,\% or less) of the dusty
(synchrotron) sources contributing to the counts. These number counts
of extragalactic dusty and synchrotron sources are an important step
towards further constraining models of galaxy SED and to including the results in more
general models of galaxy evolution (see below).

\subsection{\Planck\ Number Counts Compared with Other Datasets} 
 
\begin{figure*}[!ht] 
   \centering 
  \includegraphics[width=0.98\textwidth,angle=0]{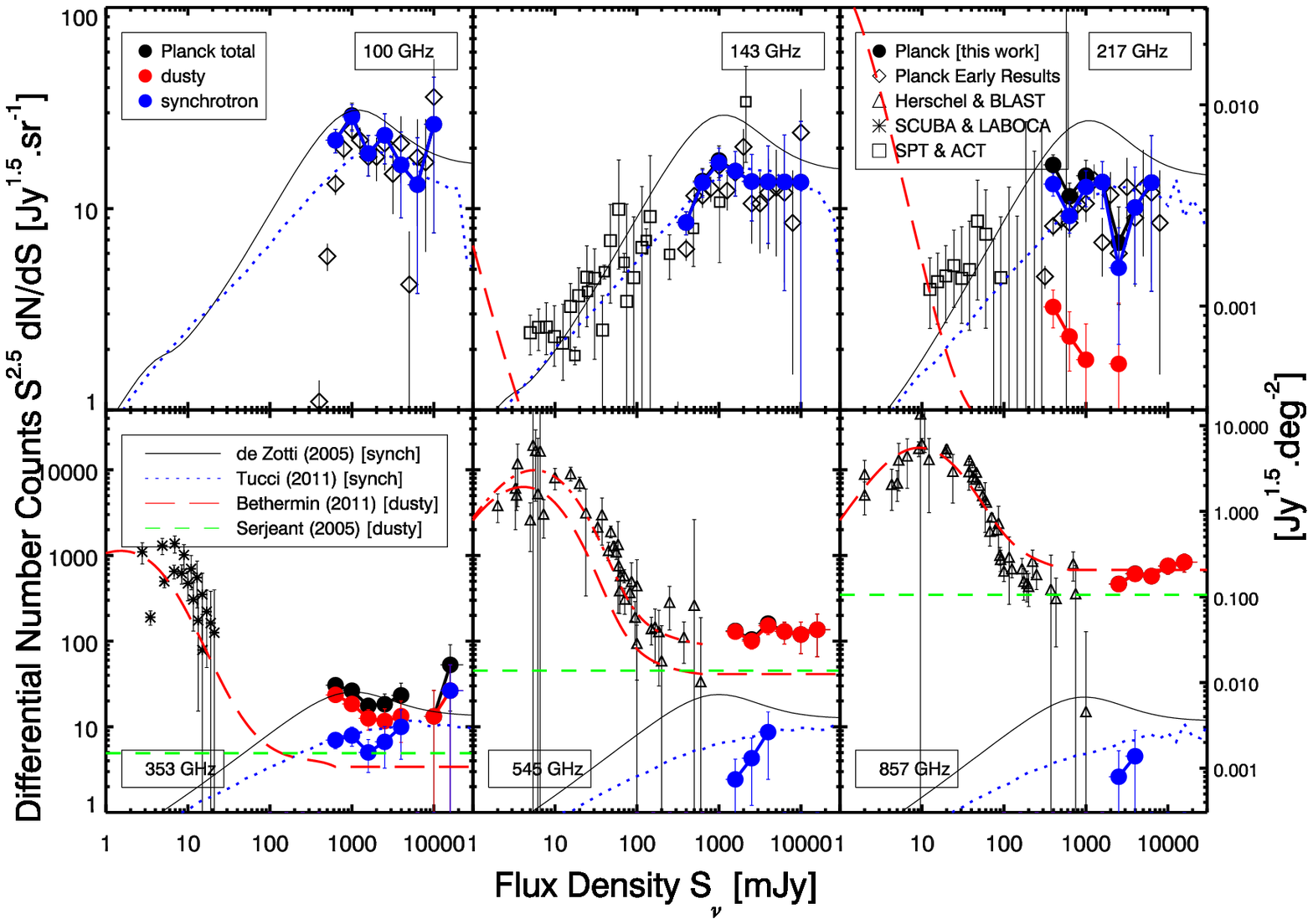} 
  \caption{Same as Fig.~\ref{fig:countsdiffall}, but on a wider flux
    density scale and with the addition of ACT \citep{marriage2011}
    and SPT \citep{vieira2010} data shown as squares at 143 and
    217\,GHz, and SCUBA and LABOCA data shown as stars at 353\, GHz
    \citep{borys2003,coppin2006,scott2006,beelen2008,weiss2009}.
    Notice that we added the model of \cite{bethermin2011a} at
    500~$\mu$m (dash-dot) to comply with the Herschel data taken at
    that wavelength (and not at 545 GHz).}
  \label{fig:countsdiffallspt}  
\end{figure*}

\begin{figure}[!h] 
   \centering 
  \includegraphics[width=0.47\textwidth,angle=0]{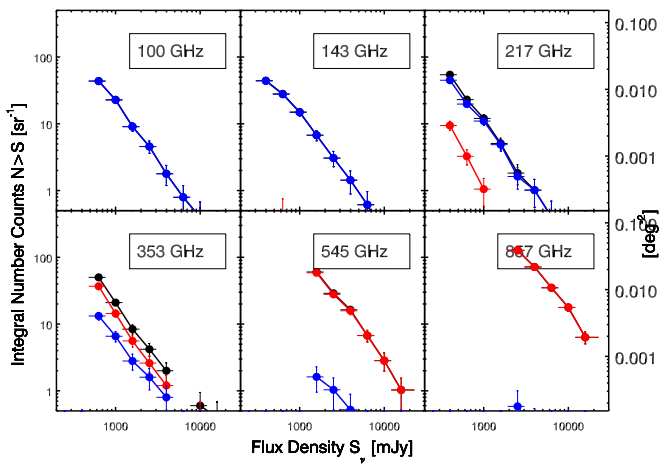} 
  \caption{\Planck\ integral number counts (filled circles); dusty 
    (red); synchrotron (blue). Vertical axes are in number per 
    steradian; right axis is in number per square degree. Counts are 
    completeness-corrected. The same faint flux density cut as for 
    differential counts is applied.} 
  \label{fig:countsintegral} 
\end{figure}

The number counts are in fairly good agreement at lower frequencies
(100 to 217\,GHz) with the counts published in the \Planck\ early
results, based on a 30\,GHz selected sample (\cite{planck2011-6.1};
represented as diamonds in Fig.~\ref{fig:countsdiffall} and
\ref{fig:countsdiffallspt}).  The effect of incompleteness in the latter
is seen in the fainter flux density bins, below about
500\,mJy. We also notice a slight disagreement around 400\,mJy at
217\,GHz, where our counts of synchrotron galaxies exceed the \Planck\
early counts by a factor of 1.7 ($13.7 \pm 1.5$ vs.~$8.2 \pm
0.9\,$Jy$^{1.5}$\,sr$^{-1}$). This discrepancy can be easily
understood: our current selection of synchrotron sources is not the
same as the one adopted in the \Planck\ Early results paper
\citet{planck2011-6.1}, in which a more restrictive criterion was used
($\alpha^{217}_{143} < 0.5$).  If we adopt the same criterion as in
that paper, we find no statistically significant difference between the
two estimates of the number counts.
 
Our estimates of counts also seem consistent with the {\it Herschel} ATLAS 
and HerMES counts \citep{clements2010,oliver2010} at high frequency 
(545 and 857\,GHz) as well as BLAST at the same two wavelengths 
\citep{bethermin2010a}, although there is no direct overlap in flux 
density and small number statistics affect the brightest {\it Herschel} 
points. 

The ACT 148\,GHz data \citep{marriage2011} and SPT 150 and 220\,GHz
\citep{vieira2010} data are also plotted in
Fig.~\ref{fig:countsdiffallspt}, together with SCUBA and LABOCA data
at 353\,GHz
\citep{borys2003,coppin2006,scott2006,beelen2008,weiss2009}.  The ACT and SPT data, 
when added to the \Planck\ data at 143\,GHz, cover
more than four orders of magnitude in flux density.
 
Finally, we checked that our counts are in agreement with the \Planck\ Sky 
Model \citep{delabrouille2012}.


\begin{figure}[!h] 
   \centering 
  \includegraphics[width=0.47\textwidth,angle=0]{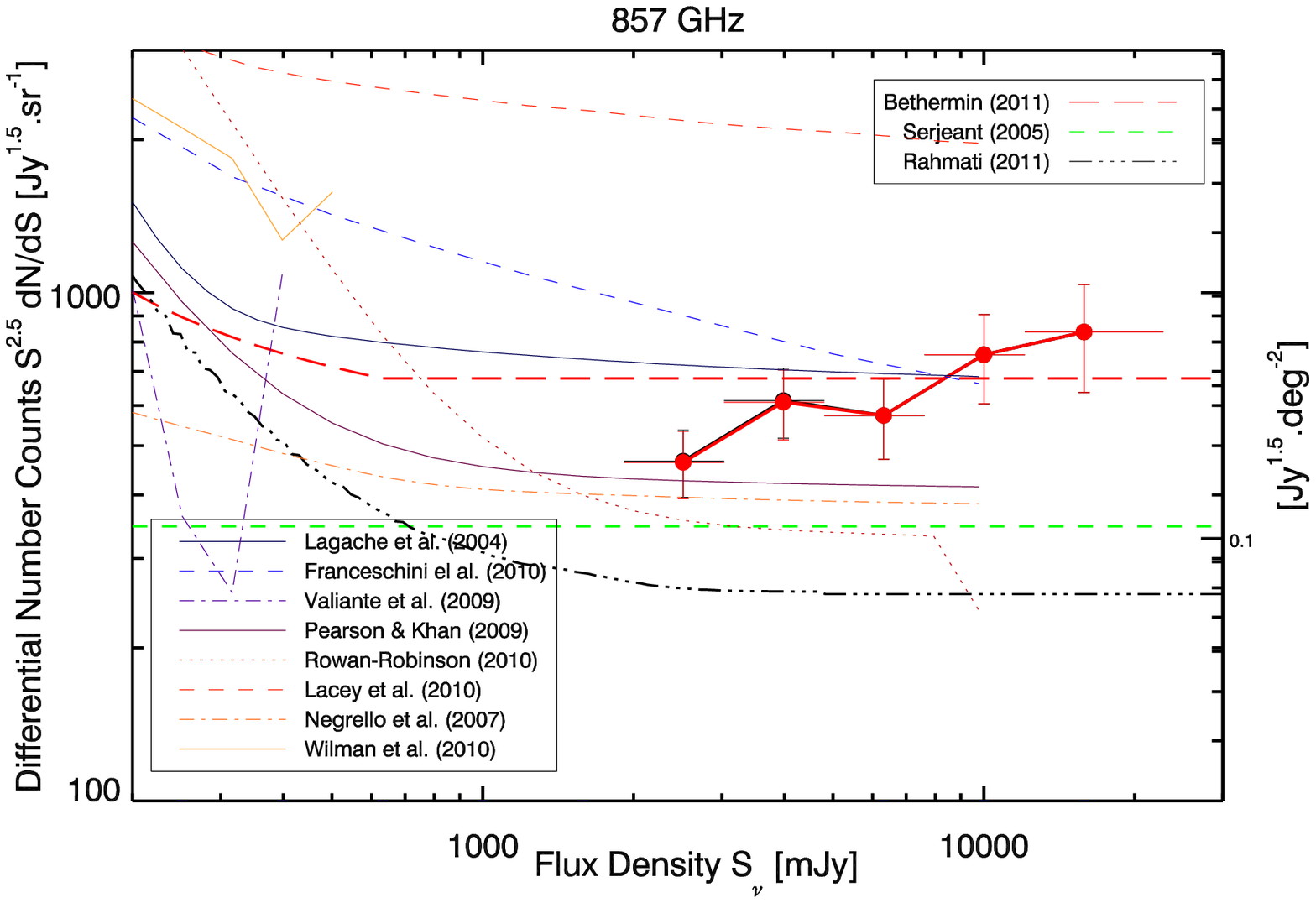} 
  \caption{\Planck\ differential number counts at 857\,GHz, normalised 
    to the Euclidean (i.e. $S^{2.5} dN/dS$). \Planck\ counts: total 
    (black filled circles); dusty galaxies (red circles). Models: 
\citet{bethermin2011a} (long red dashes -- dusty); 
    \citet{serjeant2005} (short green dashes -- dusty); 
    \citet{rahmati2011} (black dash-dot-dot-dot line).  Other: 
    \citet{lagache2004}; \citet{negrello2007}; 
    \citet{rowan-robinson2009}; \citet{valiante2009}; 
    \citet{pearson2009}; \citet{franceschini2010}; \citet{lacey2010}; 
    \citet{wilman2010}. 
} 
  \label{fig:countsdiffmodel857}  
\end{figure}

\subsection{\Planck\ Number Counts and Models}

\subsubsection{Models} 
 
Fig.~\ref{fig:countsdiffall} and \ref{fig:countsdiffallspt} display
our present estimates of number counts of extragalactic point sources,
based on ERCSC data, together with predictions from recent 
models of the numbers and evolution of extragalactic sources. 
These models focus either on radio-selected sources --
i.e. sources with spectra dominated by synchrotron radiation at
mm/submm wavelengths (``synchrotron sources''): \cite{de_zotti2005}
and \cite{tucci2011} -- or on far-IR selected sources -- i.e. sources
with spectra dominated by thermal cold dust emission at mm/submm
wavelengths (``dusty sources''): \cite{serjeant2005} and
\cite{bethermin2011a}.  Many other models exist in the literature,
among which are \cite{le_borgne2009}, \cite{negrello2007},
\cite{pearson2009}, \cite{rowan-robinson2009}, \cite{valiante2009},
\cite{franceschini2010}, \cite{lacey2010}, \cite{marsden2011},
\cite{wilman2010}, and \cite{rahmati2011}. A comparison is given with
these models in Fig.~\ref{fig:countsdiffmodel857} for 857\,GHz.

The \cite{de_zotti2005} model focusses on radio sources, both flat- and 
steep-spectrum, the latter having a component of dusty spheroidals and 
GPS (GHz peaked spectrum) sources. It includes cosmological evolution of 
extragalactic radio sources, based on an analysis of all the main source
populations at GHz frequencies.  It currently provides a good fit to all
data on number counts and on other statistics from $\sim 5\,$GHz up
to $\sim 100\,$GHz. This model adopts a simple power-law, with a very
flat spectral index ($\alpha\simeq -0.1$), for extrapolating the
spectra of the brightest extragalactic sources 
(essentially ``blazar\footnote{Blazars are jet-dominated extragalactic 
  objects, observed within a small angle of the jet axis and 
  characterized by a highly variable, non-thermal synchrotron emission 
  at GHz frequencies in which the beamed component dominates the 
  observed emission \citep{angel1980}.} sources'') to frequencies 
  above 100\,GHz. 
 
The \citet{tucci2011} models provide a description of three populations of
radio sources: steep-, flat-, and inverted-spectrum.
The flat-spectrum population is further 
divided into Flat-Spectrum Radio Quasars (FSRQ), and BL\,Lacs. The 
main novelty of these models is the statistical prediction of the 
``break'' frequency, $\nu_{\rm M}$, in the spectra of blazar jets 
modeled by classical, synchrotron-emission physics.  The most 
successful of these models, ``C2Ex'', assumes different distributions 
of break frequencies for BL Lac objects and Flat Spectrum Radio 
Quasars, with the relevant synchrotron emission coming from 
more compact regions in the jets of the former objects. This model, 
developed to fit both the Atacama Cosmology Telescope (ACT) 
data \citep{marriage2011} at 148\,GHz and the results published 
in the \Planck\ Early paper \citet{planck2011-6.1}, is able to give a 
very good fit to all published data on statistics of extragalactic 
radio sources: i.e.\ number counts and spectral index distributions.
The model ``C2Ex'' also correctly predicts the number of blazars observed
in the Herschel Astrophysical Terahertz Large Area Survey (H-ATLAS)
at 600\,GHz, as discussed in \citet{lopez-caniego2012}. 
 
The \citet{serjeant2005} model is based on the SED properties of local 
galaxies detected by {\it IRAS} and by SCUBA in the SLUGS sample 
\citep{dunne2000}. These local SEDs are used in many models, including 
the \citet{lapi2011} model (based on \citealt{lapi2006} and 
\citealt{granato2004}) which links dark matter halo masses with the mass
of black holes and the star formation rate. 
 
\citet{bethermin2011a} present a backwards evolution model, taking
into account IR galaxies, which is a parametric model fitting the fainter 
counts. It contains two families of SEDs: normal and starburst, from 
\citet{lagache2004}.

\subsubsection{Synchrotron sources} 
 
The \citet{de_zotti2005} model over-predicts the number counts of 
extragalactic “synchrotron” sources detected by \Planck\ at HFI 
frequencies. The main reason for this disagreement is the spectral 
“steepening” observed in ERCSC sources above about 70\,GHz 
\cite[]{planck2011-6.1,planck2011-6.3a}, and already suggested by 
other data sets \cite[]{gonzalez-nuevo2008,sadler2008}. 
 
The more recent ``C2Ex'' model by \cite{tucci2011} is able to give a
reasonable fit to the \Planck\ number counts on bright extragalactic
radio sources from 100 up to 545\,GHz (and marginally at 857\,GHz
where our data are noisy). However, our current data at 100 and
217\,GHz are consistently higher than the model number counts of
synchrotron sources in the faintest flux density bin probed by ERCSC
completeness-corrected data (300 and 600\,mJy, respectively). On the whole, 
however, the
``C2Ex'' model accounts well for the observed level of
bright extragalactic radio sources up to 545\,GHz.
 
\subsubsection{ Dusty sources} 
 
The \citet{serjeant2005} model performs reasonably well at 857\,GHz, but 
is lower than our observations at 545 and 353\,GHz.
The \citet{bethermin2011a} model has the same trend -- it is compatible with 
the data at 857\,GHz, but is lower than the observations by a factor of 
about 3 at 353 and 545\,GHz.  This is likely due to the limits of that 
model's validity at high flux density (typically above one Jy).
For both models, the likely origin of the discrepancy with our 
new, \Planck, high-frequency data is the models' inaccurate description 
of local SEDs.  Since the counts of bright sources at high frequency 
depend mainly on the SED of low-$z$, IR galaxies, rather than on cosmological 
evolution at higher redshifts, any discrepancy with models is telling 
us more about their accuracy in reproducing the averaged SED of the 
low-$z$ Universe than about any cosmological evolution. This effect is 
also seen as a discrepancy in the Euclidean level (Sect.~\ref{sect:euclidean} and 
Fig.~\ref{fig:euclidean}).

\subsubsection{Other models} 
 
Fig.~\ref{fig:countsdiffmodel857} shows the predictions of more models at 857\,GHz. Most 
of the models do not explicitly include the counts at such high flux 
densities (and/or are subject to numerical uncertainties, like 
\citealt{valiante2009,wilman2010}). We thus suggest that future model 
predictions extend up to a few tens of Jy in order to provide a good 
anchor for the SEDs at low redshift. At 857\,GHz, many models disagree 
with our data, e.g.\cite{negrello2007}, \cite{franceschini2010}, 
\cite{lacey2010}, \cite{rahmati2011}, \cite{rowan-robinson2010}. Other 
models agree or marginally agree with our data, 
e.g. \cite{lagache2004,pearson2009,bethermin2011a}. 
 
\subsubsection{Main results} 
The two main results from the comparison with models are: 1) the good
agreement of the \cite{tucci2011} model with our counts of
synchrotron-dominated sources, including for the first time at 353,
545 and marginally at 857\,GHz; and 2) the failure of most models to
reproduce the dusty-dominated sources between 353 and 857\,GHz. This
latter point is likely due to errors in the SEDs of local galaxies
used (i.e.\ at redshifts less than 0.1 and flux densities larger than 1\,Jy).

\begin{figure*}[!ht] 
   \centering 
  \includegraphics[width=0.98\textwidth,angle=0]{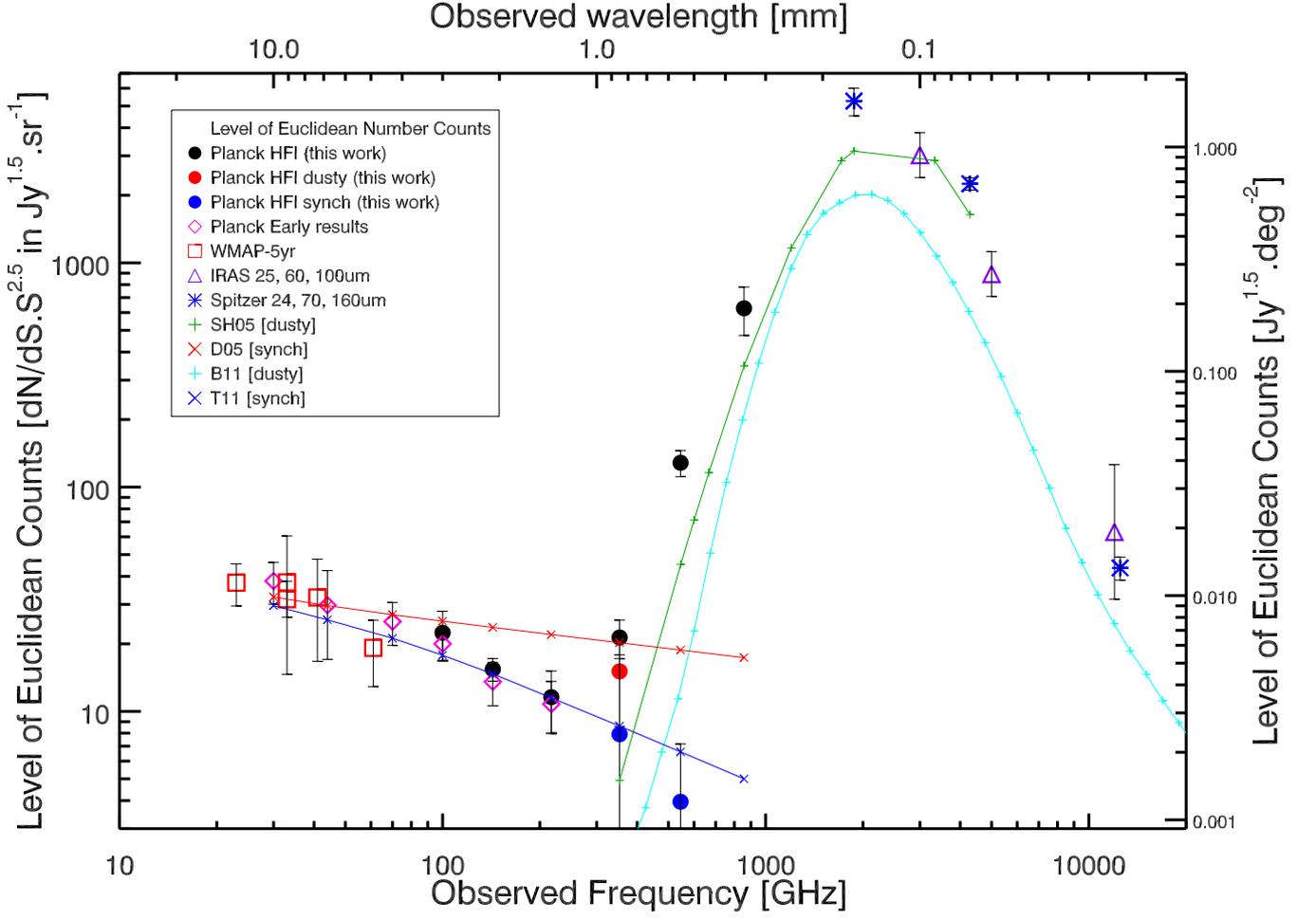} 
  \caption{Euclidean level $p$ (plateau in $ S^{2.5} dN/dS$, see
    Eq. ~\ref{eq:p}) for bright sources, expressed in number of galaxies times
    Jy$^{1.5}$\,sr$^{-1}$ (or Jy$^{1.5}$deg$^{-2}$ on the right axis)
    averaged between approximately 1 and 3\,Jy at microwave to mid-IR
    frequencies. We specifically show: our \Planck\ HFI results (black
    circles), and separately dusty sources (red circles) and synchrotron sources
    (blue circles).  The dusty 545\,GHz point is almost on top of the
    total \Planck\ point. Also shown are: \Planck\ Early results (purple diamonds)
    from \cite[]{planck2011-6.1}; {\it WMAP}-5year (red squares) from
    \cite{wright2009,massardi2009,de_zotti2010}; {\it IRAS} 25, 60 and
    100\micron\ results (purple triangles) from \cite{hacking91},
    \cite{lonsdale89}, and \cite{bertin97}; and {\it Spitzer} 24, 70 and
    160\micron\ (blue stars) from \cite{bethermin2010}.  We also plot
    the models (solid lines): \cite{serjeant2005}, based on {\it IRAS}
    and SCUBA data and dealing with dusty galaxies (SH05 green plus
    signs); \cite{bethermin2011a} (B11 light blue plus signs) dealing with
    dusty galaxies; \cite{de_zotti2005} (D05 red crosses) dealing with
    synchrotron sources; \cite{tucci2011} (T11 blue crosses) for synchrotron
    sources. Values are given in Tab.~\ref{tab:valuesp}. }
\label{fig:euclidean} 
\end{figure*}

\subsection{Beyond the number counts} 
\label{sect:euclidean} 
 
\subsubsection{\Planck\ observations of the Euclidean level} 

The Euclidean level of the number counts, described as the plateau level,
$p$, in the normalised differential number counts at high flux density, 
\begin{equation} 
  dN/dS = p\ S^{-2.5} 
\label{eq:p} 
\end{equation} 
mainly depends on the SED shape of galaxies (local galaxies in the 
case of high frequency observations).

Figure~\ref{fig:euclidean} shows $p$ over more than two orders of 
magnitude in observed frequency, from the mid-IR to the radio 
range. The values of $p$ are reported in Table~\ref{tab:valuesp}. The 
Euclidean level was determined using number counts above 1\,Jy (except 
in the case of {\it Spitzer}, where number counts at fainter flux densities were 
used). Beyond our measurements at \Planck\ HFI frequencies (total in 
black, but also shown by source type: dusty and synchrotron), we also show the 
\Planck\ early results at LFI and HFI 100\,GHz frequencies 
\cite[]{planck2011-6.1}, as well as {\it WMAP}-5year results at Ka 
\cite[]{wright2009} and in all bands 
\cite[]{massardi2009,de_zotti2010}, and finally 
{\it IRAS} 25, 60 and 100\micron\ results 
\cite[]{lonsdale89,hacking91,bertin97}. The {\it Spitzer} level at 24, 
70 and 160\micron\ comes from counts above 8, 70 and 300\,mJy,
respectively \citep{bethermin2010}.
We also plot the models of \citet{serjeant2005} (based on {\it IRAS} and
SCUBA 850\micron\ local colors), of \citet{de_zotti2005}, of
\citet{bethermin2011a}, and of \citet{tucci2011}. 
 
As expected from our data on number counts discussed above, our
current and the early \Planck\ estimates are in good agreement at
100\,GHz. Also, the \Planck\ LFI and {\it WMAP} estimates agree within
the error bars. The \Planck\ contribution is unique in disentangling
the dusty from synchrotron sources in the key spectral regime around
300\,GHz where the two populations contribute equally to the Euclidean
level.
 
Likewise, the \Planck\ measurements of synchrotron sources between 30
and 217\,GHz at LFI frequencies and lower HFI frequencies are very
well reproduced by the \citet{tucci2011} model ``C2Ex'', as is the
Euclidean level for synchrotron sources at 353\,GHz.

The \Planck\ measurements lie above the \citet{serjeant2005} and
\citet{bethermin2011a} models at the three upper HFI frequencies
between 353 and 857\,GHz. There are two explanations for this: (1) the
presence of synchrotron galaxies in equal numbers to dusty galaxies
between 217 and 353\,GHz which are not seen in the {\it IRAS}
60\micron\ selected sample; and (2) the cold dust component in the
local Universe.  Although the presence of cold dust has been known for
some time \citep{stickel98,dunne2000}, its effects may have been
underestimated, as suggested in \cite{planck2011-6.4a}. There is a
significant and largely unexplored cold ($T<20$\,K) component in many
nearby galaxies.  This excess of submm emission is statistically
confirmed here.  At 545\,GHz for instance, we measure $p=(125 \pm
16)\,$Jy$^{1.5}\,$sr$^{-1}$ for the dusty galaxies; the
\citet{serjeant2005} model predicts $45$\,Jy$^{1.5}$\,sr$^{-1}$ (a
factor of 2.7 lower) and the \citet{bethermin2011a} predicts
$10\,$Jy$^{1.5}$\,sr$^{-1}$ (a factor of 12 lower).  At 353\,GHz, we
measure $p=13 \pm 7$\,Jy$^{1.5}$\,sr$^{-1}$ for the dusty galaxies,
while the \citet{serjeant2005} model predicts
$4.92$\,Jy$^{1.5}$\,sr$^{-1}$ (a factor of 2.7 lower).  This is in
line with the cooler 60\micron:450\micron\ colour (i.e, smaller 60/450
flux ratio) found in ERCSC sources \cite[][e.g.\ their
figure.~4]{planck2011-6.4a}.  Unlike the case of the SLUGS sample
\citep{dunne2000}, \Planck\ ERCSC sources can have
60\micron:450\micron\ flux ratios up to ten times smaller.

\subsubsection{Link between the Euclidean level for dusty galaxies and 
the local luminosity density} 
\label{sect:lumdensity} 
 
In the IR and submm, the bright counts of dusty galaxies probe only 
the local Universe, which can be approximated as a Euclidean space 
filled with non-evolving populations. The volume $V_{\rm max}$ where a 
source with a luminosity density $L_\nu$ is seen with a flux density 
larger than $S_{\nu, \rm lim}$ is: 
\begin{equation} 
V_{\rm max} =  \frac{4 \pi}{3} D_{\rm max}^3 = \frac{4 \pi}{3}  \left ( \frac{L_\nu}{4 \pi S_{\nu, \rm lim}} \right ) ^{\frac{3}{2}}, 
\end{equation} 
where $D_{\rm max}$ is the maximum distance at which a source can 
be detected, and $S_{\nu, \rm lim}$ the limiting flux density at 
frequency $\nu$. The contribution of sources with $L_\nu - dL_\nu/2 < 
L_\nu < L_\nu + dL_\nu/2$ to the counts is then 
\begin{equation} 
\frac{dN(S_\nu>S_{\nu, \rm lim})}{dL_\nu} = \frac{d^2 N}{dL_\nu dV} \times \frac{L_\nu^{\frac{3}{2}}}{S_{\nu, \rm lim}^{\frac{3}{2}} 6 \sqrt{\pi}}, 
\end{equation}. 
where $N(S_\nu>S_{\nu, \rm lim})$ is the number of sources brighter 
than $S_{\nu, \rm lim}$ over the entire sky and $\frac{d^2 N}{dL_\nu 
  dV}$ the local luminosity function. The integral counts 
$dN(S_\nu>S_{\nu, \rm lim})/d\Omega$ are linked to this local 
luminosity function by: 
\begin{equation} 
\frac{dN(S_\nu>S_{\nu, \rm lim})}{d\Omega} =  \frac{1}{4 \pi} \int_{L_\nu = 0}^\infty \frac{d^2 N}{dL_\nu dV} \times \frac{L_\nu^{\frac{3}{2}}}{S_{\nu, \rm lim}^{\frac{3}{2}} 6 \sqrt{\pi}}\ dL_\nu 
\end{equation} 
The differential counts $d^2N/(dS_\nu d\Omega)$ are thus 
\begin{equation} 
\frac{d^2 N}{dS_\nu d\Omega} = \frac{S_{\nu}^{\frac{-5}{2}}}{16 \pi^{\frac{3}{2}}} \int_{L_\nu = 0}^\infty \frac{d^2 N}{dL_\nu dV} \times L_\nu^{\frac{3}{2}}\ dL_\nu 
\end{equation}, 
and the level $p$ of the Euclidean plateau is thus 
\begin{equation} 
p_\nu = \frac{1}{16 \pi^{\frac{3}{2}}} \int_{L_\nu = 0}^\infty \frac{d^2 N}{dL_\nu dV} \times L_\nu^{\frac{3}{2}}\ dL_\nu 
\end{equation}.

The local monochromatic luminosity density $\rho_\nu$ can be computed as 
\begin{equation} 
\rho_\nu = \int_{L_\nu = 0}^\infty \frac{d^2 N}{dL_\nu dV} \times L_\nu\ dL_\nu 
\end{equation} 
If we assume a single mean color $C$ between frequencies $\nu_1$ and 
$\nu_2$ (with $S_{\nu_1} = C S_{\nu_2}$) for all the sources, we simply 
have the relation 
\begin{equation} 
\frac{\rho_{\nu_2}}{\rho_{\nu_1}} = C. 
\end{equation} 
We make this assumption for simplicity.  Note, however, that the
strongly peaked distributions of spectral indices from
Figure~\ref{fig_alphas} at 857 and 545\,GHz are consistent with this
assumption.  At 353\,GHz, the lowest frequency we consider in this
analysis, the situation is complicated by the appearance of some
synchrotron sources.  Their effect, however, is small compared to
other uncertainties in the calculation of $\rho$.  If we perform the
same analysis on the level of the Euclidean plateau, we obtain
\begin{equation} 
\frac{p_{\nu_2}}{p_{\nu_1}} = C^{\frac{3}{2}}. 
\end{equation} 
The quantities $p_\nu$ and $\rho_\nu$ are thus linked by 
\begin{equation} 
\frac{\rho_{\nu_2}}{\rho_{\nu_1}} = \left ( \frac{p_{\nu_2}}{p_{\nu_1}} \right )^{\frac{2}{3}}
\end{equation}
 
We could use $\rho_{60}$ (the {\it IRAS} local luminosity density at 
60\micron) and $p_{60}$ (the Euclidean level at 60\micron) as a 
reference, in order to derive $\rho_{\nu}$, the luminosity density of 
dusty galaxies at frequency $\nu$ (with {\it IRAS} as a reference): 
 
\begin{equation} 
\rho_{\nu} = \left ( \frac{p_{\nu}}{p_{60}} \right)^{\frac{2}{3}} \rho_{60}
\label{eq:rho_nu_from_60} 
\end{equation}
 
However, the extrapolation of the dust emission from the FIR to the 
(sub-)millimetre wavelengths is uncertain (as our data show). We might 
instead want to use the luminosity density estimated at 850\micron\ 
from previous studies, and correct it to account for the excess 
observed by \Planck. We can thus use: 
\begin{equation} 
\rho_{\nu} = \left ( \frac{p_{\nu}}{p_{850}} \right)^{\frac{2}{3}} \rho_{850} 
\label{eq:rho_nu_from_850} 
\end{equation} 
 
Both the 60\micron-based and the 850\micron-based estimates are shown 
in Fig.~\ref{fig:rhoir_from_p_and_iras} and discussed in the next 
section. 
 
\subsubsection{Estimate of the local luminosity density for dusty 
  galaxies} 
\label{sect:dust} 
We use two reference wavelengths to derive $\rho_{\nu}$:  
 
\noindent$\bullet$ at 60\micron\ we use {\it IRAS} data: $\rho_{60}$ is 
estimated by \citet{soifer91} and \citet{takeuchi2006}; $p_{60}$ by 
\citet{soifer91} and \citet{bertin97}; 
 
\noindent$\bullet$ at 850\micron\ we use SCUBA SLUGS: $\rho_{850}$ is 
estimated by \citet{dunne2000} and \citet{takeuchi2006}; $p_{850}$ by 
\citet{serjeant2005}. 
 
\noindent The values from these references are: $\rho_{60}=4.08 \times 10^7 h$ L$_{\odot}$ 
Mpc$^{-3}$ and $p_{60}=891.3$\,Jy$^{1.5}$\,sr$^{-1}$ at 60\micron; 
$\rho_{850}=9.45 \times 10^4 h$ L$_{\odot}$ Mpc$^{-3}$ and 
$p_{850}=4.92$\,Jy$^{1.5}$\,sr$^{-1}$.  The use of two reference 
wavelengths is driven by the oversimplified hypothesis of a constant 
color $C$ between two frequencies (assumptions described in 
Sect.~\ref{sect:lumdensity}). Computing $\rho_{\nu}$ using two 
different reference wavelengths allows us to estimate the impact of 
this hypothesis. 
 
\begin{figure}[!t] 
   \centering 
  \includegraphics[width=0.49\textwidth,angle=0]{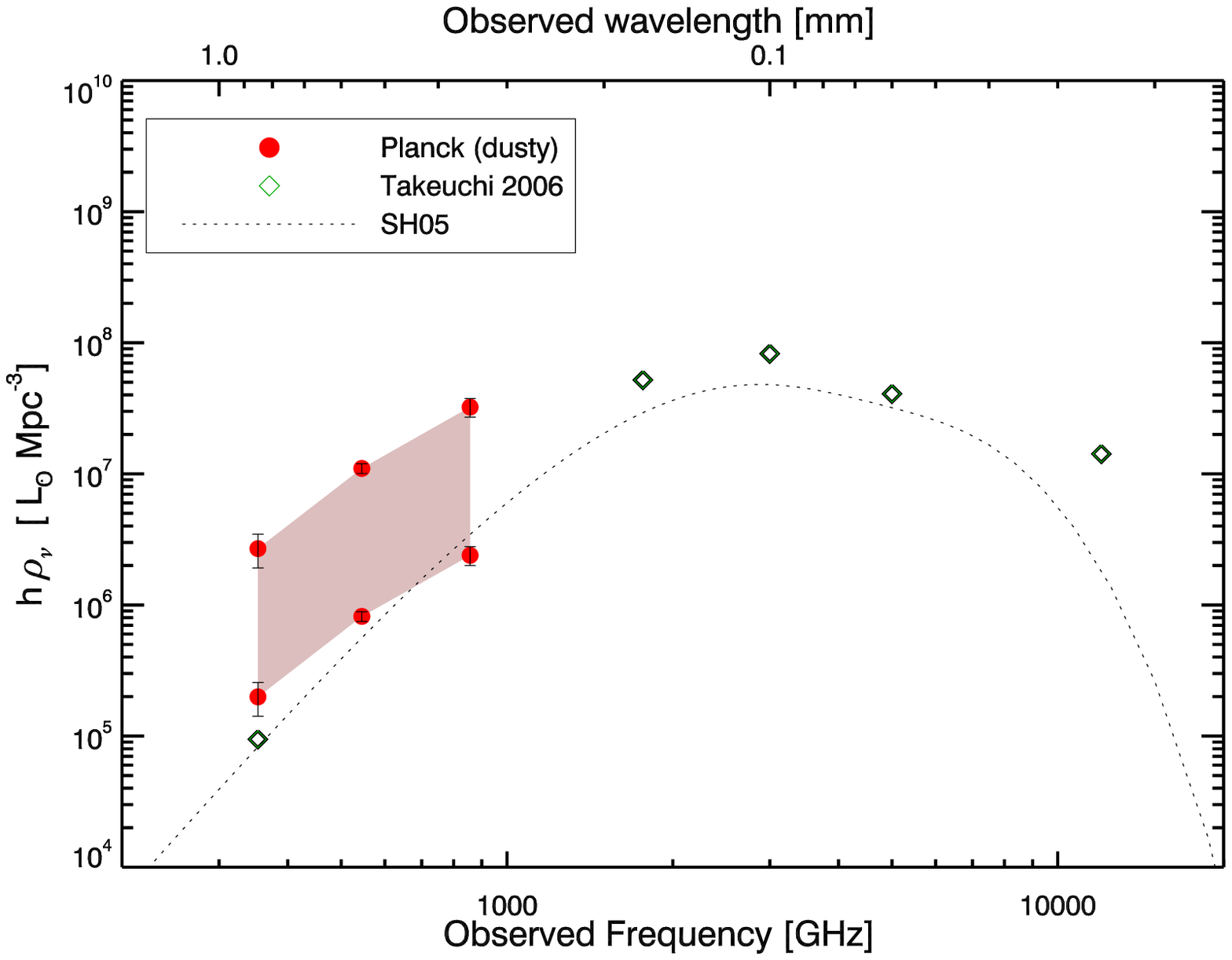} 
  \caption{Luminosity density $\rho_{\nu}$ (in units of 
    $h$ \Lsolar\, Mpc$^{-3}$) of dusty galaxies, derived from the 
    Euclidean level $p$ and scaled to the luminosity density of: SCUBA 
    850\micron\ data \cite[]{dunne2000,serjeant2005} (see 
    Eq.~\ref{eq:rho_nu_from_850}); or {\it IRAS} 60\micron\ data 
    \cite[]{soifer91,bertin97,takeuchi2003} (see 
    Eq.~\ref{eq:rho_nu_from_60}). Red circles: estimate from \Planck\ 
    for dusty galaxies (this work); green diamonds: compilation from 
    \cite{takeuchi2006}; black dots: model from \cite{serjeant2005}. 
    Values are given in Tab. ~\ref{tab:rhonu}. Our \Planck\ estimates 
    lie in the shaded area.  Note that we take $h=0.71$ here.} 
\label{fig:rhoir_from_p_and_iras} 
\end{figure}

The results of our estimated luminosity densities from this simple 
model are shown in Fig.~\ref{fig:rhoir_from_p_and_iras}: lower points 
using the SCUBA 850\micron\ reference (Eq.~\ref{eq:rho_nu_from_850}); 
upper points using the {\it IRAS} 60\micron\ reference 
(Eq.~\ref{eq:rho_nu_from_60}).  The values are given in 
Tab.~\ref{tab:rhonu}. 
 
As expected, our \Planck\ indirect upper estimate is higher than SLUGS 
at 353\,GHz if we use 850\micron\ as a reference. This is clearly consistent
with our value of $p$ being 2.7 times higher, implying a factor of 2 
(i.e. $2.7^{2/3}$) in the luminosity densities. On the other hand, our 
353\,GHz estimate using 60\micron\ as a reference falls way above the 
SCUBA estimate at 850\micron. This again illustrates that caution is required
when extrapolating FIR colors to the submm. 
 
The true luminosity density should lie between our lower and upper 
estimates; the ratio equals 13.5. At 353\,GHz, our estimate using SCUBA 
as a reference should be more appropriate to use than the {\it IRAS} 
extrapolation.

\section{Conclusion and Summary}

From the \Planck\ all-sky survey, we derive extragalactic number 
counts based on the ERCSC \cite[]{planck2011-1.10} from 100 to 857\,GHz 
(3\,mm to 350\micron). We use an 80\,\% completeness cut on three 
homogeneous zones, covering a total of about 16000\,deg${}^2$ 
($f_{\rm{sky}}\sim$0.31 to 0.40) outside a Galactic mask. We provide, 
for the first time, bright extragalactic source counts at 353, 545 and 857\,GHz
(i.e., 850, 550 and 350\micron; see Fig.~\ref{fig:countsdiffall}). 
Our counts are in the Euclidean regime, and generally agree with other 
data sets, when available (Fig.~\ref{fig:countsdiffallspt}).  
 
Using multi-frequency information to classify the sources as dusty- or 
synchrotron-dominated (and measure their spectral indices), the most 
striking result is the contribution to the number counts by each 
population. The cross-over takes place at high frequencies, between 
217 and 353\,GHz, where both populations contribute almost equally to 
the number counts. At higher or lower frequencies, counts are quickly 
dominated by one or other population. We provide for the first time 
number counts estimates of synchrotron-dominated sources at high 
frequency (353 to 857\,GHz) and dusty-dominated sources at lower 
frequencies (217 and 353\,GHz). 
 
Our counts provide new constraints on models which extend their
predictions to bright flux densities.  Existing models of
synchrotron-dominated sources are not far off from our observations,
with the model ``C2Ex'' of \cite{tucci2011} performing particularly
well at reproducing the synchrotron-dominated source counts up to 545\,GHz
(and marginally up to 857\,GHz, where our statistics become sparse).
Perhaps less expected is the failure of most models of dusty
sources to reproduce all the high-frequency counts.
The model of \citet{bethermin2011a} agrees marginally at 857\,GHz but is
too low at 545\,GHz and at lower frequencies, while the model of
\citet{serjeant2005} is marginally lower at 857\,GHz, fits the data well
at 545\,GHz, but is too low at 353\,GHz.
The likely origin of the discrepancies is an inaccurate description of the
SEDs for galaxies at low redshift in these models.
Indeed a cold dust component, seen e.g.\ by \citet{planck2011-6.4a},
is rarely included in the models at low redshift.
This failure to reproduce high-frequency counts at bright flux density
should not have any impact on the predictions at fainter flux densities
and higher redshifts, as is shown in the good fit to {\it Herschel} counts.
Nevertheless it tells us about the ubiquity of cold dust in the local
Universe, at least in statistical terms.
 
Finally, in Fig.~\ref{fig:euclidean}, we provide a review of the 
Euclidean plateau level $p$ of the number counts, spanning nearly 
three orders of magnitude in both frequency and counts. The values of 
$p$ are calculated for flux densities above 1\,Jy, except in the case 
of {\it Spitzer} where fainter objects are used.
Fig.~\ref{fig:euclidean} compares these results with some relevant models.
The $p$ value is usually not well reproduced by models (at least for 
\citealt{de_zotti2005,serjeant2005,bethermin2011a}) in the 
synchrotron- or dust-dominated regimes.
The \cite{tucci2011} model, on the contrary, reproduces our observations
of synchrotron sources, up to 545\,GHz.
This multifrequency diagnostic is a powerful tool for investigating the
SEDs of galaxies in the context of cosmological evolution -- at relatively
low redshifts for the dusty galaxies.  We derive a range of values for the local luminosity density
for dusty galaxies, based on simple considerations and using the
SCUBA 850\micron\ and {\it IRAS} 60\micron luminosity density as a reference. 
 
The \Planck\ multi-frequency all-sky survey is very rich dataset, in
particular for extragalactic studies
(e.g. \citealt{negrello2012}). The final \Planck\ catalogue of sources
will be based on five complete sky surveys, while the present work is
based on only 1.6 surveys.  With this improved data set, we expect to
provide further constraints on the synchrotron and dust-dominated
populations at all frequencies, and over a wider range in redshift.

\input{euclidean_levels_p_planck2012wg6.tex}

\input{rho_nu_planck2012wg6.tex}

\begin{acknowledgements} 
Based on observations obtained with Planck (http://www.esa.int/Planck), an ESA science mission with instruments and contributions directly funded by ESA Member States, NASA, and Canada.

The development of Planck has been supported by: ESA; CNES and
CNRS/INSU-IN2P3-INP (France); ASI, CNR, and INAF (Italy); NASA and DoE
(USA); STFC and UKSA (UK); CSIC, MICINN and JA (Spain); Tekes, AoF and
CSC (Finland); DLR and MPG (Germany); CSA (Canada); DTU Space
(Denmark); SER/SSO (Switzerland); RCN (Norway); SFI (Ireland);
FCT/MCTES (Portugal); and PRACE (EU). This research has made use of
the SIMBAD database, operated at CDS, Strasbourg, France. This
research has made use of the NASA/IPAC Extragalactic Database (NED)
which is operated by the Jet Propulsion Laboratory, California
Institute of Technology, under contract with the National Aeronautics
and Space Administration.  This research has made use of the NASA/
IPAC Infrared Science Archive, which is operated by the Jet Propulsion
Laboratory, California Institute of Technology, under contract with
the National Aeronautics and Space Administration. This publication
makes use of data products from the Wide-field Infrared Survey
Explorer, which is a joint project of the University of California,
Los Angeles, and the Jet Propulsion Laboratory/California Institute of
Technology, funded by the National Aeronautics and Space
Administration.
\end{acknowledgements} 
 

\bibliographystyle{natbib} 

\appendix

\section{Spectral classification; effect of intermediate sources and 
  photometric noise} 
\label{sect:interm} 
 
In this appendix, we investigate the fate and influence of the so-called 
``intermediate'' sources as defined in 
Sect.~\ref{sect:classification}.

\begin{figure}[!ht] 
   \centering 
  \includegraphics[width=0.5\textwidth,angle=0]{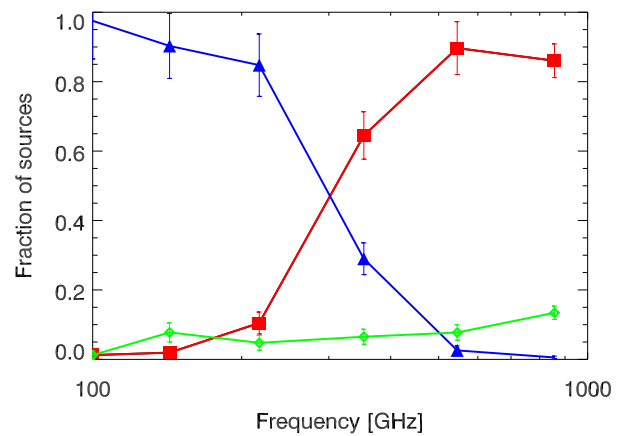} 
  \caption{As in Fig.~\ref{fig:statsed}, the fraction of source type 
    (dusty, red squares; synch, blue triangles) as a function of 
    frequency.  The difference is that we have now included the 
    ``intermediate'' population (green diamonds). We can see that at 
    most 13\,\% of our classification as dusty or synchrotron can be 
    affected by intermediate sources. Our number counts by type (above 
    80\,\% completeness) are thus almost unaffected by these 
    intermediate type sources.} 
  \label{fig:fraction_intermediate_complete} 
\end{figure}

Fig.~\ref{fig:fraction_intermediate_complete} shows the fraction of 
sources (like Fig.~\ref{fig:statsed}) by type (dusty, synchrotron, and 
now intermediate) as a function of frequency computed for sources 
above 80\,\% completeness. The fraction is at most 13\,\%, and is on 
average around 7\,\%. The intermediate source population thus has no 
impact on our number counts by type. 
 
We conclude that a genuine population of intermediate sources exist 
(i.e. having both a thermal dust emission component and a synchrotron 
component) but its contribution in number is less than typically 10\,\% 
(Fig.~\ref{fig:fraction_intermediate_complete}). Notice that a 
free-free emission can also play a role in the spectrum flattening 
around 100\,GHz \citep{peel2011}. 
 
\begin{figure}[!ht] 
   \centering 
  \includegraphics[width=0.5\textwidth,angle=0]{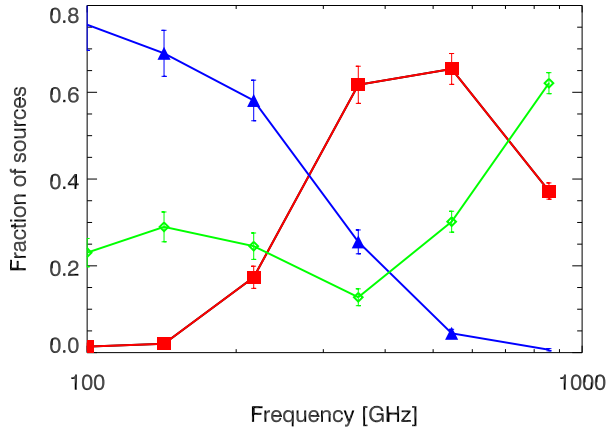} 
  \caption{Like Fig.~\ref{fig:fraction_intermediate_complete}, the 
    fraction of source type (dusty, red squares; synchrotron, blue 
    triangles; green diamonds, intermediate) as a function of 
    frequency. The difference is that we have now included the whole 
    catalogue, i.e. including sources affected by more photometric 
    noise below the 80\,\% completeness limit cut. The effect of 
    increasing noise is to induce more sources to be classified as 
    intermediate.} 
  \label{fig:fraction_intermediate_all} 
\end{figure}

We notice, however, that this intermediate populations lies at the 
faint end of the flux distribution (i.e. they usually are among the 
faintest sources of our sample).  To investigate further if the 
presence of intermediate sources is linked to the level of photometric 
noise, we performed the classification on our whole sample, 
thus including sources at fluxes below the 80\,\% completeness 
limit. The results, shown in 
Fig.~\ref{fig:fraction_intermediate_all}, indicates that the 
higher the photometric noise the more sources are classified as 
intermediate. 
 
When using the total sample (i.e. with sources fainter that the 80\,\% 
completeness cut), the fraction of intermediate sources increases, but 
those sources are always at the faint end of the flux distribution: 
the effect of the photometric noise is thus mainly responsible for the uncertain 
classification. This emphasises that we should use highly-complete 
samples for such statistical analysis, in order not to be biased 
towards mis-classification.

\section{Some peculiarities; individual sources or groups of sources} 
\label{sect:individualsources} 
 
While the SEDs of some particular sources have been published in the 
\Planck\ early papers 
\cite{planck2011-6.2,planck2011-6.3a,planck2011-6.4a}, we review here 
some specific sources detected at low or high frequency, but with 
unexpected classifications. 
 
\subsection{Low-frequency dusty galaxies} 
 
There are seven low-frequency sources (three detected at 100\,GHz and four at 
143\,GHz) that are classified as dusty galaxies. This kind of 
classification is not necessarily expected, unless we detect local 
galaxies showing both radio and infrared components. For this
reason we check them individually.\\
 
1) PLCKERC100 G062.69$-$14.07:
There is no radio identification in NVSS \& GB6 or in NED, and no 
detection at LFI frequencies.  This source is likely correctly 
classified as a dusty galaxy.\\

2) PLCKERC100 G140.41$-$17.39: This source is found with NED to be NGC 
891. There is no LFI detection, but detections in NVSS/GB6. We might 
be seeing two spectral components (dusty and synchrotron) of this nearby galaxy.\\ 
 
3) PLCKERC100 G141.42+40.57:
This is NGC3034 (M82). As above we are sensitive to both components of 
this nearby and well-studied galaxy.\\ 
 
4) PLCKERC143 G001.33$-$20.49:
No LFI detection nor radio identification. This source is correctly 
classified as a dusty galaxy.\\ 
 
5) PLCKERC143 G148.59+28.70:  This source is likely a blazar with an 
almost flat spetrcum at high frequencies and detections in NVSS and 
GB6. This source is likely incorrectly classified as dusty, because of
the small jump in flux at 545\,GHz.\\ 
 
6) PLCKERC143 G236.47$-$14.38:
No LFI detection nor radio identification. This source is correctly 
classified as a dusty galaxy.\\ 
 
7) PLCKERC143 G349.61$-$52.57:
No LFI detection. At 0.2 to 20\,GHz it is identified as a
flat-spectrum source but its high-frequency spectrum shows a clear
dusty behaviour. This source is correctly classified as a dusty
galaxy, although a clear radio component is detected.
 
\subsection{High frequency synchrotron galaxies} 
 
There are four sources classified at synchrotron sources at 857\,GHz. We 
also check them individually.\\
 
1) PLCKERC857 G206.80+35.82:
This is a confirmed blazar detected with {\it WMAP} . This 
source is correctly classified as synchrotron-dominated.\\ 
 
2) PLCKERC857 G237.75$-$48.48:
This is a confirmed blazar detected with {\it WMAP}  and {\it ATCA}.
This source is correctly classified as synchrotron-dominated.\\ 
 
3) PLCKERC857 G250.08$-$31.09:
This is a confirmed blazar detected with {\it WMAP}  and {\it ATCA}.
This source is correctly classified as synchrotron-dominated.\\ 
 
4) PLCKERC857 G148.24+52.44:
This is NGC 3408, quite faint for \Planck\ at high frequencies
(812\,mJy at 857\,GHz and not detected at 545\,GHz).
This source, although in our sample defined in Sect.~\ref{sect:sample},
was not used in the number counts because of the low completeness level
at this flux density.

\subsection{Bump at 4\,Jy at 545\,GHz in the medium zone} 
\label{sect:bump545} 
 
\begin{figure*}[!ht] 
   \centering 
  \includegraphics[width=0.95\textwidth,angle=0]{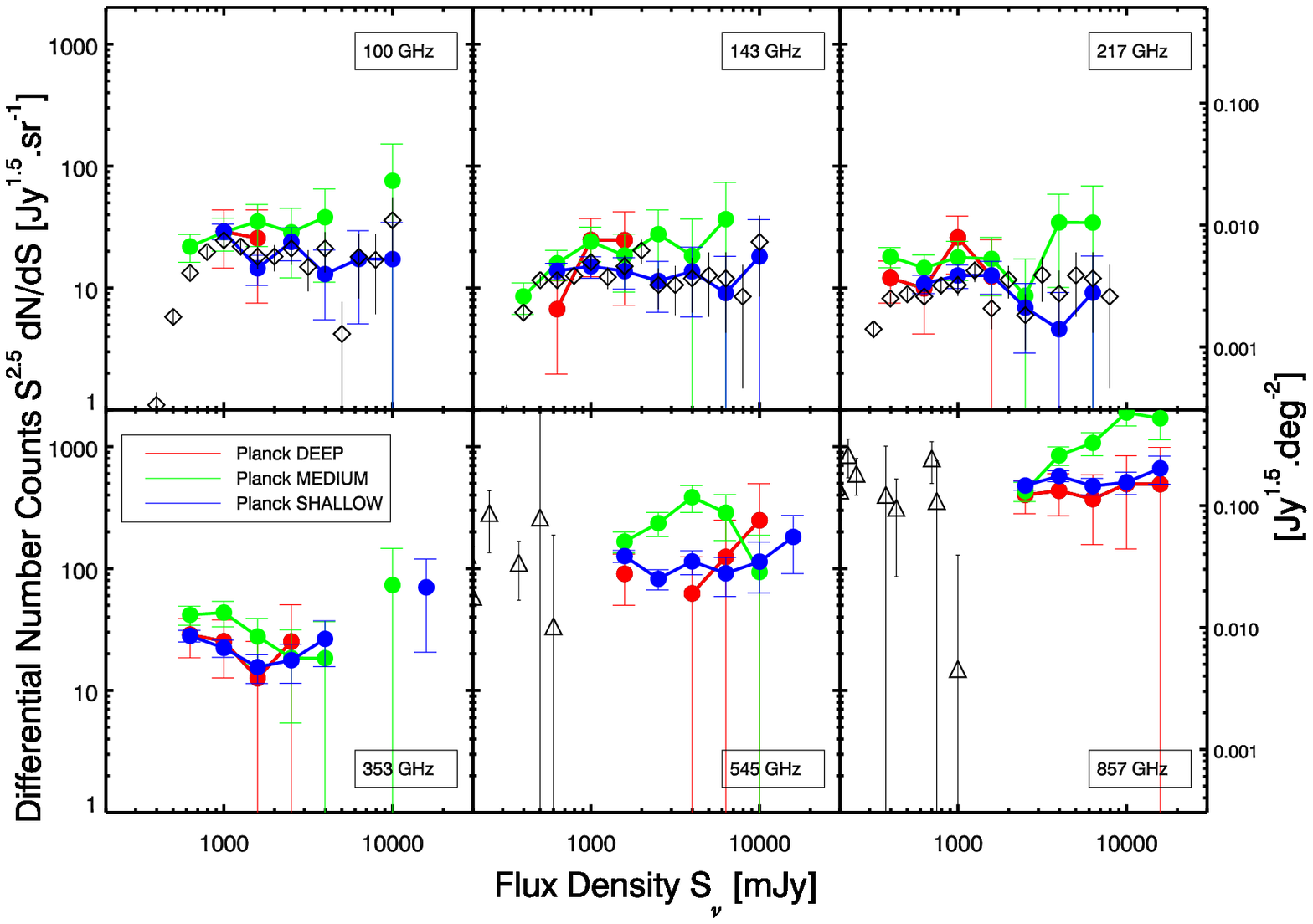} 
  \caption{\Planck\ differential number counts (total, dusty and synchrotron) at 
    6 frequencies between 100 and 857\,GHz, normalized to the 
    Euclidean, for each zone (filled circles): deep (red), medium 
    (green) and shallow (blue). Diamonds are from \Planck\ HFI 
    \cite[]{planck2011-6.1}, triangles from Herschel SPIRE 
    \citep{oliver2010,clements2010} and BLAST 
    \cite{bethermin2010a}. The bump at 4\,Jy at 545\,GHz (and at 10\,Jy 
    at 857\,GHz) in the medium zone is discussed in 
    Sect.~\ref{sect:bump545}.} 
  \label{fig:countsbyzone} 
\end{figure*}

As discussed in Sect.~\ref{sect:counts} and shown in 
Fig.~\ref{fig:countsbyzone}, there is an excess of 545\,GHz sources in the 
medium zone at 4\,Jy, which is also seen at 857\,GHz at 10\,Jy and at 353\,GHz
at 1\,Jy. This bump is created at 545 by 18 sources (between 3 and 
5\,Jy). Among the sources, we find NGC 3147, NGC 4449, NGC 4217, NGC 
3992, NGC 4088, NGC 4096, NGC 4051, NGC 3631, NGC 3938, IC 0750, NGC 
4244, NGC 3726, NGC 4214, NGC 7582 and NGC 7552.  At 857\,GHz, we find 
20 sources between 7.6 and 12\,Jy, with many in common with the list above. These 
sources are not physically associated and are spread over a large 
surface of the sky, although the majority lie around (150 deg., 60 deg.) 
in Galactic coordinates.

\section{Number counts by zone} 
\label{sect:countzone} 
 
Fig.~\ref{fig:countsbyzone} shows the number counts for each of the three 
zones: deep, medium, and shallow. This illustrates the sample
variance, as mentioned in Appendix~\ref{sect:bump545}. 

Since, with the exception of one zone at 545\,GHz, there is little
difference between the counts in different zones, this figure also
demonstrates that the weight given to each zone in calculating the
total number counts has little influence on the result.
 
\end{document}

%% file: PlanckCommands.tex
\def\setsymbol#1#2{\expandafter\def\csname #1\endcsname{#2}}
\def\getsymbol#1{\csname #1\endcsname}

\def\Planck{\textit{Planck}}

\newbox\tablebox    \newdimen\tablewidth
\def\leaderfil{\leaders\hbox to 5pt{\hss.\hss}\hfil}
\def\endPlancktable{\tablewidth=\columnwidth 
    $$\hss\copy\tablebox\hss$$
    \vskip-\lastskip\vskip -2pt}
\def\endPlancktablewide{\tablewidth=\textwidth 
    $$\hss\copy\tablebox\hss$$
    \vskip-\lastskip\vskip -2pt}
\def\tablenote#1 #2\par{\begingroup \parindent=0.8em
    \abovedisplayshortskip=0pt\belowdisplayshortskip=0pt
    \noindent
    $$\hss\vbox{\hsize\tablewidth \hangindent=\parindent \hangafter=1 \noindent
    \hbox to \parindent{$^#1$\hss}\strut#2\strut\par}\hss$$
    \endgroup}
\def\doubleline{\vskip 3pt\hrule \vskip 1.5pt \hrule \vskip 5pt}

\def\L2{\ifmmode L_2\else $L_2$\fi}

\def\DeltaT{\ifmmode \Delta T\else $\Delta T$\fi}
\def\deltat{\ifmmode \Delta t\else $\Delta t$\fi}
\def\fknee{\ifmmode f_{\rm knee}\else $f_{\rm knee}$\fi}
\def\Fmax{\ifmmode F_{\rm max}\else $F_{\rm max}$\fi}
\def\solar{\ifmmode{\rm M}_{\mathord\odot}\else${\rm M}_{\mathord\odot}$\fi}
\def\Msolar{\ifmmode{\rm M}_{\mathord\odot}\else${\rm M}_{\mathord\odot}$\fi}
\def\Lsolar{\ifmmode{\rm L}_{\mathord\odot}\else${\rm L}_{\mathord\odot}$\fi}

\def\inv{\ifmmode^{-1}\else$^{-1}$\fi}
\def\mo{\ifmmode^{-1}\else$^{-1}$\fi}
\def\sup#1{\ifmmode ^{\rm #1}\else $^{\rm #1}$\fi}
\def\expo#1{\ifmmode \times 10^{#1}\else $\times 10^{#1}$\fi}
\def\,{\thinspace}
\def\lsim{\mathrel{\raise .4ex\hbox{\rlap{$<$}\lower 1.2ex\hbox{$\sim$}}}}
\def\gsim{\mathrel{\raise .4ex\hbox{\rlap{$>$}\lower 1.2ex\hbox{$\sim$}}}}

\def\simprop{\mathrel{\raise .4ex\hbox{\rlap{$\propto$}\lower 1.2ex\hbox{$\sim$}}}}
\def\deg{\ifmmode^\circ\else$^\circ$\fi}
\def\pdeg{\ifmmode $\setbox0=\hbox{$^{\circ}$}\rlap{\hskip.11\wd0 .}$^{\circ}
          \else \setbox0=\hbox{$^{\circ}$}\rlap{\hskip.11\wd0 .}$^{\circ}$\fi}
\def\arcs{\ifmmode {^{\scriptstyle\prime\prime}}
          \else $^{\scriptstyle\prime\prime}$\fi}
\def\arcm{\ifmmode {^{\scriptstyle\prime}}
          \else $^{\scriptstyle\prime}$\fi}
\newdimen\sa  \newdimen\sb
\def\parcs{\sa=.07em \sb=.03em
     \ifmmode \hbox{\rlap{.}}^{\scriptstyle\prime\kern -\sb\prime}\hbox{\kern -\sa}
     \else \rlap{.}$^{\scriptstyle\prime\kern -\sb\prime}$\kern -\sa\fi}
\def\parcm{\sa=.08em \sb=.03em
     \ifmmode \hbox{\rlap{.}\kern\sa}^{\scriptstyle\prime}\hbox{\kern-\sb}
     \else \rlap{.}\kern\sa$^{\scriptstyle\prime}$\kern-\sb\fi}
\def\ra[#1 #2 #3.#4]{#1\sup{h}#2\sup{m}#3\sup{s}\llap.#4}
\def\dec[#1 #2 #3.#4]{#1\deg#2\arcm#3\arcs\llap.#4}
\def\deco[#1 #2 #3]{#1\deg#2\arcm#3\arcs}
\def\rra[#1 #2]{#1\sup{h}#2\sup{m}}

\def\dots{\relax\ifmmode \ldots\else $\ldots$\fi}
\def\WHzsr{\ifmmode $W\,Hz\mo\,sr\mo$\else W\,Hz\mo\,sr\mo\fi}
\def\mHz{\ifmmode $\,mHz$\else \,mHz\fi}
\def\GHz{\ifmmode $\,GHz$\else \,GHz\fi}
\def\mKs{\ifmmode $\,mK\,s$^{1/2}\else \,mK\,s$^{1/2}$\fi}
\def\muKs{\ifmmode \,\mu$K\,s$^{1/2}\else \,$\mu$K\,s$^{1/2}$\fi}
\def\muKRJs{\ifmmode \,\mu$K$_{\rm RJ}$\,s$^{1/2}\else \,$\mu$K$_{\rm RJ}$\,s$^{1/2}$\fi}
\def\muKHz{\ifmmode \,\mu$K\,Hz$^{-1/2}\else \,$\mu$K\,Hz$^{-1/2}$\fi}
\def\MJysr{\ifmmode \,$MJy\,sr\mo$\else \,MJy\,sr\mo\fi}
\def\MJysrmK{\ifmmode \,$MJy\,sr\mo$\,mK$_{\rm CMB}\mo\else \,MJy\,sr\mo\,mK$_{\rm CMB}\mo$\fi}
\def\microns{\ifmmode \,\mu$m$\else \,$\mu$m\fi}
\def\micron{\microns}
\def\muK{\ifmmode \,\mu$K$\else \,$\mu$\hbox{K}\fi}
\def\microK{\ifmmode \,\mu$K$\else \,$\mu$\hbox{K}\fi}
\def\muW{\ifmmode \,\mu$W$\else \,$\mu$\hbox{W}\fi}
\def\kms{\ifmmode $\,km\,s$^{-1}\else \,km\,s$^{-1}$\fi}
\def\kmsMpc{\ifmmode $\,\kms\,Mpc\mo$\else \,\kms\,Mpc\mo\fi}
\setsymbol{LFI:center:frequency:70GHz:units}{70.3\,GHz}
\setsymbol{LFI:center:frequency:44GHz:units}{44.1\,GHz}
\setsymbol{LFI:center:frequency:30GHz:units}{28.5\,GHz}

\setsymbol{LFI:center:frequency:70GHz}{70.3}
\setsymbol{LFI:center:frequency:44GHz}{44.1}
\setsymbol{LFI:center:frequency:30GHz}{28.5}

\setsymbol{LFI:center:frequency:LFI18:Rad:M:units}{71.7\GHz}
\setsymbol{LFI:center:frequency:LFI19:Rad:M:units}{67.5\GHz}
\setsymbol{LFI:center:frequency:LFI20:Rad:M:units}{69.2\GHz}
\setsymbol{LFI:center:frequency:LFI21:Rad:M:units}{70.4\GHz}
\setsymbol{LFI:center:frequency:LFI22:Rad:M:units}{71.5\GHz}
\setsymbol{LFI:center:frequency:LFI23:Rad:M:units}{70.8\GHz}
\setsymbol{LFI:center:frequency:LFI24:Rad:M:units}{44.4\GHz}
\setsymbol{LFI:center:frequency:LFI25:Rad:M:units}{44.0\GHz}
\setsymbol{LFI:center:frequency:LFI26:Rad:M:units}{43.9\GHz}
\setsymbol{LFI:center:frequency:LFI27:Rad:M:units}{28.3\GHz}
\setsymbol{LFI:center:frequency:LFI28:Rad:M:units}{28.8\GHz}
\setsymbol{LFI:center:frequency:LFI18:Rad:S:units}{70.1\GHz}
\setsymbol{LFI:center:frequency:LFI19:Rad:S:units}{69.6\GHz}
\setsymbol{LFI:center:frequency:LFI20:Rad:S:units}{69.5\GHz}
\setsymbol{LFI:center:frequency:LFI21:Rad:S:units}{69.5\GHz}
\setsymbol{LFI:center:frequency:LFI22:Rad:S:units}{72.8\GHz}
\setsymbol{LFI:center:frequency:LFI23:Rad:S:units}{71.3\GHz}
\setsymbol{LFI:center:frequency:LFI24:Rad:S:units}{44.1\GHz}
\setsymbol{LFI:center:frequency:LFI25:Rad:S:units}{44.1\GHz}
\setsymbol{LFI:center:frequency:LFI26:Rad:S:units}{44.1\GHz}
\setsymbol{LFI:center:frequency:LFI27:Rad:S:units}{28.5\GHz}
\setsymbol{LFI:center:frequency:LFI28:Rad:S:units}{28.2\GHz}

\setsymbol{LFI:center:frequency:LFI18:Rad:M}{71.7}
\setsymbol{LFI:center:frequency:LFI19:Rad:M}{67.5}
\setsymbol{LFI:center:frequency:LFI20:Rad:M}{69.2}
\setsymbol{LFI:center:frequency:LFI21:Rad:M}{70.4}
\setsymbol{LFI:center:frequency:LFI22:Rad:M}{71.5}
\setsymbol{LFI:center:frequency:LFI23:Rad:M}{70.8}
\setsymbol{LFI:center:frequency:LFI24:Rad:M}{44.4}
\setsymbol{LFI:center:frequency:LFI25:Rad:M}{44.0}
\setsymbol{LFI:center:frequency:LFI26:Rad:M}{43.9}
\setsymbol{LFI:center:frequency:LFI27:Rad:M}{28.3}
\setsymbol{LFI:center:frequency:LFI28:Rad:M}{28.8}
\setsymbol{LFI:center:frequency:LFI18:Rad:S}{70.1}
\setsymbol{LFI:center:frequency:LFI19:Rad:S}{69.6}
\setsymbol{LFI:center:frequency:LFI20:Rad:S}{69.5}
\setsymbol{LFI:center:frequency:LFI21:Rad:S}{69.5}
\setsymbol{LFI:center:frequency:LFI22:Rad:S}{72.8}
\setsymbol{LFI:center:frequency:LFI23:Rad:S}{71.3}
\setsymbol{LFI:center:frequency:LFI24:Rad:S}{44.1}
\setsymbol{LFI:center:frequency:LFI25:Rad:S}{44.1}
\setsymbol{LFI:center:frequency:LFI26:Rad:S}{44.1}
\setsymbol{LFI:center:frequency:LFI27:Rad:S}{28.5}
\setsymbol{LFI:center:frequency:LFI28:Rad:S}{28.2}

\setsymbol{LFI:white:noise:sensitivity:70GHz:units}{134.7\muKs}
\setsymbol{LFI:white:noise:sensitivity:44GHz:units}{164.7\muKs}
\setsymbol{LFI:white:noise:sensitivity:30GHz:units}{143.4\muKs}

\setsymbol{LFI:white:noise:sensitivity:70GHz}{134.7}
\setsymbol{LFI:white:noise:sensitivity:44GHz}{164.7}
\setsymbol{LFI:white:noise:sensitivity:30GHz}{143.4}

\setsymbol{LFI:white:noise:sensitivity:LFI18:Rad:M:units}{512.0\muKs}
\setsymbol{LFI:white:noise:sensitivity:LFI19:Rad:M:units}{581.4\muKs}
\setsymbol{LFI:white:noise:sensitivity:LFI20:Rad:M:units}{590.8\muKs}
\setsymbol{LFI:white:noise:sensitivity:LFI21:Rad:M:units}{455.2\muKs}
\setsymbol{LFI:white:noise:sensitivity:LFI22:Rad:M:units}{492.0\muKs}
\setsymbol{LFI:white:noise:sensitivity:LFI23:Rad:M:units}{507.7\muKs}
\setsymbol{LFI:white:noise:sensitivity:LFI24:Rad:M:units}{462.2\muKs}
\setsymbol{LFI:white:noise:sensitivity:LFI25:Rad:M:units}{413.6\muKs}
\setsymbol{LFI:white:noise:sensitivity:LFI26:Rad:M:units}{478.6\muKs}
\setsymbol{LFI:white:noise:sensitivity:LFI27:Rad:M:units}{277.7\muKs}
\setsymbol{LFI:white:noise:sensitivity:LFI28:Rad:M:units}{312.3\muKs}
\setsymbol{LFI:white:noise:sensitivity:LFI18:Rad:S:units}{465.7\muKs}
\setsymbol{LFI:white:noise:sensitivity:LFI19:Rad:S:units}{555.6\muKs}
\setsymbol{LFI:white:noise:sensitivity:LFI20:Rad:S:units}{623.2\muKs}
\setsymbol{LFI:white:noise:sensitivity:LFI21:Rad:S:units}{564.1\muKs}
\setsymbol{LFI:white:noise:sensitivity:LFI22:Rad:S:units}{534.4\muKs}
\setsymbol{LFI:white:noise:sensitivity:LFI23:Rad:S:units}{542.4\muKs}
\setsymbol{LFI:white:noise:sensitivity:LFI24:Rad:S:units}{399.2\muKs}
\setsymbol{LFI:white:noise:sensitivity:LFI25:Rad:S:units}{392.6\muKs}
\setsymbol{LFI:white:noise:sensitivity:LFI26:Rad:S:units}{418.6\muKs}
\setsymbol{LFI:white:noise:sensitivity:LFI27:Rad:S:units}{302.9\muKs}
\setsymbol{LFI:white:noise:sensitivity:LFI28:Rad:S:units}{285.3\muKs}

\setsymbol{LFI:white:noise:sensitivity:LFI18:Rad:M}{512.0}
\setsymbol{LFI:white:noise:sensitivity:LFI19:Rad:M}{581.4}
\setsymbol{LFI:white:noise:sensitivity:LFI20:Rad:M}{590.8}
\setsymbol{LFI:white:noise:sensitivity:LFI21:Rad:M}{455.2}
\setsymbol{LFI:white:noise:sensitivity:LFI22:Rad:M}{492.0}
\setsymbol{LFI:white:noise:sensitivity:LFI23:Rad:M}{507.7}
\setsymbol{LFI:white:noise:sensitivity:LFI24:Rad:M}{462.2}
\setsymbol{LFI:white:noise:sensitivity:LFI25:Rad:M}{413.6}
\setsymbol{LFI:white:noise:sensitivity:LFI26:Rad:M}{478.6}
\setsymbol{LFI:white:noise:sensitivity:LFI27:Rad:M}{277.7}
\setsymbol{LFI:white:noise:sensitivity:LFI28:Rad:M}{312.3}
\setsymbol{LFI:white:noise:sensitivity:LFI18:Rad:S}{465.7}
\setsymbol{LFI:white:noise:sensitivity:LFI19:Rad:S}{555.6}
\setsymbol{LFI:white:noise:sensitivity:LFI20:Rad:S}{623.2}
\setsymbol{LFI:white:noise:sensitivity:LFI21:Rad:S}{564.1}
\setsymbol{LFI:white:noise:sensitivity:LFI22:Rad:S}{534.4}
\setsymbol{LFI:white:noise:sensitivity:LFI23:Rad:S}{542.4}
\setsymbol{LFI:white:noise:sensitivity:LFI24:Rad:S}{399.2}
\setsymbol{LFI:white:noise:sensitivity:LFI25:Rad:S}{392.6}
\setsymbol{LFI:white:noise:sensitivity:LFI26:Rad:S}{418.6}
\setsymbol{LFI:white:noise:sensitivity:LFI27:Rad:S}{302.9}
\setsymbol{LFI:white:noise:sensitivity:LFI28:Rad:S}{285.3}

\setsymbol{LFI:knee:frequency:70GHz:units}{29.5\mHz}
\setsymbol{LFI:knee:frequency:44GHz:units}{56.2\mHz}
\setsymbol{LFI:knee:frequency:30GHz:units}{113.7\mHz}

\setsymbol{LFI:knee:frequency:70GHz}{29.5}
\setsymbol{LFI:knee:frequency:44GHz}{56.2}
\setsymbol{LFI:knee:frequency:30GHz}{113.7}

\setsymbol{LFI:knee:frequency:LFI18:Rad:M:units}{16.3\mHz}
\setsymbol{LFI:knee:frequency:LFI19:Rad:M:units}{15.1\mHz}
\setsymbol{LFI:knee:frequency:LFI20:Rad:M:units}{18.7\mHz}
\setsymbol{LFI:knee:frequency:LFI21:Rad:M:units}{37.2\mHz}
\setsymbol{LFI:knee:frequency:LFI22:Rad:M:units}{12.7\mHz}
\setsymbol{LFI:knee:frequency:LFI23:Rad:M:units}{34.6\mHz}
\setsymbol{LFI:knee:frequency:LFI24:Rad:M:units}{46.2\mHz}
\setsymbol{LFI:knee:frequency:LFI25:Rad:M:units}{24.9\mHz}
\setsymbol{LFI:knee:frequency:LFI26:Rad:M:units}{67.6\mHz}
\setsymbol{LFI:knee:frequency:LFI27:Rad:M:units}{187.4\mHz}
\setsymbol{LFI:knee:frequency:LFI28:Rad:M:units}{122.2\mHz}
\setsymbol{LFI:knee:frequency:LFI18:Rad:S:units}{17.7\mHz}
\setsymbol{LFI:knee:frequency:LFI19:Rad:S:units}{22.0\mHz}
\setsymbol{LFI:knee:frequency:LFI20:Rad:S:units}{8.7\mHz}
\setsymbol{LFI:knee:frequency:LFI21:Rad:S:units}{25.9\mHz}
\setsymbol{LFI:knee:frequency:LFI22:Rad:S:units}{15.8\mHz}
\setsymbol{LFI:knee:frequency:LFI23:Rad:S:units}{129.8\mHz}
\setsymbol{LFI:knee:frequency:LFI24:Rad:S:units}{100.9\mHz}
\setsymbol{LFI:knee:frequency:LFI25:Rad:S:units}{38.9\mHz}
\setsymbol{LFI:knee:frequency:LFI26:Rad:S:units}{58.9\mHz}
\setsymbol{LFI:knee:frequency:LFI27:Rad:S:units}{104.4\mHz}
\setsymbol{LFI:knee:frequency:LFI28:Rad:S:units}{40.7\mHz}

\setsymbol{LFI:knee:frequency:LFI18:Rad:M}{16.3}
\setsymbol{LFI:knee:frequency:LFI19:Rad:M}{15.1}
\setsymbol{LFI:knee:frequency:LFI20:Rad:M}{18.7}
\setsymbol{LFI:knee:frequency:LFI21:Rad:M}{37.2}
\setsymbol{LFI:knee:frequency:LFI22:Rad:M}{12.7}
\setsymbol{LFI:knee:frequency:LFI23:Rad:M}{34.6}
\setsymbol{LFI:knee:frequency:LFI24:Rad:M}{46.2}
\setsymbol{LFI:knee:frequency:LFI25:Rad:M}{24.9}
\setsymbol{LFI:knee:frequency:LFI26:Rad:M}{67.6}
\setsymbol{LFI:knee:frequency:LFI27:Rad:M}{187.4}
\setsymbol{LFI:knee:frequency:LFI28:Rad:M}{122.2}
\setsymbol{LFI:knee:frequency:LFI18:Rad:S}{17.7}
\setsymbol{LFI:knee:frequency:LFI19:Rad:S}{22.0}
\setsymbol{LFI:knee:frequency:LFI20:Rad:S}{8.7}
\setsymbol{LFI:knee:frequency:LFI21:Rad:S}{25.9}
\setsymbol{LFI:knee:frequency:LFI22:Rad:S}{15.8}
\setsymbol{LFI:knee:frequency:LFI23:Rad:S}{129.8}
\setsymbol{LFI:knee:frequency:LFI24:Rad:S}{100.9}
\setsymbol{LFI:knee:frequency:LFI25:Rad:S}{38.9}
\setsymbol{LFI:knee:frequency:LFI26:Rad:S}{58.9}
\setsymbol{LFI:knee:frequency:LFI27:Rad:S}{104.4}
\setsymbol{LFI:knee:frequency:LFI28:Rad:S}{40.7}

\setsymbol{LFI:slope:70GHz:units}{$-1.03$\mHz}
\setsymbol{LFI:slope:44GHz:units}{$-0.89$\mHz}
\setsymbol{LFI:slope:30GHz:units}{$-0.87$\mHz}

\setsymbol{LFI:slope:70GHz}{$-1.03$}
\setsymbol{LFI:slope:44GHz}{$-0.89$}
\setsymbol{LFI:slope:30GHz}{$-0.87$}

\setsymbol{LFI:slope:LFI18:Rad:M:units}{$-1.04$\mHz}
\setsymbol{LFI:slope:LFI19:Rad:M:units}{$-1.09$\mHz}
\setsymbol{LFI:slope:LFI20:Rad:M:units}{$-0.69$\mHz}
\setsymbol{LFI:slope:LFI21:Rad:M:units}{$-1.56$\mHz}
\setsymbol{LFI:slope:LFI22:Rad:M:units}{$-1.01$\mHz}
\setsymbol{LFI:slope:LFI23:Rad:M:units}{$-0.96$\mHz}
\setsymbol{LFI:slope:LFI24:Rad:M:units}{$-0.83$\mHz}
\setsymbol{LFI:slope:LFI25:Rad:M:units}{$-0.91$\mHz}
\setsymbol{LFI:slope:LFI26:Rad:M:units}{$-0.95$\mHz}
\setsymbol{LFI:slope:LFI27:Rad:M:units}{$-0.87$\mHz}
\setsymbol{LFI:slope:LFI28:Rad:M:units}{$-0.88$\mHz}
\setsymbol{LFI:slope:LFI18:Rad:S:units}{$-1.15$\mHz}
\setsymbol{LFI:slope:LFI19:Rad:S:units}{$-1.00$\mHz}
\setsymbol{LFI:slope:LFI20:Rad:S:units}{$-0.95$\mHz}
\setsymbol{LFI:slope:LFI21:Rad:S:units}{$-0.92$\mHz}
\setsymbol{LFI:slope:LFI22:Rad:S:units}{$-1.01$\mHz}
\setsymbol{LFI:slope:LFI23:Rad:S:units}{$-0.95$\mHz}
\setsymbol{LFI:slope:LFI24:Rad:S:units}{$-0.73$\mHz}
\setsymbol{LFI:slope:LFI25:Rad:S:units}{$-1.16$\mHz}
\setsymbol{LFI:slope:LFI26:Rad:S:units}{$-0.79$\mHz}
\setsymbol{LFI:slope:LFI27:Rad:S:units}{$-0.82$\mHz}
\setsymbol{LFI:slope:LFI28:Rad:S:units}{$-0.91$\mHz}

\setsymbol{LFI:slope:LFI18:Rad:M}{$-1.04$}
\setsymbol{LFI:slope:LFI19:Rad:M}{$-1.09$}
\setsymbol{LFI:slope:LFI20:Rad:M}{$-0.69$}
\setsymbol{LFI:slope:LFI21:Rad:M}{$-1.56$}
\setsymbol{LFI:slope:LFI22:Rad:M}{$-1.01$}
\setsymbol{LFI:slope:LFI23:Rad:M}{$-0.96$}
\setsymbol{LFI:slope:LFI24:Rad:M}{$-0.83$}
\setsymbol{LFI:slope:LFI25:Rad:M}{$-0.91$}
\setsymbol{LFI:slope:LFI26:Rad:M}{$-0.95$}
\setsymbol{LFI:slope:LFI27:Rad:M}{$-0.87$}
\setsymbol{LFI:slope:LFI28:Rad:M}{$-0.88$}
\setsymbol{LFI:slope:LFI18:Rad:S}{$-1.15$}
\setsymbol{LFI:slope:LFI19:Rad:S}{$-1.00$}
\setsymbol{LFI:slope:LFI20:Rad:S}{$-0.95$}
\setsymbol{LFI:slope:LFI21:Rad:S}{$-0.92$}
\setsymbol{LFI:slope:LFI22:Rad:S}{$-1.01$}
\setsymbol{LFI:slope:LFI23:Rad:S}{$-0.95$}
\setsymbol{LFI:slope:LFI24:Rad:S}{$-0.73$}
\setsymbol{LFI:slope:LFI25:Rad:S}{$-1.16$}
\setsymbol{LFI:slope:LFI26:Rad:S}{$-0.79$}
\setsymbol{LFI:slope:LFI27:Rad:S}{$-0.82$}
\setsymbol{LFI:slope:LFI28:Rad:S}{$-0.91$}

\setsymbol{LFI:FWHM:70GHz:units}{13\parcm01}
\setsymbol{LFI:FWHM:44GHz:units}{27\parcm92}
\setsymbol{LFI:FWHM:30GHz:units}{32\parcm65}

\setsymbol{LFI:FWHM:70GHz}{13.01}
\setsymbol{LFI:FWHM:44GHz}{27.92}
\setsymbol{LFI:FWHM:30GHz}{32.65}

\setsymbol{LFI:FWHM:LFI18:units}{13\parcm39}
\setsymbol{LFI:FWHM:LFI19:units}{13\parcm01}
\setsymbol{LFI:FWHM:LFI20:units}{12\parcm75}
\setsymbol{LFI:FWHM:LFI21:units}{12\parcm74}
\setsymbol{LFI:FWHM:LFI22:units}{12\parcm87}
\setsymbol{LFI:FWHM:LFI23:units}{13\parcm27}
\setsymbol{LFI:FWHM:LFI24:units}{22\parcm98}
\setsymbol{LFI:FWHM:LFI25:units}{30\parcm46}
\setsymbol{LFI:FWHM:LFI26:units}{30\parcm31}
\setsymbol{LFI:FWHM:LFI27:units}{32\parcm65}
\setsymbol{LFI:FWHM:LFI28:units}{32\parcm66}

\setsymbol{LFI:FWHM:LFI18}{13.39}
\setsymbol{LFI:FWHM:LFI19}{13.01}
\setsymbol{LFI:FWHM:LFI20}{12.75}
\setsymbol{LFI:FWHM:LFI21}{12.74}
\setsymbol{LFI:FWHM:LFI22}{12.87}
\setsymbol{LFI:FWHM:LFI23}{13.27}
\setsymbol{LFI:FWHM:LFI24}{22.98}
\setsymbol{LFI:FWHM:LFI25}{30.46}
\setsymbol{LFI:FWHM:LFI26}{30.31}
\setsymbol{LFI:FWHM:LFI27}{32.65}
\setsymbol{LFI:FWHM:LFI28}{32.66}

\setsymbol{LFI:FWHM:uncertainty:LFI18:units}{0.170\arcm}
\setsymbol{LFI:FWHM:uncertainty:LFI19:units}{0.174\arcm}
\setsymbol{LFI:FWHM:uncertainty:LFI20:units}{0.170\arcm}
\setsymbol{LFI:FWHM:uncertainty:LFI21:units}{0.156\arcm}
\setsymbol{LFI:FWHM:uncertainty:LFI22:units}{0.164\arcm}
\setsymbol{LFI:FWHM:uncertainty:LFI23:units}{0.171\arcm}
\setsymbol{LFI:FWHM:uncertainty:LFI24:units}{0.652\arcm}
\setsymbol{LFI:FWHM:uncertainty:LFI25:units}{1.075\arcm}
\setsymbol{LFI:FWHM:uncertainty:LFI26:units}{1.131\arcm}
\setsymbol{LFI:FWHM:uncertainty:LFI27:units}{1.266\arcm}
\setsymbol{LFI:FWHM:uncertainty:LFI28:units}{1.287\arcm}

\setsymbol{LFI:FWHM:uncertainty:LFI18}{0.170}
\setsymbol{LFI:FWHM:uncertainty:LFI19}{0.174}
\setsymbol{LFI:FWHM:uncertainty:LFI20}{0.170}
\setsymbol{LFI:FWHM:uncertainty:LFI21}{0.156}
\setsymbol{LFI:FWHM:uncertainty:LFI22}{0.164}
\setsymbol{LFI:FWHM:uncertainty:LFI23}{0.171}
\setsymbol{LFI:FWHM:uncertainty:LFI24}{0.652}
\setsymbol{LFI:FWHM:uncertainty:LFI25}{1.075}
\setsymbol{LFI:FWHM:uncertainty:LFI26}{1.131}
\setsymbol{LFI:FWHM:uncertainty:LFI27}{1.266}
\setsymbol{LFI:FWHM:uncertainty:LFI28}{1.287}

\setsymbol{HFI:center:frequency:100GHz:units}{100\,GHz}
\setsymbol{HFI:center:frequency:143GHz:units}{143\,GHz}
\setsymbol{HFI:center:frequency:217GHz:units}{217\,GHz}
\setsymbol{HFI:center:frequency:353GHz:units}{353\,GHz}
\setsymbol{HFI:center:frequency:545GHz:units}{545\,GHz}
\setsymbol{HFI:center:frequency:857GHz:units}{857\,GHz}

\setsymbol{HFI:center:frequency:100GHz}{100}
\setsymbol{HFI:center:frequency:143GHz}{143}
\setsymbol{HFI:center:frequency:217GHz}{217}
\setsymbol{HFI:center:frequency:353GHz}{353}
\setsymbol{HFI:center:frequency:545GHz}{545}
\setsymbol{HFI:center:frequency:857GHz}{857}

\setsymbol{HFI:Ndetectors:100GHz}{8}
\setsymbol{HFI:Ndetectors:143GHz}{11}
\setsymbol{HFI:Ndetectors:217GHz}{12}
\setsymbol{HFI:Ndetectors:353GHz}{12}
\setsymbol{HFI:Ndetectors:545GHz}{3}
\setsymbol{HFI:Ndetectors:857GHz}{4}

\setsymbol{HFI:FWHM:Maps:100GHz:units}{9\parcm88}
\setsymbol{HFI:FWHM:Maps:143GHz:units}{7\parcm18}
\setsymbol{HFI:FWHM:Maps:217GHz:units}{4\parcm87}
\setsymbol{HFI:FWHM:Maps:353GHz:units}{4\parcm65}
\setsymbol{HFI:FWHM:Maps:545GHz:units}{4\parcm72}
\setsymbol{HFI:FWHM:Maps:857GHz:units}{4\parcm39}
\setsymbol{HFI:FWHM:Maps:100GHz}{9.88}
\setsymbol{HFI:FWHM:Maps:143GHz}{7.18}
\setsymbol{HFI:FWHM:Maps:217GHz}{4.87}
\setsymbol{HFI:FWHM:Maps:353GHz}{4.65}
\setsymbol{HFI:FWHM:Maps:545GHz}{4.72}
\setsymbol{HFI:FWHM:Maps:857GHz}{4.39}

\setsymbol{HFI:beam:ellipticity:Maps:100GHz}{1.15}
\setsymbol{HFI:beam:ellipticity:Maps:143GHz}{1.01}
\setsymbol{HFI:beam:ellipticity:Maps:217GHz}{1.06}
\setsymbol{HFI:beam:ellipticity:Maps:353GHz}{1.05}
\setsymbol{HFI:beam:ellipticity:Maps:545GHz}{1.14}
\setsymbol{HFI:beam:ellipticity:Maps:857GHz}{1.19}

\setsymbol{HFI:FWHM:Mars:100GHz:units}{9\parcm37}
\setsymbol{HFI:FWHM:Mars:143GHz:units}{7\parcm04}
\setsymbol{HFI:FWHM:Mars:217GHz:units}{4\parcm68}
\setsymbol{HFI:FWHM:Mars:353GHz:units}{4\parcm43}
\setsymbol{HFI:FWHM:Mars:545GHz:units}{3\parcm80}
\setsymbol{HFI:FWHM:Mars:857GHz:units}{3\parcm67}

\setsymbol{HFI:FWHM:Mars:100GHz}{9.37}
\setsymbol{HFI:FWHM:Mars:143GHz}{7.04}
\setsymbol{HFI:FWHM:Mars:217GHz}{4.68}
\setsymbol{HFI:FWHM:Mars:353GHz}{4.43}
\setsymbol{HFI:FWHM:Mars:545GHz}{3.80}
\setsymbol{HFI:FWHM:Mars:857GHz}{3.67}

\setsymbol{HFI:beam:ellipticity:Mars:100GHz}{1.18}
\setsymbol{HFI:beam:ellipticity:Mars:143GHz}{1.03}
\setsymbol{HFI:beam:ellipticity:Mars:217GHz}{1.14}
\setsymbol{HFI:beam:ellipticity:Mars:353GHz}{1.09}
\setsymbol{HFI:beam:ellipticity:Mars:545GHz}{1.25}
\setsymbol{HFI:beam:ellipticity:Mars:857GHz}{1.03}

\setsymbol{HFI:CMB:relative:calibration:100GHz}{$\lsim 1\%$}
\setsymbol{HFI:CMB:relative:calibration:143GHz}{$\lsim 1\%$}
\setsymbol{HFI:CMB:relative:calibration:217GHz}{$\lsim 1\%$}
\setsymbol{HFI:CMB:relative:calibration:353GHz}{$\lsim 1\%$}
\setsymbol{HFI:CMB:relative:calibration:545GHz}{}
\setsymbol{HFI:CMB:relative:calibration:857GHz}{}

\setsymbol{HFI:CMB:absolute:calibration:100GHz}{$\lsim 2\%$}
\setsymbol{HFI:CMB:absolute:calibration:143GHz}{$\lsim 2\%$}
\setsymbol{HFI:CMB:absolute:calibration:217GHz}{$\lsim 2\%$}
\setsymbol{HFI:CMB:absolute:calibration:353GHz}{$\lsim 2\%$}
\setsymbol{HFI:CMB:absolute:calibration:545GHz}{}
\setsymbol{HFI:CMB:absolute:calibration:857GHz}{}

\setsymbol{HFI:FIRAS:gain:calibration:accuracy:statistical:100GHz}{}
\setsymbol{HFI:FIRAS:gain:calibration:accuracy:statistical:143GHz}{}
\setsymbol{HFI:FIRAS:gain:calibration:accuracy:statistical:217GHz}{}
\setsymbol{HFI:FIRAS:gain:calibration:accuracy:statistical:353GHz}{2.5\%}
\setsymbol{HFI:FIRAS:gain:calibration:accuracy:statistical:545GHz}{1\%}
\setsymbol{HFI:FIRAS:gain:calibration:accuracy:statistical:857GHz}{0.5\%}

\setsymbol{HFI:FIRAS:gain:calibration:accuracy:systematic:100GHz}{}
\setsymbol{HFI:FIRAS:gain:calibration:accuracy:systematic:143GHz}{}
\setsymbol{HFI:FIRAS:gain:calibration:accuracy:systematic:217GHz}{}
\setsymbol{HFI:FIRAS:gain:calibration:accuracy:systematic:353GHz}{}
\setsymbol{HFI:FIRAS:gain:calibration:accuracy:systematic:545GHz}{7\%}
\setsymbol{HFI:FIRAS:gain:calibration:accuracy:systematic:857GHz}{7\%}

\setsymbol{HFI:FIRAS:zero:point:accuracy:100GHz:units}{0.8\MJysr}
\setsymbol{HFI:FIRAS:zero:point:accuracy:143GHz:units}{}
\setsymbol{HFI:FIRAS:zero:point:accuracy:217GHz:units}{}
\setsymbol{HFI:FIRAS:zero:point:accuracy:353GHz:units}{1.4\MJysr}
\setsymbol{HFI:FIRAS:zero:point:accuracy:545GHz:units}{2.2\MJysr}
\setsymbol{HFI:FIRAS:zero:point:accuracy:857GHz:units}{1.7\MJysr}

\setsymbol{HFI:FIRAS:zero:point:accuracy:100GHz}{0.8}
\setsymbol{HFI:FIRAS:zero:point:accuracy:143GHz}{}
\setsymbol{HFI:FIRAS:zero:point:accuracy:217GHz}{}
\setsymbol{HFI:FIRAS:zero:point:accuracy:353GHz}{1.4}
\setsymbol{HFI:FIRAS:zero:point:accuracy:545GHz}{2.2}
\setsymbol{HFI:FIRAS:zero:point:accuracy:857GHz}{1.7}

\setsymbol{HFI:unit:conversion:100GHz:units}{0.2415\MJysrmK}
\setsymbol{HFI:unit:conversion:143GHz:units}{0.3694\MJysrmK}
\setsymbol{HFI:unit:conversion:217GHz:units}{0.4811\MJysrmK}
\setsymbol{HFI:unit:conversion:353GHz:units}{0.2883\MJysrmK}
\setsymbol{HFI:unit:conversion:545GHz:units}{0.05826\MJysrmK}
\setsymbol{HFI:unit:conversion:857GHz:units}{0.002238\MJysrmK}

\setsymbol{HFI:unit:conversion:100GHz}{0.2415}
\setsymbol{HFI:unit:conversion:143GHz}{0.3694}
\setsymbol{HFI:unit:conversion:217GHz}{0.4811}
\setsymbol{HFI:unit:conversion:353GHz}{0.2883}
\setsymbol{HFI:unit:conversion:545GHz}{0.05826}
\setsymbol{HFI:unit:conversion:857GHz}{0.002238}

\setsymbol{HFI:colour:correction:alpha=-2:V1.01:100GHz}{0.9893}
\setsymbol{HFI:colour:correction:alpha=-2:V1.01:143GHz}{0.9759}
\setsymbol{HFI:colour:correction:alpha=-2:V1.01:217GHz}{1.0007}
\setsymbol{HFI:colour:correction:alpha=-2:V1.01:353GHz}{1.0028}
\setsymbol{HFI:colour:correction:alpha=-2:V1.01:545GHz}{1.0019}
\setsymbol{HFI:colour:correction:alpha=-2:V1.01:857GHz}{0.9889}

\setsymbol{HFI:colour:correction:alpha=0:V1.01:100GHz}{1.0008}
\setsymbol{HFI:colour:correction:alpha=0:V1.01:143GHz}{1.0148}
\setsymbol{HFI:colour:correction:alpha=0:V1.01:217GHz}{0.9909}
\setsymbol{HFI:colour:correction:alpha=0:V1.01:353GHz}{0.9888}
\setsymbol{HFI:colour:correction:alpha=0:V1.01:545GHz}{0.9878}
\setsymbol{HFI:colour:correction:alpha=0:V1.01:857GHz}{1.0014}

%% file: PIP_38_authors_and_institutes.tex
\author{\small
Planck Collaboration:
P.~A.~R.~Ade\inst{85}
\and
N.~Aghanim\inst{60}
\and
F.~Arg\"{u}eso\inst{20}
\and
M.~Arnaud\inst{74}
\and
M.~Ashdown\inst{71, 6}
\and
F.~Atrio-Barandela\inst{18}
\and
J.~Aumont\inst{60}
\and
C.~Baccigalupi\inst{84}
\and
A.~Balbi\inst{37}
\and
A.~J.~Banday\inst{89, 9}
% ...
}
\institute{\small
APC, AstroParticule et Cosmologie, Universit\'{e} Paris Diderot, CNRS/IN2P3, CEA/lrfu, Observatoire de Paris, Sorbonne Paris Cit\'{e}, 10, rue Alice Domon et L\'{e}onie Duquet, 75205 Paris Cedex 13, France\\
\and
Aalto University Mets\"{a}hovi Radio Observatory, Mets\"{a}hovintie 114, FIN-02540 Kylm\"{a}l\"{a}, Finland\\
\and
African Institute for Mathematical Sciences, 6-8 Melrose Road, Muizenberg, Cape Town, South Africa\\
\and
Agenzia Spaziale Italiana Science Data Center, c/o ESRIN, via Galileo Galilei, Frascati, Italy\\
\and
Agenzia Spaziale Italiana, Viale Liegi 26, Roma, Italy\\
\and
Astrophysics Group, Cavendish Laboratory, University of Cambridge, J J Thomson Avenue, Cambridge CB3 0HE, U.K.\\
\and
Atacama Large Millimeter/submillimeter Array, ALMA Santiago Central Offices, Alonso de Cordova 3107, Vitacura, Casilla 763 0355, Santiago, Chile\\
\and
CITA, University of Toronto, 60 St. George St., Toronto, ON M5S 3H8, Canada\\
\and
CNRS, IRAP, 9 Av. colonel Roche, BP 44346, F-31028 Toulouse cedex 4, France\\
\and
California Institute of Technology, Pasadena, California, U.S.A.\\
\and
Centre of Mathematics for Applications, University of Oslo, Blindern, Oslo, Norway\\
\and
Centro de Estudios de F\'{i}sica del Cosmos de Arag\'{o}n (CEFCA), Plaza San Juan, 1, planta 2, E-44001, Teruel, Spain\\
\and
Computational Cosmology Center, Lawrence Berkeley National Laboratory, Berkeley, California, U.S.A.\\
\and
Consejo Superior de Investigaciones Cient\'{\i}ficas (CSIC), Madrid, Spain\\
\and
DSM/Irfu/SPP, CEA-Saclay, F-91191 Gif-sur-Yvette Cedex, France\\
\and
DTU Space, National Space Institute, Technical University of Denmark, Elektrovej 327, DK-2800 Kgs. Lyngby, Denmark\\
\and
D\'{e}partement de Physique Th\'{e}orique, Universit\'{e} de Gen\`{e}ve, 24, Quai E. Ansermet,1211 Gen\`{e}ve 4, Switzerland\\
\and
Departamento de F\'{\i}sica Fundamental, Facultad de Ciencias, Universidad de Salamanca, 37008 Salamanca, Spain\\
\and
Departamento de F\'{\i}sica, Universidad de Oviedo, Avda. Calvo Sotelo s/n, Oviedo, Spain\\
\and
Departamento de Matem\'{a}ticas, Universidad de Oviedo, Avda. Calvo Sotelo s/n, Oviedo, Spain\\
\and
Department of Astrophysics/IMAPP, Radboud University Nijmegen, P.O. Box 9010, 6500 GL Nijmegen, The Netherlands\\
\and
Department of Physics \& Astronomy, University of British Columbia, 6224 Agricultural Road, Vancouver, British Columbia, Canada\\
\and
Department of Physics and Astronomy, Dana and David Dornsife College of Letter, Arts and Sciences, University of Southern California, Los Angeles, CA 90089, U.S.A.\\
\and
Department of Physics and Astronomy, Tufts University, Medford, MA 02155, U.S.A.\\
\and
Department of Physics, Gustaf H\"{a}llstr\"{o}min katu 2a, University of Helsinki, Helsinki, Finland\\
\and
Department of Physics, Princeton University, Princeton, New Jersey, U.S.A.\\
\and
Department of Physics, University of California, Berkeley, California, U.S.A.\\
\and
Department of Physics, University of California, One Shields Avenue, Davis, California, U.S.A.\\
\and
Department of Physics, University of California, Santa Barbara, California, U.S.A.\\
\and
Department of Physics, University of Illinois at Urbana-Champaign, 1110 West Green Street, Urbana, Illinois, U.S.A.\\
\and
Department of Statistics, Purdue University, 250 N. University Street, West Lafayette, Indiana, U.S.A.\\
\and
Dipartimento di Fisica e Astronomia G. Galilei, Universit\`{a} degli Studi di Padova, via Marzolo 8, 35131 Padova, Italy\\
\and
Dipartimento di Fisica e Scienze della Terra, Universit\`{a} di Ferrara, Via Saragat 1, 44122 Ferrara, Italy\\
\and
Dipartimento di Fisica, Universit\`{a} La Sapienza, P. le A. Moro 2, Roma, Italy\\
\and
Dipartimento di Fisica, Universit\`{a} degli Studi di Milano, Via Celoria, 16, Milano, Italy\\
\and
Dipartimento di Fisica, Universit\`{a} degli Studi di Trieste, via A. Valerio 2, Trieste, Italy\\
\and
Dipartimento di Fisica, Universit\`{a} di Roma Tor Vergata, Via della Ricerca Scientifica, 1, Roma, Italy\\
\and
Dipartimento di Matematica, Universit\`{a} di Roma Tor Vergata, Via della Ricerca Scientifica, 1, Roma, Italy\\
\and
Discovery Center, Niels Bohr Institute, Blegdamsvej 17, Copenhagen, Denmark\\
\and
Dpto. Astrof\'{i}sica, Universidad de La Laguna (ULL), E-38206 La Laguna, Tenerife, Spain\\
\and
European Southern Observatory, ESO Vitacura, Alonso de Cordova 3107, Vitacura, Casilla 19001, Santiago, Chile\\
\and
European Space Agency, ESAC, Planck Science Office, Camino bajo del Castillo, s/n, Urbanizaci\'{o}n Villafranca del Castillo, Villanueva de la Ca\~{n}ada, Madrid, Spain\\
\and
European Space Agency, ESTEC, Keplerlaan 1, 2201 AZ Noordwijk, The Netherlands\\
\and
Haverford College Astronomy Department, 370 Lancaster Avenue, Haverford, Pennsylvania, U.S.A.\\
\and
Helsinki Institute of Physics, Gustaf H\"{a}llstr\"{o}min katu 2, University of Helsinki, Helsinki, Finland\\
\and
INAF - Osservatorio Astronomico di Padova, Vicolo dell'Osservatorio 5, Padova, Italy\\
\and
INAF - Osservatorio Astronomico di Roma, via di Frascati 33, Monte Porzio Catone, Italy\\
\and
INAF - Osservatorio Astronomico di Trieste, Via G.B. Tiepolo 11, Trieste, Italy\\
\and
INAF Istituto di Radioastronomia, Via P. Gobetti 101, 40129 Bologna, Italy\\
\and
INAF/IASF Bologna, Via Gobetti 101, Bologna, Italy\\
\and
INAF/IASF Milano, Via E. Bassini 15, Milano, Italy\\
\and
INFN, Sezione di Roma 1, Universit`{a} di Roma Sapienza, Piazzale Aldo Moro 2, 00185, Roma, Italy\\
\and
IPAG: Institut de Plan\'{e}tologie et d'Astrophysique de Grenoble, Universit\'{e} Joseph Fourier, Grenoble 1 / CNRS-INSU, UMR 5274, Grenoble, F-38041, France\\
\and
ISDC Data Centre for Astrophysics, University of Geneva, ch. d'Ecogia 16, Versoix, Switzerland\\
\and
IUCAA, Post Bag 4, Ganeshkhind, Pune University Campus, Pune 411 007, India\\
\and
Imperial College London, Astrophysics group, Blackett Laboratory, Prince Consort Road, London, SW7 2AZ, U.K.\\
\and
Infrared Processing and Analysis Center, California Institute of Technology, Pasadena, CA 91125, U.S.A.\\
\and
Institut N\'{e}el, CNRS, Universit\'{e} Joseph Fourier Grenoble I, 25 rue des Martyrs, Grenoble, France\\
\and
Institut Universitaire de France, 103, bd Saint-Michel, 75005, Paris, France\\
\and
Institut d'Astrophysique Spatiale, CNRS (UMR8617) Universit\'{e} Paris-Sud 11, B\^{a}timent 121, Orsay, France\\
\and
Institut d'Astrophysique de Paris, CNRS (UMR7095), 98 bis Boulevard Arago, F-75014, Paris, France\\
\and
Institut de Ci\`{e}ncies de l'Espai, CSIC/IEEC, Facultat de Ci\`{e}ncies, Campus UAB, Torre C5 par-2, Bellaterra 08193, Spain\\
\and
Institute for Space Sciences, Bucharest-Magurale, Romania\\
\and
Institute of Astronomy and Astrophysics, Academia Sinica, Taipei, Taiwan\\
\and
Institute of Astronomy, University of Cambridge, Madingley Road, Cambridge CB3 0HA, U.K.\\
\and
Institute of Theoretical Astrophysics, University of Oslo, Blindern, Oslo, Norway\\
\and
Instituto de Astrof\'{\i}sica de Canarias, C/V\'{\i}a L\'{a}ctea s/n, La Laguna, Tenerife, Spain\\
\and
Instituto de F\'{\i}sica de Cantabria (CSIC-Universidad de Cantabria), Avda. de los Castros s/n, Santander, Spain\\
\and
Jet Propulsion Laboratory, California Institute of Technology, 4800 Oak Grove Drive, Pasadena, California, U.S.A.\\
\and
Jodrell Bank Centre for Astrophysics, Alan Turing Building, School of Physics and Astronomy, The University of Manchester, Oxford Road, Manchester, M13 9PL, U.K.\\
\and
Kavli Institute for Cosmology Cambridge, Madingley Road, Cambridge, CB3 0HA, U.K.\\
\and
LAL, Universit\'{e} Paris-Sud, CNRS/IN2P3, Orsay, France\\
\and
LERMA, CNRS, Observatoire de Paris, 61 Avenue de l'Observatoire, Paris, France\\
\and
Laboratoire AIM, IRFU/Service d'Astrophysique - CEA/DSM - CNRS - Universit\'{e} Paris Diderot, B\^{a}t. 709, CEA-Saclay, F-91191 Gif-sur-Yvette Cedex, France\\
\and
Laboratoire Traitement et Communication de l'Information, CNRS (UMR 5141) and T\'{e}l\'{e}com ParisTech, 46 rue Barrault F-75634 Paris Cedex 13, France\\
\and
Laboratoire de Physique Subatomique et de Cosmologie, Universit\'{e} Joseph Fourier Grenoble I, CNRS/IN2P3, Institut National Polytechnique de Grenoble, 53 rue des Martyrs, 38026 Grenoble cedex, France\\
\and
Laboratoire de Physique Th\'{e}orique, Universit\'{e} Paris-Sud 11 \& CNRS, B\^{a}timent 210, 91405 Orsay, France\\
\and
Lawrence Berkeley National Laboratory, Berkeley, California, U.S.A.\\
\and
Max-Planck-Institut f\"{u}r Astrophysik, Karl-Schwarzschild-Str. 1, 85741 Garching, Germany\\
\and
National University of Ireland, Department of Experimental Physics, Maynooth, Co. Kildare, Ireland\\
\and
Niels Bohr Institute, Blegdamsvej 17, Copenhagen, Denmark\\
\and
Observational Cosmology, Mail Stop 367-17, California Institute of Technology, Pasadena, CA, 91125, U.S.A.\\
\and
Optical Science Laboratory, University College London, Gower Street, London, U.K.\\
\and
SISSA, Astrophysics Sector, via Bonomea 265, 34136, Trieste, Italy\\
\and
School of Physics and Astronomy, Cardiff University, Queens Buildings, The Parade, Cardiff, CF24 3AA, U.K.\\
\and
Space Sciences Laboratory, University of California, Berkeley, California, U.S.A.\\
\and
Stanford University, Dept of Physics, Varian Physics Bldg, 382 Via Pueblo Mall, Stanford, California, U.S.A.\\
\and
UPMC Univ Paris 06, UMR7095, 98 bis Boulevard Arago, F-75014, Paris, France\\
\and
Universit\'{e} de Toulouse, UPS-OMP, IRAP, F-31028 Toulouse cedex 4, France\\
% ...
}

%% file: table_source_identifications_new.tex
%
\begin{table*}[tmb]
\begingroup
\newdimen\tblskip \tblskip=5pt
\caption{Percentage and number of \Planck\ source identifications 
using the SIMBAD and NED databases.}
\label{tab:ids}
\nointerlineskip
\vskip -3mm
\footnotesize
\setbox\tablebox=\vbox{
   \newdimen\digitwidth
   \setbox0=\hbox{\rm 0}
    \digitwidth=\wd0
    \catcode`*=\active
    \def*{\kern\digitwidth}
    \newdimen\signwidth
    \setbox0=\hbox{+}
    \signwidth=\wd0
    \catcode`!=\active
    \def!{\kern\signwidth}
\halign{\hbox to 3cm{#\leaderfil}\tabskip=3em&
    \hfil#\hfil\tabskip=1.5em&
    \hfil#\hfil\tabskip=3em&
    \hfil#\hfil\tabskip=1.5em&
    \hfil#\hfil\tabskip=3em&
    \hfil#\hfil\tabskip=1.5em&
    \hfil#\hfil\tabskip=3em&
    \hfil#\hfil\tabskip=1.5em&
    \hfil#\hfil\tabskip=3em&
    \hfil#\hfil\tabskip=1.5em&
    \hfil#\hfil\tabskip=0pt\cr
\noalign{\doubleline}
\omit&&&\multispan4\hfil G{\sc alactic}\hfil\cr
\noalign{\vskip -3pt}
\omit&&&\multispan4\hrulefill\cr
\noalign{\vskip 1pt}
\omit&\multispan2\hfil E{\sc xtragal}\hfil   & \multispan2\hfil 
Insecure\hfil& \multispan2\hfil Secure\hfil&\multispan2\hfil U{\sc 
nident}\hfil&\multispan2\hfil T{\sc otal}\hfil\cr
\noalign{\vskip -4pt}
\omit&\multispan2\hrulefill&\multispan2\hrulefill&\multispan2\hrulefill&\multispan2\hrulefill&\multispan2\hrulefill\cr
\omit\hfil Z{\sc one}\hfil&  \% &$N$ &  \% &$N$ &  \% &$N$ &  \% &$N$ 
& \% &$N$ \cr
\noalign{\vskip 3pt\hrule\vskip 5pt}
857 deep&    *91.2& *73& 2.5& *2& *1.2& *1& *5.0& *4& 100& *80\cr
*** medium&  *95.5& 255& 1.9& *5& *0.0& *0& *2.6& *7& 100& 267\cr
*** shallow &*94.6& 697& 1.6& 12& *0.8& *6& *3.0& 22& 100& 737\cr
\noalign{\vskip 4pt}
545 deep &   *76.5& *39& 7.8& *4& *2.0& *1& 13.7& *7& 100& *51\cr
*** medium & *91.1& 143& 2.5& *4& *0.0& *0& *6.4& 10& 100& 157\cr
*** shallow &*91.8& 301& 2.1& *7& *0.9& *3& *5.2& 17& 100& 328\cr
\noalign{\vskip 4pt}
353 deep &   *81.6& *31& 2.6& *1& *5.3& *2& 10.5& *4& 100& *38\cr
*** medium & *87.0& *94& 0.9& *1& *2.8& *3& *9.3& 10& 100& 108\cr
*** shallow &*78.0& 170& 4.6& 10& *4.6& 10& 12.8& 28& 100& 218\cr
\noalign{\vskip 4pt}
217 deep &   *77.3& *17& 0.0& *0& 22.7& *5& *0.0& *0& 100& *22\cr
*** medium & *92.5& *62& 1.5& *1& *1.5& *1& *4.5& *3& 100& *67\cr
*** shallow &*88.5& 170& 0.5& *1& *4.2& *8& *6.8& 13& 100& 192\cr
\noalign{\vskip 4pt}
143 deep &   *86.7& *13& 0.0& *0& 13.3& *2& *0.0& *0& 100& *15\cr
*** medium & 100.0& *48& 0.0& *0& *0.0& *0& *0.0& *0& 100& *48\cr
*** shallow &*96.8& 182& 0.5& *1& *1.6& *3& *1.1& *2& 100& 188\cr
\noalign{\vskip 4pt}
100 deep &   *77.8& *14& 0.0& *0& 22.2& *4& *0.0& *0& 100& *18\cr
*** medium & 100.0& *45& 0.0& *0& *0.0& *0& *0.0& *0& 100& *45\cr
*** shallow &*93.9& 154& 0.0& *0& *3.7& *6& *2.4& *4& 100& 164\cr
\noalign{\vskip 5pt\hrule\vskip 3pt}}}
\endPlancktablewide                 
\endgroup
\end{table*}                        

%% file: table_source_zone_stat_planck_numbercounts_new.tex
%
\begin{table*}[tmb]
\begingroup
\newdimen\tblskip \tblskip=5pt
\caption{Number of extragalactic sources by zone before (after) the 
completeness cut, and the surface area of the zones.}
\label{tab:sourcenumber}
\nointerlineskip
\vskip -3mm
\footnotesize
\setbox\tablebox=\vbox{
   \newdimen\digitwidth
   \setbox0=\hbox{\rm 0}
    \digitwidth=\wd0
    \catcode`*=\active
    \def*{\kern\digitwidth}
    \newdimen\signwidth
    \setbox0=\hbox{+}
    \signwidth=\wd0
    \catcode`!=\active
    \def!{\kern\signwidth}
\halign{\hbox to 2cm{#\leaderfil}\tabskip=2em&
     \hfil#\hfil\tabskip=1.5em&
     \hfil#\hfil&
     \hfil#\hfil&
     \hfil#\hfil\tabskip=2em&
     \hfil#\hfil\tabskip=1em&
     \hfil#\hfil&
     \hfil#\hfil&
     \hfil#\hfil\cr
\noalign{\doubleline}
\omit\hfil$\nu$\hfil&\multispan4\hfil $N$ B{\sc efore} (A{\sc fter}) 
C{\sc ompleteness} C{\sc ut}\hfil&\multispan4\hfil S{\sc urface} 
A{\sc rea} [deg$^2$]\hfil\cr
\noalign{\vskip -3pt}
\omit&\multispan4\hrulefill&\multispan4\hrulefill\cr
\omit\hfil[GHz]\hfil&Deep&Medium&Shallow&Total&Deep&Medium&Shallow&Total\cr
\noalign{\vskip 3pt\hrule\vskip 5pt}
857& 77 (24)& 262 (115)& 719 (313)& 1058 (452)& *880& 2288& *9800& 12969\cr
545& 46 (*8)& 153 (*69)& 318 (143)& *517 (220)& *874& 2324& *9551& 12749\cr
353& 35 (14)& 104 (*59)& 198 (151)& *337 (224)& 1086& 2971& 12373& 16431\cr
217& 17 (15)& *65 (*57)& 183 (*71)& *265 (143)& 1104& 3169& 11900& 16174\cr
143& 13 (*8)& *48 (*44)& 184 (*90)& *245 (142)& 1111& 2972& 11977& 16061\cr
100& 14 (*6)& *45 (*39)& 158 (*77)& *217 (122)& 1072& 2870& 12611& 16554\cr
\noalign{\vskip 5pt\hrule\vskip 3pt}}}
\endPlancktablewide                 
\endgroup
\end{table*}                        

%% file: table_numbers_planck_numbercounts_new.tex
%
\begin{table*}[tmb]
\begingroup
\newdimen\tblskip \tblskip=5pt
\caption{Number of extragalactic sources used in the deep (D), medium 
(M), and shallow (S) number counts, and the corresponding number of 
unidentified sources.}
\label{tab:numbers}
\nointerlineskip
\vskip -3mm
\footnotesize
\setbox\tablebox=\vbox{
   \newdimen\digitwidth
   \setbox0=\hbox{\rm 0}
    \digitwidth=\wd0
    \catcode`*=\active
    \def*{\kern\digitwidth}
    \newdimen\signwidth
    \setbox0=\hbox{+}
    \signwidth=\wd0
    \catcode`!=\active
    \def!{\kern\signwidth}
\halign{#\hfil\tabskip=0.5em&
     \hfil#\hfil\tabskip=1em&
     \hfil#\hfil\tabskip=0.5em&
     \hfil#\hfil&
     \hfil#\hfil\tabskip=1.0em&
     \hfil#\hfil\tabskip=0.5em&
     \hfil#\hfil&
     \hfil#\hfil\tabskip=1.0em&
     \hfil#\hfil\tabskip=0.5em&
     \hfil#\hfil&
     \hfil#\hfil\tabskip=1.0em&
     \hfil#\hfil\tabskip=0.5em&
     \hfil#\hfil&
     \hfil#\hfil\tabskip=1.0em&
     \hfil#\hfil\tabskip=0.5em&
     \hfil#\hfil&
     \hfil#\hfil\tabskip=1.0em&
     \hfil#\hfil\tabskip=0.5em&
     \hfil#\hfil&
     \hfil#\hfil\tabskip=0pt\cr
\noalign{\doubleline}
\omit&&\multispan{18}\hfil $N_{\rm used}$/$N_{\rm unidentified}$\hfil\cr
\noalign{\vskip -3pt}
\omit&&\multispan{18}\hrulefill\cr
\multispan2\hfil $S_\nu$ 
[Jy]\hfil&\multispan3\hfil857\,GHz\hfil&\multispan3\hfil545\,GHz\hfil&\multispan3\hfil353\,GHz\hfil& 
\multispan3\hfil217\,GHz\hfil&\multispan3\hfil143\,GHz\hfil&\multispan3\hfil100\,GHz\hfil\cr
\noalign{\vskip -3pt}
\omit&&\multispan3\hrulefill&\multispan3\hrulefill&\multispan3\hrulefill&\multispan3\hrulefill&\multispan3\hrulefill&\multispan3\hrulefill\cr
\omit&&D&M&S&D&M&S&D&M&S&D&M&S&D&M&S&D&M&S\cr
\noalign{\vskip 3pt\hrule\vskip 5pt}
*0.398 
&*0.303--*0.480&\dots&\dots&\dots&\dots&\dots&\dots&\dots&\dots& 
\dots&  7/0& 28/1& 92/9&  3/0& 12/0& 81/1&\dots&\dots&\dots\cr
*0.631 &*0.480--*0.762&\dots&\dots&\dots&\dots&\dots&\dots&  8/0& 
31/4& 83/14&  3/0& 13/1& 34/2&  2/0& 13/0& 43/0&  4/0& 15/0& 68/3\cr
*1.000 &*0.762--*1.207&\dots&\dots&\dots&\dots&\dots&\dots&  4/0& 
18/2& 38/*7&  4/0& *8/0& 21/1&  4/0& 10/0& 25/0&  4/0& 11/0& 47/1\cr
*1.585 &*1.207--*1.913&\dots&\dots&\dots& *5/0& 26/6& 78/9&  1/1& 
*6/0& 14/*0&  1/0& *4/0& 11/0&  2/0& *4/0& 12/0&  2/0& *7/0& 13/0\cr
*2.512 &*1.913--*3.032& 11/0& 31/1&139/4& *0/0& 20/1& 28/1&  1/0& 
*2/0& *8/*0&\dots& *1/0& *3/0&\dots& *3/0& *5/0&\dots& *3/0& 11/0\cr
*3.981 &*3.032--*4.805& *7/0& 33/2&*95/4& *1/0& 16/1& 20/1&\dots& 
*1/0& *6/*0&\dots& *2/0& *1/0&\dots& *1/0& *3/0&\dots& *2/0& *3/0\cr
*6.310 &*4.805--*7.615& *3/0& 22/2&*41/7& *1/0& *6/0& 
*8/1&\dots&\dots& \dots&\dots& *1/0& *1/0&\dots& *1/0& 
*1/0&\dots&\dots& *2/0\cr
10.000 &*7.615--12.069& *2/0& 20/1&*23/1& *1/0& *1/0& *5/0&\dots& 
*1/0& \dots&\dots&\dots&\dots&\dots&\dots&\dots&\dots& *1/0& *1/0\cr
15.849 &12.069--22.801& *1/0& *9/0&*15/0& *0/0& *0/0& 
*4/0&\dots&\dots& 
*2/*0&\dots&\dots&\dots&\dots&\dots&\dots&\dots&\dots&\dots\cr
\noalign{\vskip 5pt\hrule\vskip 3pt}}}
\endPlancktablewide                 
\endgroup
\end{table*}                        

%% file: table_number_counts_highfreq_new.tex
%
\begin{table*}[tmb]
\begingroup
\newdimen\tblskip \tblskip=5pt
\caption{\Planck\ number counts at 353, 545, and 857\,GHz.}
\label{tab:number_counts_highfreq}
\nointerlineskip
\vskip -3mm
\footnotesize
\setbox\tablebox=\vbox{
   \newdimen\digitwidth
   \setbox0=\hbox{\rm 0}
    \digitwidth=\wd0
    \catcode`*=\active
    \def*{\kern\digitwidth}
    \newdimen\signwidth
    \setbox0=\hbox{+}
    \signwidth=\wd0
    \catcode`!=\active
    \def!{\kern\signwidth}
\halign{#\hfil\tabskip=1em&
     \hfil#\hfil\tabskip=2.5em&
     \hfil#\hfil\tabskip=1.2em&
     \hfil#\hfil\tabskip=2.5em&
     \hfil#\hfil\tabskip=1.2em&
     \hfil#\hfil\tabskip=2.5em&
     \hfil#\hfil\tabskip=1.2em&
     \hfil#\hfil\tabskip=0pt\cr
\noalign{\doubleline}
\omit&&\multispan2\hfil 857\,GHz\hfil&\multispan2\hfil 
545\,GHz\hfil&\multispan2\hfil 353\,GHz\hfil\cr
\noalign{\vskip -3pt}
\omit&&\multispan2\hrulefill&\multispan2\hrulefill&\multispan2\hrulefill\cr
\noalign{\vskip 1pt}
\multispan2\hfil$S_\nu$ [Jy]\hfil&${dN\over 
dS_\nu}S_\nu^{2.5}$&$N{>}S_\nu$&${dN\over 
dS_\nu}S_\nu^{2.5}$&$N{>}S_\nu$&${dN\over 
dS_\nu}S_\nu^{2.5}$&$N{>}S_\nu$\cr
\noalign{\vskip -3pt}
\multispan2\hrulefill\cr
\omit\hfil Mean\hfil&Range&    [Jy$^{1.5}$\,sr$^{-1}$] & [sr$^{-1}$] 
&[Jy$^{1.5}$\,sr$^{-1}]$ &[sr$^{-1}$]&[Jy$^{1.5}$\,sr$^{-1}$] 
&[sr$^{-1}$]  \cr
\noalign{\vskip 3pt\hrule\vskip 5pt}
*0.631& *0.480--*0.762&           \dots&         \dots& 
\dots&        \dots& $30.6\pm*3.7$& $50.1\pm3.3$\cr
*1.000& *0.762--*1.207&           \dots&         \dots& 
\dots&        \dots& $26.4\pm*4.0$& $21.0\pm2.1$\cr
*1.585& *1.207--*1.913&           \dots&         \dots& 
$131.4\pm21.4$& $60.3\pm4.1$& $17.6\pm*4.1$& $*8.4\pm1.3$\cr
*2.512& *1.913--*3.032& $466.2\pm*70.5$& $127.7\pm6.0$& 
$104.6\pm20.4$& $28.9\pm2.7$& $18.3\pm*5.7$& $*4.2\pm0.9$\cr
*3.981& *3.032--*4.805& $613.6\pm*96.6$& $*71.8\pm4.4$& 
$160.2\pm33.8$& $16.3\pm2.1$& $23.3\pm*9.0$& $*2.0\pm0.6$\cr
*6.310& *4.805--*7.615& $573.4\pm103.4$& $*35.0\pm3.0$& 
$129.4\pm37.5$& $*6.7\pm1.3$&         \dots&        \dots\cr
10.000& *7.615--12.069& $755.2\pm150.3$& $*17.7\pm2.1$& 
$119.5\pm47.8$& $*2.8\pm0.9$& $13.2\pm13.3$& $*0.6\pm0.3$\cr
15.849& 12.069--22.801& $837.1\pm200.5$& $**6.3\pm1.3$& 
$136.2\pm70.4$& $*1.0\pm0.5$& $52.9\pm37.6$& $*0.4\pm0.3$\cr
\noalign{\vskip 5pt\hrule\vskip 3pt}}}
\endPlancktablewide                 
\endgroup
\end{table*}                        

%% file: table_number_counts_lowfreq_new.tex
%
\begin{table*}[tmb]
\begingroup
\newdimen\tblskip \tblskip=5pt
\caption{\Planck\ number counts at 100, 143, and 217 GHz.}
\label{tab:number_counts_lowfreq}
\nointerlineskip
\vskip -3mm
\footnotesize
\setbox\tablebox=\vbox{
   \newdimen\digitwidth
   \setbox0=\hbox{\rm 0}
    \digitwidth=\wd0
    \catcode`*=\active
    \def*{\kern\digitwidth}
    \newdimen\signwidth
    \setbox0=\hbox{+}
    \signwidth=\wd0
    \catcode`!=\active
    \def!{\kern\signwidth}
\halign{#\hfil\tabskip=1em&
     \hfil#\hfil\tabskip=2.5em&
     \hfil#\hfil\tabskip=1.2em&
     \hfil#\hfil\tabskip=2.5em&
     \hfil#\hfil\tabskip=1.2em&
     \hfil#\hfil\tabskip=2.5em&
     \hfil#\hfil\tabskip=1.2em&
     \hfil#\hfil\tabskip=0pt\cr
\noalign{\doubleline}
\omit&&\multispan2\hfil 217\,GHz\hfil&\multispan2\hfil 
143\,GHz\hfil&\multispan2\hfil 100\,GHz\hfil\cr
\noalign{\vskip -3pt}
\omit&&\multispan2\hrulefill&\multispan2\hrulefill&\multispan2\hrulefill\cr
\noalign{\vskip 1pt}
\multispan2\hfil$S_\nu$ [Jy]\hfil&${dN\over 
dS_\nu}S_\nu^{2.5}$&$N{>}S_\nu$&${dN\over 
dS_\nu}S_\nu^{2.5}$&$N{>}S_\nu$&${dN\over 
dS_\nu}S_\nu^{2.5}$&$N{>}S_\nu$\cr
\noalign{\vskip -3pt}
\multispan2\hrulefill\cr
\omit\hfil Mean\hfil&Range&    [Jy$^{1.5}$\,sr$^{-1}$]& 
[sr$^{-1}$]&[Jy$^{1.5}$\,sr$^{-1}]$&[sr$^{-1}$]&[Jy$^{1.5}$\,sr$^{-1}$]&[sr$^{-1}$] 
\cr
\noalign{\vskip 3pt\hrule\vskip 5pt}
*0.398& 0.303--*0.480& $16.5\pm2.0 $& $54.3\pm3.54$& $*8.5\pm*1.1$& 
$44.3\pm2.95$&       \dots &       \dots  \cr
*0.631& 0.480--*0.762& $11.5\pm1.9 $& $23.0\pm2.21$& $13.7\pm*2.1$& 
$28.1\pm2.46$& $21.9\pm*3.0$& $43.7\pm3.14$\cr
*1.000& 0.762--*1.207& $14.6\pm2.8 $& $12.0\pm1.58$& $17.4\pm*3.1$& 
$15.0\pm1.77$& $29.1\pm*4.4$& $22.9\pm2.22$\cr
*1.585& 1.207--*1.913& $13.6\pm3.6 $& $*5.1\pm1.01$& $15.4\pm*3.8$& 
$*6.7\pm1.17$& $18.8\pm*4.3$& $*9.1\pm1.35$\cr
*2.512& 1.913--*3.032& $*6.8\pm3.4 $& $*1.8\pm0.61$& $13.6\pm*5.0$& 
$*3.1\pm0.79$& $23.2\pm*6.5$& $*4.6\pm0.95$\cr
*3.981& 3.032--*4.805& $10.1\pm5.9 $& $*1.0\pm0.45$& $13.6\pm*6.9$& 
$*1.4\pm0.54$& $16.5\pm*7.5$& $*1.8\pm0.59$\cr
*6.310& 4.805--*7.615& $13.5\pm9.6 $& $*0.4\pm0.29$& $13.6\pm*9.7$& 
$*0.6\pm0.35$& $13.2\pm*9.4$& $*0.8\pm0.40$\cr
10.000& 7.615--12.069&       \dots &         \dots & $13.6\pm13.6$& 
$*0.2\pm0.20$& $26.3\pm18.7$& $*0.4\pm0.28$\cr
\noalign{\vskip 5pt\hrule\vskip 3pt}}}
\endPlancktablewide                 
\endgroup
\end{table*}                        

%% file: table_number_counts_highfreq_dusty_new.tex
%
\begin{table*}[tmb]
\begingroup
\newdimen\tblskip \tblskip=5pt
\caption{\Planck\ number counts of dusty galaxies at 217, 353, 545, 
and 857\,GHz.}
\label{tab:number_counts_dusty}
\nointerlineskip
\vskip -3mm
\footnotesize
\setbox\tablebox=\vbox{
   \newdimen\digitwidth
   \setbox0=\hbox{\rm 0}
    \digitwidth=\wd0
    \catcode`*=\active
    \def*{\kern\digitwidth}
    \newdimen\signwidth
    \setbox0=\hbox{+}
    \signwidth=\wd0
    \catcode`!=\active
    \def!{\kern\signwidth}
\halign{#\hfil\tabskip=1em&
     \hfil#\hfil\tabskip=2.5em&
     \hfil#\hfil\tabskip=1.2em&
     \hfil#\hfil&
     \hfil#\hfil&
     \hfil#\hfil\tabskip=0pt\cr
\noalign{\doubleline}
\multispan2\hfil $S_\nu$ [Jy]\hfil&\multispan4\hfil ${dN\over dS_\nu} 
S_\nu^{2.5}$ [Jy$^{1.5}$\,sr\mo]\hfil\cr
\noalign{\vskip -3pt}
\multispan2\hrulefill&\multispan4\hrulefill\cr
\noalign{\vskip 2pt}
\omit\hfil Mean\hfil&Range&857\,GHz&545\,GHz&353\,GHz&217\,GHz\cr
\noalign{\vskip 3pt\hrule\vskip 5pt}
*0.398& *0.303--*0.480&           \dots&          \dots& 
\dots& $3.2\pm0.7$\cr
*0.631& *0.480--*0.762&           \dots&          \dots& 
$23.6\pm*3.1$& $2.3\pm0.8$\cr
*1.000& *0.762--*1.207&           \dots&          \dots& 
$18.5\pm*3.2$& $1.8\pm0.9$\cr
*1.585& *1.207--*1.913&           \dots& $129.0\pm21.1$& 
$12.5\pm*3.4$&       \dots\cr
*2.512& *1.913--*3.032& $463.6\pm*70.2$& $100.3\pm19.8$& 
$11.7\pm*4.5$& $1.7\pm1.7$\cr
*3.981& *3.032--*4.805& $609.0\pm**6.0$& $151.5\pm32.5$& 
$13.3\pm*6.7$&       \dots\cr
*6.310& *4.805--*7.615& $573.4\pm103.4$& $129.4\pm37.5$& 
\dots&       \dots\cr
10.000& *7.615--12.069& $755.2\pm150.3$& $119.5\pm47.8$& 
$13.2\pm13.3$&       \dots\cr
15.849& 12.069--22.801& $837.1\pm200.5$& $136.2\pm70.4$& 
$26.4\pm26.5$&       \dots\cr
\noalign{\vskip 5pt\hrule\vskip 3pt}}}
\endPlancktablewide                 
\endgroup
\end{table*}

%% file: table_number_counts_lowfreq_synch_new.tex
%
\begin{table*}[tmb]
\begingroup
\newdimen\tblskip \tblskip=5pt
\caption{\Planck\ number counts of synchrotron galaxies at 100, 143, 
217, 353, 545, and 857\,GHz.}
\label{tab:number_counts_synch}
\nointerlineskip
\vskip -3mm
\footnotesize
\setbox\tablebox=\vbox{
   \newdimen\digitwidth
   \setbox0=\hbox{\rm 0}
    \digitwidth=\wd0
    \catcode`*=\active
    \def*{\kern\digitwidth}
    \newdimen\signwidth
    \setbox0=\hbox{+}
    \signwidth=\wd0
    \catcode`!=\active
    \def!{\kern\signwidth}
\halign{#\hfil\tabskip=1em&
     \hfil#\hfil\tabskip=2.5em&
     \hfil#\hfil\tabskip=1.2em&
     \hfil#\hfil&
     \hfil#\hfil&
     \hfil#\hfil&
     \hfil#\hfil&
     \hfil#\hfil\tabskip=0pt\cr
\noalign{\doubleline}
\multispan2\hfil $S_\nu$ [Jy]\hfil&\multispan6\hfil ${dN\over dS_\nu} 
S_\nu^{2.5}$ [Jy$^{1.5}$\,sr\mo]\hfil\cr
\noalign{\vskip -3pt}
\multispan2\hrulefill&\multispan6\hrulefill\cr
\noalign{\vskip 2pt}
\omit\hfil 
Mean\hfil&Range&857\,GHz&545\,GHz&353\,GHz&217\,GHz&143\,GHz&100\,GHz\cr
\noalign{\vskip 3pt\hrule\vskip 5pt}
*0.398& *0.303--*0.480&        \dots&        \dots&         \dots& 
$13.3\pm1.7$& $*8.5\pm*1.1$&  \dots \cr
*0.631& *0.480--*0.762&        \dots&        \dots& $*7.0\pm*1.4$& 
$*9.2\pm1.6$& $13.5\pm*2.1$& $21.9\pm*3.0$\cr
*1.000& *0.762--*1.207&        \dots&        \dots& $*7.9\pm*2.0$& 
$12.8\pm2.6$& $16.9\pm*3.1$& $28.6\pm*4.3$\cr
*1.585& *1.207--*1.913&        \dots& $2.4\pm1.7 $& $*5.0\pm*2.1$& 
$13.6\pm3.6$& $15.4\pm*3.8$& $18.8\pm*4.3$\cr
*2.512& *1.913--*3.032& $2.6\pm2.6 $& $4.3\pm3.1 $& $*6.7\pm*3.4$& 
$*5.1\pm3.0$& $13.6\pm*5.0$& $23.2\pm*6.5$\cr
*3.981& *3.032--*4.805& $4.5\pm4.6 $& $8.7\pm6.2 $& $10.0\pm*5.8$& 
$10.1\pm5.9$& $13.6\pm*6.9$& $16.5\pm*7.5$\cr
*6.310& *4.805--*7.615&        \dots&        \dots&         \dots& 
$13.5\pm9.6$& $13.6\pm*9.7$& $13.2\pm*9.4$\cr
10.000& *7.615--12.069&        \dots&        \dots&         \dots& 
\dots& $13.6\pm13.6$& $26.3\pm18.7$\cr
15.849& 12.069--22.801&        \dots&        \dots& $26.4\pm26.5$& 
\dots&         \dots&         \dots\cr
\noalign{\vskip 5pt\hrule\vskip 3pt}}}
\endPlancktablewide                 
\endgroup
\end{table*}                        

%% file: euclidean_levels_p_planck2012wg6.tex
%
%
\begin{table}[tmb]                 
\begingroup
\newdimen\tblskip \tblskip=5pt
\caption{Values of $p$, the Euclidean plateau level (in Jy$^{1.5}$\,sr$^{-1}$) from and \Planck\ and other satellite data. The column "flag" indicates the nature of the sources (a=all; d=dusty; s=synch).}                          
\label{tab:valuesp}                            
\nointerlineskip
\vskip -3mm
\footnotesize
\setbox\tablebox=\vbox{
  \newdimen\digitwidth 
  \setbox0=\hbox{\rm 0} 
   \digitwidth=\wd0 
   \catcode`*=\active 
   \def*{\kern\digitwidth}
   \newdimen\signwidth 
   \setbox0=\hbox{+} 
   \signwidth=\wd0 
   \catcode`!=\active 
   \def!{\kern\signwidth}
\halign{ # & # & # & # & #\cr                        
\noalign{\doubleline}
     $\nu$  & $p$  & Flag & Experiment & Reference \cr
      $[$GHz$]$   & [Jy$^{1.5}$\,sr$^{-1}]$ &  & & \cr
\noalign{\vskip 3pt\hrule\vskip 5pt}
  100 &     22 $\pm$     5 & s & Planck & PlanckCollab2012 \cr
  143 &     15 $\pm$     1 & s & Planck & PlanckCollab2012 \cr
  217 &     11 $\pm$     3 & s & Planck & PlanckCollab2012 \cr
  353 &     21 $\pm$     4 & a & Planck & PlanckCollab2012 \cr
  545 &    128 $\pm$    17 & a & Planck & PlanckCollab2012 \cr
  857 &    627 $\pm$   152 & d & Planck & PlanckCollab2012 \cr
  353 &     15 $\pm$     6 & d & Planck & PlanckCollab2012 \cr
  545 &    125 $\pm$    15 & d & Planck & PlanckCollab2012 \cr
  353 &      7 $\pm$     9 & s & Planck & PlanckCollab2012 \cr
  545 &      3 $\pm$     3 & s & Planck & PlanckCollab2012 \cr
   30 &     38 $\pm$     8 & s & Planck & PlanckCollab2011 \cr
   44 &     29 $\pm$    12 & s & Planck & PlanckCollab2011 \cr
   70 &     25 $\pm$     5 & s & Planck & PlanckCollab2011 \cr
  100 &     20 $\pm$     3 & s & Planck & PlanckCollab2011 \cr
  143 &     13 $\pm$     2 & s & Planck & PlanckCollab2011 \cr
  217 &     10 $\pm$     2 & s & Planck & PlanckCollab2011 \cr
   33 &     31 $\pm$     1 & s & WMAP & Wright2009 \cr
   23 &     37 $\pm$     7 & s & WMAP & Massardi2009 \cr
   33 &     37 $\pm$    22 & s & WMAP & Massardi2009 \cr
   41 &     32 $\pm$    15 & s & WMAP & Massardi2009 \cr
   61 &     19 $\pm$     6 & s & WMAP & Massardi2009 \cr
12000 &     63 $\pm$     1 & d & IRAS & Soifer91Bertin97 \cr
 5000 &    891 $\pm$     1 & d & IRAS & Soifer91Bertin97 \cr
 3000 &   3019 $\pm$     1 & d & IRAS & Soifer91Bertin97 \cr
12500 &     43 $\pm$     5 & d & Spitzer & Bethermin2010 \cr
 4285 &   2252 $\pm$   143 & d & Spitzer & Bethermin2010 \cr
 1875 &   5261 $\pm$   743 & d & Spitzer & Bethermin2010 \cr
\noalign{\vskip 5pt\hrule\vskip 3pt}}}
\endPlancktable                    
\endgroup
\end{table}                        

%% file: rho_nu_planck2012wg6.tex
%
%
\begin{table}[tmb]                 
\begingroup
\newdimen\tblskip \tblskip=5pt
\caption{Values of $\rho_{\nu}$, the luminosity density of dusty galaxies (in $h$\,L$_{\odot}$\,Mpc$^{-3}$), inferred from the Euclidean plateau level $p$ and scaled to the luminosity density at 850~$\mu$m (upper values) and 60~$\mu$m (lower values).}                          
\label{tab:rhonu}                            
\nointerlineskip
\vskip -3mm
\footnotesize
\setbox\tablebox=\vbox{
  \newdimen\digitwidth 
  \setbox0=\hbox{\rm 0} 
   \digitwidth=\wd0 
   \catcode`*=\active 
   \def*{\kern\digitwidth}
   \newdimen\signwidth 
   \setbox0=\hbox{+} 
   \signwidth=\wd0 
   \catcode`!=\active 
   \def!{\kern\signwidth}
\halign{ # & # & # \cr                        
\noalign{\doubleline}
$\nu$  & $\rho_{\nu}$ [scaled 850\,$\mu$m]  & $\rho_{\nu}$ [scaled 60\,$\mu$m]  \cr
$[$GHz$]$   & [$h$\,L$_{\odot}$\,Mpc$^{-3}$] & [$h$\,L$_{\odot}$\,Mpc$^{-3}$] \cr
\noalign{\vskip 3pt\hrule\vskip 5pt}
353 &  $( 1.99 \pm 0.57) \times 10^5$ & $ ( 2.69 \pm 0.77) \times 10^6$ \cr
545 &  $( 8.18 \pm 0.69) \times 10^5$ & $ ( 1.10 \pm 0.09) \times 10^7$ \cr
857 &  $( 2.39 \pm 0.39) \times 10^6$ & $ ( 3.23 \pm 0.52) \times 10^7$ \cr
\noalign{\vskip 5pt\hrule\vskip 3pt}}}
\endPlancktable                    
\endgroup
\end{table}                        